\documentclass{aa}

\usepackage[utf8]{inputenc}
\usepackage[english]{babel}
\usepackage{stmaryrd}
\usepackage[dvipsnames,hyperref]{xcolor}
\usepackage{tikz}
\usepackage{xspace}
\usepackage{natbib}
\bibpunct{(}{)}{;}{a}{}{,}
\usepackage{amsmath}
\usepackage{amsfonts}
\usepackage{amssymb}
\usepackage{mathtools}
\usepackage{graphicx}
\usepackage[colorinlistoftodos]{todonotes}
\usepackage[colorlinks=true, allcolors=blue]{hyperref}
\usepackage{mathrsfs}
\usepackage{bigints}
\usepackage[normalem]{ulem}
\usepackage{subcaption}
\usepackage{pdflscape}
\usepackage{nidanfloat}
\usepackage{caption}

\DeclarePairedDelimiterX{\Paren}[1]{(}{)}{#1}
\DeclarePairedDelimiterX{\Brace}[1]{\{}{\}}{#1}
\DeclarePairedDelimiterX{\Brack}[1]{[}{]}{#1}
\DeclarePairedDelimiterX{\Abs}[1]{\rvert}{\lvert}{#1}
\DeclarePairedDelimiterX{\Norm}[1]{\lVert}{\rVert}{#1}
\DeclarePairedDelimiterX{\Group}[1]{\lgroup}{\rgroup}{#1}
\DeclarePairedDelimiterX{\Avg}[1]{\langle}{\rangle}{#1}
\DeclarePairedDelimiterX{\Round}[1]{\lfloor}{\rceil}{#1}
\DeclarePairedDelimiterX{\Floor}[1]{\lfloor}{\rfloor}{#1}
\DeclarePairedDelimiterX{\Ceil}[1]{\lceil}{\rceil}{#1}
\DeclarePairedDelimiterX{\Inner}[2]{\langle}{\rangle}{#1,#2}
\DeclarePairedDelimiterXPP{\IndexedList}[3]{}\{\}{_{#2,\ldots,#3}}{#1}
\DeclarePairedDelimiterX{\IntRange}[1]{\llbracket}{\rrbracket}{#1}

\DeclarePairedDelimiterXPP{\List}[3]{}{\{}{\}}{_{#2,\ldots,#3}}{#1}

\newcommand*{\delimsize}{}
\newcommand*{\SuchThat}{\:\delimsize\vert\:} 

\newcommand*{\V}[1]{\boldsymbol{#1}} 
\newcommand*{\M}[1]{\mathbf{#1}} 
\newcommand*{\Tag}[1]{{\text{#1}}} 
\newcommand*{\TransposeLetter}{\top}
\newcommand*{\T}{^{\TransposeLetter}} 

\makeatletter
\newcommand*{\Tr}[1]{#1_\Tag{gt}}
\newcommand*{\Es}[1]{\widehat{#1}}
\newcommand*{\Cvlv}[1]{#1}
 
\newcommand*{\To}[3]{#1 \in \IntRange{#2,#3}}

\newcommand*{\Var}{\textrm{Var}}
\newcommand*{\Expect}{\mathbb{E}}

\newcommand*{\pol}{\Tag{p}}   
\newcommand*{\unpol}{\Tag{u}} 

\newcommand*{\Iu}{I^\unpol}
\newcommand*{\Ip}{I^\pol}

\newcommand*{\Is}{I}
\newcommand*{\Qs}{Q}
\newcommand*{\Us}{U}
\newcommand*{\Vs}{V}

\newcommand*{\VIu}{\V{I}^\unpol}
\newcommand*{\VIp}{\V{I}^\pol}
\newcommand*{\VIs}{\V{I}}
\newcommand*{\VQs}{\V{Q}}
\newcommand*{\VUs}{\V{U}}

\newcommand*{\X}{X}
\newcommand*{\VX}{\V{\X}} 

\newcommand*{\Rotation}[1]{\M{T}\Paren*{#1}}
\newcommand*{\Mueller}[1]{\M{M}^\Tag{#1}}

\newcommand*{\ParallacticAngle}{\theta^\Tag{par}}
\newcommand*{\AltitudeAngle}{\theta^\Tag{alt}}

\newcommand*{\DerotatorAngle}{\theta^\Tag{der}}

\newcommand*{\AllAngles}{\Theta}

\newcommand*{\from}{\,{:}\:}

\newcommand*{\CompactFrac}[2]{\ensuremath{^{#1}\!/\!_{#2}}}

\newcommand*{\FeasibleSet}{\mathcal{C}}

\newcommand*{\R}{\mathbb{R}}

\newcommand*{\irm}{\mathrm{i}}
\newcommand*{\erm}{\mathrm{e}}

\newcommand*{\sro}{\sigma_{\text{ro}}}

\DeclareMathOperator*{\argmin}{arg\,min}

\DeclareMathOperator{\Diag}{diag}

\newcommand*{\ie}{\emph{i.e.}\xspace}
\newcommand*{\eg}{\emph{e.g.}\xspace}
\newcommand*{\etc}{\emph{etc.}\xspace}
\newcommand*{\hide}[1]{}

\newcommand*{\smallfrac}[2]{{\textstyle\frac{#1}{#2}}}

\newcommand{\LDrevised}[1]{#1}%
\newcommand{\LD}[1]{#1}
\newcommand{\CHANGE}[1]{\textcolor{OrangeRed}{#1}}

\newcommand*{\NewBeg}{\color{black}}
\newcommand*{\NewEnd}{\color{black}}

\begin{document}

\title{RHAPSODIE:  Reconstruction of High-contrAst Polarized SOurces and Deconvolution for cIrcumstellar Environments}
\author{L. Denneulin$^{1,2}$
	\and M. Langlois$^1$
	\and É. Thiébaut$^1$
	\and N. Pustelnik$^{2}$}

\institute{Univ Lyon, Univ Lyon1, Ens de Lyon, CNRS, Centre de Recherche Astrophysique de Lyon, UMR5574, F-69230 Saint-Genis-Laval, France
            \and Univ Lyon, ENS de Lyon, Univ Claude Bernard Lyon 1, CNRS, Laboratoire de Physique, F-69342 Lyon, France }

\abstract {Polarimetric imaging is one of the most effective techniques for high-contrast imaging and characterization of circumstellar environments. These environments can be characterized through direct-imaging polarimetry at near-infrared wavelengths. The Spectro-Polarimetric High-contrast Exoplanet REsearch (SPHERE)/IRDIS instrument installed on the Very Large Telescope in its dual-beam polarimetric imaging (DPI) mode, offers the capability to acquire polarimetric images at high contrast and high angular resolution. However dedicated image processing is needed to get rid of the contamination by the stellar light, of instrumental polarization effects, and of the blurring by the instrumental point spread function.}
{We aim to reconstruct and deconvolve the near-infrared polarization signal from circumstellar environments.}
{We use observations of these environments obtained with the high-contrast imaging infrared polarimeter SPHERE-IRDIS at the Very Large Telescope (VLT). We developed a new method to extract the polarimetric signal using an inverse approach method that benefits from the added knowledge of the detected signal formation process. The method includes weighted data fidelity term, smooth penalization\LD{,} and takes into account instrumental polarization} {The  method enables to accurately measure the polarized intensity and angle of linear polarization of circumstellar disks by taking into account the noise statistics and the convolution \LD{
by the instrumental point spread function}. It has the capability to use incomplete polarimetry cycles which enhance the sensitivity of the observations. The method improves the overall performances in particular for low SNR/small polarized flux compared to standard methods.}
{By increasing the sensitivity and including deconvolution, our method  will allow for more accurate studies of these disks morphology, especially in the innermost regions. It also will enable more accurate measurements of the angle of linear polarization at low SNR, which would lead to in-depth studies of dust properties. Finally, the method will enable more accurate measurements of the polarized intensity which is critical to construct scattering phase functions.}


\titlerunning{Reconstruction of high-contrast polarized circumstellar environments}
\authorrunning{Denneulin et al.}

\maketitle
\section{Introduction}

With the adaptive-optics-fed high-contrast imaging instruments GPI \citep{macintosh2014first} and SPHERE-IRDIS \citep{beuzit_sphere:_2019, dohlen2008infra}, we now have access to the spatial resolution and sensitivity required to observe in the near-infrared (NIR) circumstellar matter at small angular separations. Along with the Integral Field Spectrograph \citep[IFS;][]{claudi2008sphere} and the Zürich IMaging POLarimeter \citep[ZIMPOL;][]{schmid2018sphere} which can also be used to observe circumstellar environments in polarimetry, the IRDIS instrument is one of the three SPHERE instruments \citep{beuzit_sphere:_2019}. SPHERE/IRDIS is able to acquire two simultaneous images at two different  wavelengths, in a so-called Dual Band Imaging (DBI) mode \citep{vigan2014sphere}, or for two different polarizations, in a so-called Dual Polarimetry Imaging mode \citep{langlois2014high, de_boer_polarimetric_2020}. 
Both circumstellar disks and self-luminous giant exoplanets or companion brown dwarfs can be characterized by these new instruments in direct-imaging at these wavelengths.

The NIR polarimetric mode of SPHERE/IRDIS at the Very Large Telescope (VLT), which is described in \cite{beuzit_sphere:_2019, de_boer_polarimetric_2020, van_holstein_polarimetric_2020}, has proven to be very successful  for the detection of circumstellar disks in scattered light \citep{garufi2017three} and shows much promise for the characterization \LD{of} brown dwarfs \citep{van_holstein_combining_2017} and exoplanets \citep[][in prep.]{hostein2020prep} when they are surrounded by circumsubstellar disks.

Three particular types of circumstellar disks are studied: protoplanetary
disks, transition disks\LD{,} and debris disks. The observations of the
protoplanetary and transition disks morphology linked to hydrodynamical
simulations allows for the study of their formation scenario as in the study of
HD~142527 \citep{price_circumbinary_2018}, IM~Lup, RU~Lup
\citep{avenhaus_disks_2018}, GSC~07396-759 \citep{sissa2018new}. Their
observations are valuable because their shapes can be the signposts for the
formation of one or several exoplanets. In fact, during their formation, the
planets ``clean'' the dust off their orbits, creating gaps without dust such as
for RXJ~1615, MY~Lup, PDS~66 \citep{avenhaus_disks_2018}, PDS~70
\citep{keppler_discovery_2018, keppler_highly_2019,  haffert_two_2019}. The
planet formation scenario can be explained with hydrodynamical simulations as the
cases of HL~Tau \citep{dipierro_planet_2015}, HD~163296
\citep{pinte_kinematic_2019}. Due to gravity, exoplanets can also create spiral
arms, as in RY~Lup \citep{langlois_first_2018} and MWC~758
\citep{benisty2015asymmetric} where the presence of exoplanets is \LD{predicted by}
hydrodynamical simulations. Debris  disks are the oldest step in the evolution
of circumstellar disks, when there is already one or several planets in the
system and the gas is almost completely consumed. These disks are composed \LD{of}
dust and grain never accreted into the planets, as in the case of HR~4796A
\citep{perrin_polarimetry_2015, milli_optical_nodate}.

These environments can be observed with SPHERE in the near-infrared and in the
visible. Yet, such observations are difficult because of the high contrast
between the light of the environment and the residual light from the host star.
As a result, when acquiring images, the light of the environment is
contaminated by the diffraction stains from the host star. Two methods can be
used to disentangle the light of the disk from that of the star: Angular
Differential Imaging \citep[ADI;][]{marois2006angular} and Differential
Polarimetric Imaging \citep[DPI;][]{van_holstein_polarimetric_2020}. The ADI
technique uses the fact that the stellar residuals are fixed in the pupil plane
and the object of interest artificially rotates. The \LD{resulting} diversity makes
it possible to disentangle the light of the object of interest from the
residual light of the star. Yet, such a method does not allow a good
reconstruction of the disk morphology \LD{as it} is impacted by artifacts due to
self-subtraction. Moreover\LD{, the method fails} when the environment is \LD{nearly}
rotation invariant. The DPI observations allow \LD{having} access to the \LD{morphology of the disks},
without the artifacts, by using the difference of polarization state\LD{s}
between the light scattered by the environment and the light of the host star.

The state-of-the-art methods to process datasets in polarimetry, apart from the
calibration, are ``\emph{step-by-step}'' methods. First, the data are
transformed with the required translations and rotations to be easier to
process and the bad pixels are interpolated. \LD{Such interpolations introduce correlations that are not taken into account in the following processing}. Second, the \LD{interpolated} data are reduced to the
Stokes parameters \LD{which are directly related to the different polarization
states}. These reductions can be done with the double difference or the double
ratio \citep{tinbergen_astronomical_2005, avenhaus_structures_2014}. If the
double ratio takes into account the possibility of multiplicative instrumental
effects, none of these methods deal with the \LD{noise statistics}. This results in some
limitations in sensitivity in \LD{the} case of low Signal-to-noise Ratio (SNR). Last, 
\LD{deconvolution may be performed to get rid of the blurring by the instrumental point spread function.  As this is done without accounting for the noise statistics after all previous processing, the results are not optimal given the available data}. Still, these
state-of-the-art methods have proven over the years to be sufficiently
efficient to produce good quality results. However, studies of circumstellar
disks are often limited to analyses of the orientation (position angle and
inclination) and morphology (rings, gaps, cavities, and spiral arms) of the
disks \citep{muto2012discovery, quanz2013gaps, ginski2016direct,
de2016multiple}. Quantitative polarimetric measurements of circumstellar disks
and substellar companions are currently very challenging, because existing
data-reduction methods do not estimate properly the \LD{sources of the errors} from both
noise and detector calibration. They also \LD{require} complete polarimetric
cycles and do not account for the instrument convolution. For observations of
circumstellar disks \citep{van_holstein_polarimetric_2020}, calibrating  the
instrumental polarization effects with a sufficiently high accuracy already
yield to \LD{several} improvements.

Over the last decades in image processing, it has been proven that the
reconstruction of parameters of interest benefits from a global \LD{\emph{inverse
problems} approach}, taking into account \LD{the noise statistics and all instrumental effects}, rather
than ``\emph{step-by-step}'' procedures. \LD{In this framework, image restoration
methods rely on a physically grounded model of the data as a function of
the parameters of interest and express the estimated parameters of interest as
the constrained minimum of an objective function. This objective function is
generally the sum of a data fidelity term and of regularization terms
introduced to favor known priors. Depending on the convexity and on the
smoothness of the objective function, several numerical algorithms with
guarantees of convergence may be considered to seek the minimum}
\citep{nocedal_numerical_1999, Combettes_P_2007_inbook,
webster_wavelet-based_2016}. Such methods have been used over
decades in astrophysics for the physical parameters estimation
\citep{Titterington-1985-regularization}, mostly in adaptive optics
\citep{borde_highcontrast_2006} and radio-interferometry with the well known
algorithm CLEAN~\citep{hogbom1974aperture}. This last method has been the
starting point of a wide variety of algorithms, such as the algorithms SARA
\citep{carrillo_sparsity_2012} and Polca-SARA \citep{birdi_polca_2019} in
polarimetric radio-interferometry using more \LD{sophisticated} tools as non-smooth
penalizations. A non-smooth method was also used for images denoising with
curvelets~\citep{starck2003astronomical}. The minimization of a
co-log-likelihood was also used in the blind deconvolution of images \LD{convolved
by an unknown PSF with aberrations} \citep{thiebaut1995strict}. 
\LD{Learning methods have} been used more recently for the estimation of
the CMB \citep{planck_collaboration_planck_2016} and the imaging of the
supermassive blackhole \citep{akiyama2019first}.
In high contrast imaging, the
use of inverse problem methods is more recent. It has been used to perform
auto-calibration of the data \citep{berdeu_pic_2020} with the IFS/SPHERE. It
has also been used to reconstruct extended objects in total intensity by using 
ADI data \citep{pairet2019iterative, flasseur2019expaco} with the SPHERE
instrument. Yet, such reconstruction methods have not been used in polarimetric
high contrast direct imaging.

In the present work\LD{,} we describe in details the method and the benefits of the
use of an \emph{inverse problem} formalism for the reconstruction of
circumstellar environments observed in polarimetry with the instrument ESO/VLT
SPHERE IRDIS. In Section~\ref{sec:SOA}, we \LD{develop} the physical model of the
data obtained with the ESO/VLT SPHERE IRDIS instrument. This includes the
polarimetric \LD{parametrization, the convolution by
the PSF and the observing sequence}. In Section~\ref{sec:mymethod}, we describe RHAPSODIE\footnote{\LDrevised{The code of RHAPSODIE is available online at \url{https://github.com/LaurenceDenneulin/Rhapsodie.jl}.}} (Reconstruction
of High-contrAst Polarized Sources and Deconvolution for near-Infrared
Environments), \LD{the method we developed}. \LD{Section~4  is dedicated to the calibration of the detector, the instrument, and the instrumental polarization.} Finally, in
Section~\ref{sec:application}, we present the results obtained with \LD{RHAPSODIE}
on both simulated and astrophysical data.

\section{Modeling polarimetric data}
\label{sec:SOA}

\LD{The principal parameters of interest for studying circumstellar environments in
polarimetry are the intensity $\Ip$ of the linearly polarized light and the
corresponding polarization angle $\theta$ which are caused by the reflection of
the stellar light onto the circumstellar dust. By modulating the orientation
of the instrumental polarization, these parameters can be disentangled from the
intensity $\Iu$ of the unpolarized light received from the star and its
environment.  Without ADI observations, it is not possible to unravel the
contributions by the star and by its environment from the unpolarized light
$\Iu$.}

The estimation of the parameters $(\Iu,\Ip,\theta)$ \LD{from polarimetric data is
the objective} of the present contribution. \LD{Stokes parameters are however more
suitable to account for the effects of the instrument on the observable
polarization as the model of the data happens to be a simple linear combination
of these parameters.} Stokes parameters \LD{account} for the total light, the
linearly polarized light, and the circularly polarized light. Since circular
polarization is mostly generated by magnetic interactions and double
scattering, \LD{it is often negligible in the case of circumstellar environments}
and thus not measured by the SPHERE or GPI instruments. In the end, it is
possible to reconstruct the parameters of interest $\Iu$, $\Ip$, and $\theta$
from a combination of the Stokes parameters.

\begin{figure}[t]
    \centering
    \begin{tikzpicture}
        \begin{scope}[scale=0.8]
            \fill[white!80!black, opacity=0.8]  (1.5,0.4) rectangle (2.4, 1.6);
            
            \fill[white!80!black, opacity=0.8] (3.55,0.4) rectangle (3.95,1.6);
            
            \fill[white!80!black, opacity=0.8]  (5.1, 0.4) rectangle (7, 1.6);
            
            \fill[white!80!black, opacity=0.8]  (7.6,1.1) rectangle (8,1.5);
            \fill[white!80!black, opacity=0.8]  (7.6,0.5) rectangle (8,0.9);
            
            \fill[white!80!black, opacity=0.8] (8.6,1) rectangle (9.2,1.6);
            \fill[white!80!black, opacity=0.8] (8.6,0.4) rectangle (9.2,1);
            
            \draw[ultra thick] (0.8,1) -- (7.3,1);
            \draw[ultra thick] (8.9,1.3) -- (7.3,1.3) -- (7.3,0.7)--(8.9,0.7);
            
            \draw [very thick, white!40!black] (1.5,0.4) rectangle (2.4, 1.6);
            
            \draw[very thick, white!40!black] (3.55,0.4) rectangle (3.95,1.6);
            
            \draw[very thick, white!40!black] (5.1, 0.4) rectangle (7, 1.6);
            
            \draw[very thick, white!40!black] (7.6,1.1) rectangle (8,1.5);
            \draw[very thick, white!40!black] (7.6,0.5) rectangle (8,0.9);
            
            \draw[very thick, white!40!black] (8.6,1) rectangle (9.2,1.6);
            \draw[very thick, white!40!black] (8.6,0.4) rectangle (9.2,1);

            \draw [line width=3pt,opacity=0.8]  plot [domain=0:pi, samples=80, 
            smooth] ({0.3*cos(\x r)+0.2*sin(\x r)}, {1+0.05*cos(\x 
            r)+0.2*sin(\x r)});
            \draw [ultra thick] (0,1) node{$\bigstar$};
            \draw [line width=3pt,opacity=0.8]  plot [domain=pi:2*pi, 
            samples=80, smooth] ({0.3*cos(\x r)+0.2*sin(\x r)}, {1+0.05*cos(\x 
            r)+0.2*sin(\x r)});
            
            \draw[ultra thick, dashed] (0,1) -- (0.6,1);
            
            
            \draw (2,0) node{\footnotesize Optical};
            \draw (2,-0.4) node{\footnotesize devices};
            
            \draw (3.75,-0.2) node{\footnotesize HWP};
            
            \draw (6,0) node{\footnotesize Optical};
            \draw (6,-0.4) node{\footnotesize devices};
            
            \draw (7.8,-0.2) node{\footnotesize Analysers};
            \draw (7.8,0.3) node{\tiny $\psi_1=0^{\circ}$};
            \draw (7.8,1.7) node{\tiny $\psi_2=90^{\circ}$};
            
            \draw (9.2,1.8) node{\tiny $j=1$};
            \draw (9.2,0.2) node{\tiny $j=2$};
            
            \draw (-0.4,0) node{\small ${\Iu_n}$};
            \draw (0.2,0) node{\small $\Ip_n$};
            \draw (0.8,0) node{\small $\theta_n$};
            
            \draw (-0.1, -0.27) node[rotate=90]{\resizebox{0.4cm}{0.6cm}{$\{$ 
            }};
            \draw (-0.2, -0.4) node[below]{$\Is$};
            
            \draw (0.5, -0.4) node[rotate=90]{\resizebox{0.4cm}{0.6cm}{$\{$ }};
            \draw (0.55, -0.45) node[below]{$\Qs$, $\Us$};

            \draw [white!40!black] (8.9,1.3) node{\Large $\times$};
            \draw [white!40!black] (8.9,0.7) node{\Large $\times$};
            
            \draw [<-] (9,1.3) -- (9.5,1.3) -- (9.5,1.1) ;
            \draw (9.4,1) node[above, rotate=-90]{\small $\V{d}_{j,k,m}$};
            \draw [<-] (9,0.7) -- (9.5,0.7)-- (9.5,0.9) ;

            \draw[->]  (-0.4,0.3) -- (0,1);
            \draw[->]  (0.2,0.3) -- (0.3,1);
            \draw[->]  (0.8,0.3) -- (0.3,1);
            \draw (5., -0.8) node[rotate=90]{\resizebox{0.5cm}{4cm}{$\{$ }};
            \draw (5., -1.) node[below]{$\nu_{j,k,\ell}$};
            
            \draw[->, thick] (3.45, 1) arc(180:135:0.3) node[above left] {\tiny $0^{\circ}$} 
            node {\tiny $\bullet$} arc(135:45:0.3) 
            node[above right] {\tiny $45^{\circ}$} node {\tiny $\bullet$}  
            arc(45:-45:0.3) node[below right] {\tiny $22,5^{\circ}$} node 
            {\tiny $\bullet$}  arc(-45:-135:0.3) 
            node[below left] {\tiny $77,5^{\circ}$} node {\tiny $\bullet$}  
            arc(-135:-175:0.3) ;
            \draw (3.36,1.05) node[below left]{\tiny $\alpha_k$};
        \end{scope}
    \end{tikzpicture}
    \caption{\LD{Schematic view of the instrument ESO/VLT-SPHERE IRDIS showing the
    various optical parts that can induce polarization effects.  The same 
    notations as in the text are used (\eg, $n$ is the pixel index in the 
    restored maps, $k$ is the sequence index and $j$ is the polarizer index of 
    the analyzer set).}} \label{fig:irdis_simple}
\end{figure}
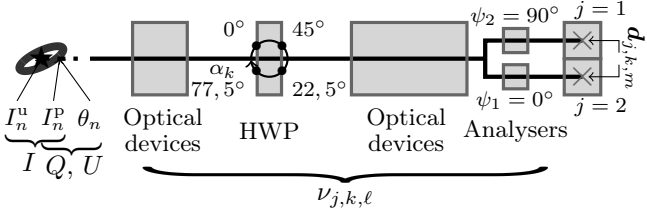

\subsection{Polarization effects}

The four Stokes parameters $S = (\Is,\Qs,\Us,\Vs) \in \R^4$ describe the state
of polarized light: $\Is$ is the total intensity accounting for the polarized
and unpolarized light, $\Qs$ and $\Us$ are the intensities of the light
linearly polarized along 2 directions at $45^\circ$ to each other, and $\Vs$ is
the intensity of the circularly polarized light.  Under this formalism,
polarization effects by an instrument like SPHERE/IRDIS (see
Fig.~\ref{fig:irdis_simple}) can be modeled by \citep[see Eq.~(17) of
][]{van_holstein_polarimetric_2020}:
\begin{align}
  \Cvlv{S}^\Tag{det}_{\!j}
  &= \Mueller{pol}_{j}\,
    \Rotation{-\DerotatorAngle}\,\Mueller{der}\,\Rotation{\DerotatorAngle}\,
    \Rotation{-\alpha}\,\Mueller{HWP}\,\Rotation{\alpha}\,\notag\\
  &\quad\Mueller{M4}\,\Rotation{\AltitudeAngle}\,
    \Mueller{UT}\,\Rotation{\ParallacticAngle}\, \Cvlv{S}
    \label{eq:polarization-effects}
\end{align}
where $S^\Tag{det}_{\!j}$ are the Stokes parameters on the detector after the
left ($j=1$) or right ($j=2$) polarizer of the analyzer set while $S$ are the
Stokes parameters at the entrance of the telescope.  In the above equation, $\Mueller{}_{}$ denotes a Mueller matrix accounting for the polarization effects of a specific
part of the instrument: $\Mueller{pol}_{j}$ for the left or right polarizer of
the analyzer set, $\Mueller{der}$ for the optical derotator, $\Mueller{HWP}$ is
for the half-wave plate (HWP), $\Mueller{M4}$ for the 4-th mirror of the
telescope and $\Mueller{UT}$ for the 3 mirrors (M1 to M3) constituting the
telescope.  The term $\Rotation{\theta}$ denotes a rotation matrix of the
polarization axes by an angle $\theta$: $\DerotatorAngle$ is the derotator
angle, $\alpha$ is the HWP angle, $\AltitudeAngle$ is the altitude angle, and $\ParallacticAngle$ is the parallactic angle of the pointing of the alt-azimuthal telescope.

In the optical, only the total intensity ${\Is_{j}}^{\!\Tag{det}}$ out of the
Stokes parameters ${S_{j}}^{\!\Tag{det}}$ can be measured by any \LD{existing}
detector. \LD{It then follows from Eq.~\eqref{eq:polarization-effects} that the
quantity measured by the detector is a simple linear combination of the input
Stokes parameters:}
\begin{align}
  \Is^\Tag{det}_{j}
  = \sum_{\ell=1}^{4}
  \nu_{j,\ell}(\AllAngles)\,S_\ell
  \label{eq:detected-intensity-general-model}
\end{align}
where, for every $\To{\ell}{1}{4}$,
$S_\ell$ denotes the $\ell$-th component of the input Stokes parameters
$S$ and $\nu_{j,\ell}(\AllAngles)$, for $j \in \Brace{1,2}$, are real coefficients depending on all involved
angles
$\AllAngles = (\DerotatorAngle,\AltitudeAngle,\ParallacticAngle,\alpha)$.

Even though the \LD{observables} are restricted to the component
$\Is^\Tag{det}_{j}$, rotating the angle $\alpha$ of the HWP introduces a
modulation of the contribution of the Stokes parameters $\Qs$ and $\Us$ in
$\Is^\Tag{det}_{j}$, which can be exploited to disentangle the Stokes
parameters $\Is$, $\Qs$ and $\Us$.  
The Stokes parameter $\Vs$ characterizing the circularly polarized light cannot
be measured with an instrument such as SPHERE/IRDIS (a modulation by a
quarter-wave plate would have been required to do so).  \hide{For most
astronomical targets of interest, the circularly polarized light is negligible,
thus this limitation is not an issue in practice \CHANGE{(ref.)}. } In the
following, we therefore neglect the circularly polarized light and only
consider the unpolarized and linearly polarized light characterized by the
\hide{reduced} Stokes parameters $S = (\Is,\Qs,\Us)$.  As a direct
simplification, the sum in the right hand side of
Eq.~\eqref{eq:detected-intensity-general-model} is reduced to its first three
terms.  For a sequence of acquisitions with different angles of the HWP, the
detected intensities follow:
\begin{align}
  \Is^\Tag{det}_{j,k} = \sum_{\ell=1}^{3}\nu_{j,k,\ell}\,\Cvlv{S}_\ell
  \label{eq:detected-intensity-linear-model}
\end{align}
where $\nu_{j,k,\ell} = \nu_{j,\ell}(\AllAngles_{k})$ with $\AllAngles_{k}$ the
set of angles during the $k$-th acquisition.

The \LD{instrumental} polarization effects being reduced to \LD{those caused by} the
analyzers and the HWP and \LD{ignoring the rotation due to the altitude and
parallactic angles,} the detected intensity writes
\citep{van_holstein_polarimetric_2020}:
\begin{equation}
  \Is^\Tag{det}_{j,k}
  = \smallfrac{1}{2}\,\Cvlv{\Is}
  + \smallfrac{1}{2}\,\cos\bigl(4\,\alpha_{k} + 2\,\psi_{j}\bigr)\,\Cvlv{\Qs}
  + \smallfrac{1}{2}\,\sin\bigl(4\,\alpha_{k} + 2\,\psi_{j}\bigr)\,\Cvlv{\Us}\,,
  \label{eq:prefect-instrument}
\end{equation}
where $\psi_{j}$ is the orientation angle of the left/right polarizer while
$\alpha_{k}$ is the HWP angle during the $k$-th acquisition.
Table~\ref{tab:Vvalues} lists the values of the \LD{linear coefficients
$\nu_{j,k,\ell}$} for a typical set of HWP angles.

\begin{table}[t]
  \centering
  \begin{tabular}{llrrr}
    \hline
    \multicolumn{1}{c}{\LD{$\alpha_k$ (HWP)}} &
    \multicolumn{1}{c}{\LD{$\psi_j$ (Analyzer)}} &
    \multicolumn{1}{c}{$\nu_{j,k, 1}$} &
    \multicolumn{1}{c}{$\nu_{j,k, 2}$} &
    \multicolumn{1}{c}{$\nu_{j,k, 3}$} \\
    \hline
    \hline
    \LD{$0^\circ$}    &  \LD{$0^\circ$} (left : $j=1$)  & \CompactFrac{1}{2} & $\CompactFrac{ 1}{2}$ & $0$  \\
    \LD{$0^\circ$}    &  \LD{$90^\circ$} (right : $j=2$) & \CompactFrac{1}{2} & $\CompactFrac{-1}{2}$ & $0$ \\
    \LD{$45^\circ$}   &  \LD{$0^\circ$} (left : $j=1$)  & \CompactFrac{1}{2} & $\CompactFrac{-1}{2}$ & $0$ \\
    \LD{$45^\circ$}   &  \LD{$90^\circ$} (right : $j=2$) & \CompactFrac{1}{2} & $\CompactFrac{ 1}{2}$ & $0$  \\
    \LD{$22.5^\circ$} &  \LD{$0^\circ$} (left : $j=1$)  & \CompactFrac{1}{2} & $0$ & $\CompactFrac{ 1}{2}$ \\
    \LD{$22.5^\circ$} &  \LD{$90^\circ$} (right : $j=2$) & \CompactFrac{1}{2} & $0$ & $\CompactFrac{-1}{2}$ \\
    \LD{$77.5^\circ$} &  \LD{$0^\circ$} (left : $j=1$)  & \CompactFrac{1}{2} & $0$ & $\CompactFrac{-1}{2}$ \\
    \LD{$77.5^\circ$} &  \LD{$90^\circ$} (right : $j=2$) & \CompactFrac{1}{2} & $0$ & $\CompactFrac{ 1}{2}$ \\
    \hline
  \end{tabular}
  \caption{\LD{This table lists the positions of HWP $\alpha_k$ and the orientations of the analyzer $\psi_j$ and the corresponding values of the coefficients $\nu_{j,k, \ell}$ assuming no
  instrumental polarization and ignoring field rotation. It is computed from Eq.~\eqref{eq:prefect-instrument}.}} \label{tab:Vvalues}
\end{table}

\subsection{Parameters of interest}
\label{sec:parameters}

The model of the detected intensity given in
Eq.~\eqref{eq:detected-intensity-linear-model} is linear in the Stokes
parameters $S = (\Is,\Qs,\Us)$, which makes its formulation suited to inverse
problem solving. However, to study circumstellar environments in polarimetry,
the knowledge of the linearly polarized light $\Ip$ and the polarization angle
$\theta$ is crucial. Both set of parameters are related as it follows:
\begin{equation}
    \begin{cases}
    \Is = \Iu + \Ip\\
    \Qs = \Ip  \, \cos(2\,\theta)\\
    \Us = \Ip \, \sin(2\,\theta)
    \end{cases}
    \label{eq:Jones-to-Stokes}
\end{equation}
and conversely by\footnote{In this representation there is a $\pm180^\circ$ 
degeneracy for the linear polarization angle \LD{$\theta$}.}:
\begin{equation}
  \begin{cases}
    \Ip = \sqrt{\Qs^2 + \Us^2}\\
    \theta = \left(1/2\right)\arctan{\left(\Us/\Qs\right)} \mod \pi\\
    \Iu = \Is - \LD{\sqrt{\Qs^2 + \Us^2}}.
  \end{cases}    
  \label{eq:Stokes-to-Jones}
\end{equation}
\LD{We assume that these relations hold} independently at any position of the field
of view (FOV). \LD{From Eqs.~\eqref{eq:detected-intensity-linear-model} and
\eqref{eq:Jones-to-Stokes}, the direct model of the detected intensity is a
non-linear function of the parameters $\Iu$, $\Ip$ and $\theta$}:
\begin{align}
  \Is^\Tag{det}_{j,k}
  &= \nu_{j,k,1}\,\Iu + \Paren[\big]{
    \nu_{j,k,1} +
    \nu_{j,k,2}\,\cos(2\,\theta) +
    \nu_{j,k,3}\,\sin(2\,\theta)
  }\,\Ip \, .
  \label{eq:detected-intensity-nonlinear-model}
\end{align}
For a perfect SPHERE/IRDIS-like instrument and not considering field rotation,
\LD{combining Eqs.~\eqref{eq:prefect-instrument} and \eqref{eq:Jones-to-Stokes}}
yields:
\begin{equation}
  \Is^\Tag{det}_{j,k} = \smallfrac{1}{2}\,\Iu +
  \Ip\,\cos^2\bigl(\theta -2\,\alpha_{k} - \psi_{j}\bigr) \,,
\end{equation}
which is the Malus law.


\begin{table}
\centering
\NewBeg
\begin{tabular}{ll}
\hline
Notation & Description \\
\hline
\hline
  $d_{j,k,m}$
  & Measured data, Eq.\eqref{eq:data-approx-model} \\
  $\Is^\Tag{det}_{j,k,m}$
  & Model of the data, Eq.~\eqref{eq:direct-linear-model}\\
  $ H_{j,k,m,n}$
  & Instrumental PSF, Eq.~\eqref{eq:PSF-factorization}\\
  $\Sigma_{j,k,m} $
  & Variance of data, Eq.~\eqref{eq:weights} \\
  $\nu_{j,k,\ell}$
  & Polarization effects,  Eq.~\eqref{eq:detected-intensity-linear-model}\\
  $S=(I,Q,U)$ 
  & Stokes parameters, Eq.~\eqref{eq:Jones-to-Stokes}\\
  $\Iu$
  & Unpolarized intensity, Eq.~\eqref{eq:Stokes-to-Jones}\\
  $\Ip$
  & Polarized intensity, Eq.~\eqref{eq:Stokes-to-Jones}\\
  $\theta$
  &angle of polarization, Eq.~\eqref{eq:Stokes-to-Jones}\\
  $\VX$ 
  & Parameters of interest, Eq.~\eqref{eq:constrained-problem}\\
  $\Es{\VX}$
  & Estimated parameters, Eq.~\eqref{eq:constrained-problem}\\
  $\VX_\Tag{gt}$
  & Ground truth, Eq.~\eqref{eq:EQMparametermulticomp}\\
  $f_\Tag{data}$
  & Data fidelity term, Eq.~\eqref{eq:datafidelity}\\
  $\M W_{j,k,m}$ 
  & Weights, Eq.~\eqref{eq:weights}\\
  $f_\rho$
  & Regularization term, Eq.~\eqref{eq:regul-Iu}-\eqref{eq:regul-Q+U}\\
  $\lambda_\rho$
  & Regul. contribution, Eq.~\eqref{eq:regul-Iu}-\eqref{eq:regul-Q+U}\\
  $\mu_\rho$
  & Regularization threshold, Eq.~\eqref{eq:regul-Iu}-\eqref{eq:regul-Q+U}\\
  $\M D_n$ 
  & 2D spatial gradient, Eq.~\eqref{eq:spatialgradient}\\
  $\mathcal{C}$
  & Positivity constraint, Eq.~\eqref{eq:feasible-set-Iu-Ip-theta} \\
\hline
  \multicolumn{2}{c}{Indices}\\
  $j \in \Brace{1,2}$ & Polarizer of the analyser set\\
  $k \in \IntRange{1,K}$ & Data frame in the sequence\\
  $\ell \in \IntRange{1,L}$ & Component the \LDrevised{parameter} of interest.\\
  $m \in \IntRange{1,M}$ & Pixel in data sub-image\\
  $n, n' \in \IntRange{1,N}$ & Pixel in restored model maps\\
  $\rho$ & Param. to regularize $(I,\Iu, \Ip, Q+U)$\\
\hline
\end{tabular}
\NewEnd
\caption{\LD{Notations.}}
\label{tab:notations}
\end{table}

\subsection{Accounting for the instrumental spatial PSF}

In polarimetric imaging, each polarimetric parameter is a function of the
2-dimensional FOV.  \LD{We consider} that the Stokes parameters are represented by
\emph{images} of $N$ pixels each \LD{and denote as $S_{\ell,n}$} the value of the
$n$-th pixel in the map of the $\ell$-th Stokes parameter.

Provided that polarization effects apply uniformly across the FOV and that the
instrumental spatial point spread function (PSF) does not depend on the
polarization of light, \LD{all} Stokes parameters of a spatially incoherent source
are independently and identically affected by the spatial PSF
\citep[\eg,][]{birdi_sparse_2018, smirnov2011revisiting, denneulin2020theses}.
As the effects of the spatial PSF are linear, we can write the detected
intensity for a given detector pixel as:
\begin{align}
  \Is^\Tag{det}_{j,k,m}
  &= \sum_{\ell=1}^{3}\sum_{n=1}^N \nu_{j,k,\ell}\, H_{j,k,m,n}\,
  S_{\ell,n} \, ,
  \label{eq:direct-linear-model}
\end{align}
where $\Is^\Tag{det}_{j,k,m}$ is the \LD{intensity measured during the $k$-th
acquisition by the $m$-th pixel of the sub-image corresponding to the $j$-th
polarizer of the analyzer set.  The coefficients $\nu_{j,k,\ell}$ accounting
for the instrumental polarization are defined in
Eq.~\eqref{eq:detected-intensity-linear-model} and $H_{j,k,m,n}$ denotes a
given entry of the discretized spatial PSF of the instrument.  Here $j \in
\Brace{1,2}$, $k \in \IntRange{1,K}$ and $m \in \IntRange{1,M}$.
Table~\ref{tab:notations} summarizes the main notations  used in this
paper.}

It follows from our assumption on spatial and polarization effects applying
independently, that the discretized spatial PSF does not depend on the
polarization index $\ell$. Consequently, the spatial and polarization effects
in Eq.~\eqref{eq:direct-linear-model} mutually commute.


Figure~\ref{fig:irdis_simple} and Eq.~\eqref{eq:polarization-effects} provide a
representation of the instrument from which we build a model of the spatial
effects of the instrument. Accordingly, an image representing the spatial
distribution of the light as \LD{the} input of the instrument should undergo a
succession of image transformations before reaching the detector.  These
transformations are either geometrical transformations (\eg, the rotation
depending on the parallactic angle) or blurring transformations (\eg, by the
telescope).  Except in the neighborhood of the coronographic mask, the effects
of the 
\LD{blur can be assumed to be shift-invariant and can thus be modeled by convolution with a shift-invariant PSF.}
Geometrical transforms and convolutions
do not commute but their order may be changed provided the shift-invariant PSFs
are appropriately rotated and/or shifted. Thanks to this property and without
loss of generality, we can model the spatial effects of the instrument by a
single convolution accounting for all shift-invariant blurs followed by a
single geometrical transform accounting for all the rotations but also possible
geometrical translations and/or spatial (de)magnification\footnote{\LD{\LDrevised{If} the pixel
size of the polarimetric maps is not chosen to be equal to the angular size of
the detector pixels}}.  Following this analysis, our model of the spatial PSF is
given by:
\begin{equation}
  H_{j,k,m,n}
  = \sum_{n' =1}^N \Paren[\big]{\M T_{j,k}}_{m,n'}\,\Paren[\big]{\M A_k}_{n',n} \, ,
  \label{eq:PSF-factorization}
\end{equation}
where $\M A_k\from\R^N\to\R^N$ implements the shift-invariant blur of the input
model maps while $\M T_{j,k}\from\R^N\to\R^{M_j}$ performs the geometrical
transform of the blurred model maps for the $j$-th polarizer of the analyzer
set during the $k$-th acquisition. $N$ is the number of pixels in the model
maps and $M_j$ is the number of pixels of the detector (or sub-image)
corresponding to the $j$-th output polarizer.

\LD{In our implementation of the PSF model, the geometrical transform of images is
performed using interpolation by Catmull-Rom splines.  The blurring due
to the instrument and the turbulence is applied by:
\begin{equation}
   \M A_k = \M F^{-1}\,\Diag(\tilde{\V p}_k)\,\M F
\end{equation}
where $\M F$ denotes a \emph{Fast Fourier Transform} (FFT) operator of suitable
size and $\Diag(\tilde{\V p}_k)$ implements the frequency-wise multiplication
by $\tilde{\V p}_k = \M F\,\V p_k$\LDrevised{,} the discrete Fourier transform of $\V p_k$\LDrevised{,}
the shift-invariant PSF. Note that $\V p_k$ must be specified in the same
reference frame as the FOV. If the shift-invariant PSF is calibrated from
empirical images of, \eg, the host star, acquired by the detector, the inverse
(or pseudo-inverse) of $\M T_{j,k}$ must be applied to the empirical images.}

%

\subsection{Polarimetric data}
\label{sec:polarimetric-data}

During a sequence of observations with SPHERE/IRDIS in DPI mode, the HWP is
rotated several times along a given cycle of angles $\alpha \in \{0^\circ,
45^\circ, 22.5^\circ, 67.5^\circ\}$.  Besides, the two polarizers of the
analyzer set of SPHERE/IRDIS are imaged on two disjoint parts of the same
detector.  The resulting dataset consists in
\LD{$K$ frames composed of two,} 
left and right, sub-images, each with a different position of the HWP. \LD{Typical 
values of $K$ can go from $32$ to more than $512$ depending of the observed target.}
After pre-processing of the raw images to compensate for the bias and the uneven
sensitivity of the detector and to extract the two sub-images, the available
data are modeled by:
\begin{equation}
  d_{j,k,m} \approx \Is^\Tag{det}_{j,k,m}
  \label{eq:data-approx-model}
\end{equation}
where $\Is^\Tag{det}_{j,k,m}$ is given in Eq.~\eqref{eq:direct-linear-model}
for $k \in \IntRange{1,K}$ the index of the acquisition, $j \in \{1,2\}$
indicating the left/right polarizer of the analyzer, and $m \in \IntRange{1,M}$
the pixel index in the corresponding left/right sub-image. 

The $\approx$ sign in Eq.~\eqref{eq:data-approx-model} is to account for an
unknown random perturbation term due to the noise. Noise in the pre-processed
images can be assumed centered and independent between two different pixels or
frames because the pre-processing suppresses the bias and treats pixels
separately thus introducing no statistical correlations between pixels. There
are \LD{many} sources of noise: \LD{shot noise for the light sources and the dark current}, detector read-out noise, \etc For most actual data, the shot noise is
the most important contribution and the number of electrons (photo-electrons
plus dark current) integrated by a pixel is large enough to approximate the
statistics of the data by an independent non-uniform Gaussian distribution
whose mean is given by the right-hand-side term of
Eq.~\eqref{eq:data-approx-model} and whose variance
$\Sigma_{j,k,m}=\Var(d_{j,k,m})$ \LD{ is estimated in} the calibration stage (see
subsection~\ref{subsec:detectorcalib}).


It is worth noticing that since the proposed model is not valid in the
neighborhood of the coronographic mask, the data pixels in this region have to
be discarded.  Moreover, the detector contains \LD{defective} pixels (\eg,
dead pixels, non-linear pixels, saturated pixels) which must be also discarded.
\LD{This} is achieved by \LD{assuming} that their variance is infinite \LD{which amounts to
setting their respective} weights to zero in the data fidelity term of the
objective function described in Section~\ref{sec:data-fidelity}.

\section{RHAPSODIE : Reconstruction of High-contrAst Polarized SOurces and Deconvolution for cIrcumstellar Environments}
\label{sec:mymethod}

\subsection{Inverse problems approach}
\label{sec:inverse-problems}

In polarimetric imaging, one is interested in recovering sampled maps 
of the polarimetric parameters, which can be the Stokes parameters \LD{$(\Is,\Qs,\Us)$} or the
intensities of unpolarized and linearly polarized light and the angle of the
linear polarization \LD{$(\Iu, \Ip, \theta)$} or some mixture of these parameters.  To remain as general
as possible, we denote by $\VX \in \R^{N\times L}$ the set of parameters of
interest to be recovered \LD{and by $\V X_{\ell} \in \R^N$, with $\ell \in
\IntRange{1,L}$, the $\ell$-th parameter component which is a $N$-pixel map}.

Given the direct model of the pre-processed data developed in the previous
section, we propose to recover the parameters of interest $\VX$ by a penalized
maximum likelihood approach.  This approach is customary in the solving of
inverse problems \citep{Titterington-1985-regularization, tarantola2005inverse}
and amounts to \LD{defining} the estimated parameters $\Es{\VX}$ as the ones that
minimize a given objective function $f(\VX)$ possibly under constraints
expressed as $\VX \in \FeasibleSet$ with $\FeasibleSet$ the set of acceptable
solutions. The objective function takes the form of the sum of a data-fidelity
term $f_\Tag{data}(\VX)$ and of regularization terms $f_\rho(\VX)$:
\begin{equation}
  \Es{\VX} = \argmin_{\VX \in \FeasibleSet} \Brace[\Big]{
    f(\VX) = f_\Tag{data}(\VX) +
    \underbrace{
      \sum\nolimits_{\rho} \lambda_{\rho}\,f_\rho(\VX)
    }_{\displaystyle f_\Tag{prior}(\VX)}
  }
  \label{eq:constrained-problem}
\end{equation}
where $\lambda_\rho \ge 0$ ($\forall\rho$) are so-called hyperparameters
introduced to tune the relative importance of the regularization terms.  The
data-fidelity term $f_\Tag{data}(\VX)$ imposes the direct model be as
close as possible to the acquired data while the regularization terms $f_\rho(\VX)$
enforce the components of the model to remain regular (\eg, smooth).
Regularization must be introduced to lift degeneracies and avoid artifacts
caused by the data noise and the ill-conditioning of the inverse problem.
Additional strict constraints may be imposed \LD{on} the sought parameters via the
feasible set $\FeasibleSet$, \eg to account for the requirement that
intensities \LD{are} non-negative quantities.  These different terms and constraints
are detailed in the following sub-sections.


\subsection{Data fidelity}
\label{sec:data-fidelity}

Knowing the sufficient statistics for the pre-processed data, agreement of the
model with the data is 
\LD{properly insured} by the co-log-likelihood of the data
\citep{tarantola2005inverse} or equivalently by the following criterion:
\begin{equation}
  f_\Tag{data}(\VX) = 
  \sum_{j,k} \Norm*{\V d_{j,k} - \VIs^\Tag{det}_{j,k}(\VX)}_{\M{W}_{j,k}}^2,
  \label{eq:datafidelity}
\end{equation}
where $\Norm*{\cdot}^2_{\M{W}} = \Inner{\cdot}{\M W\cdot}$ denotes
\citet{mahalanobis1936generalized} squared norm, $\V d_{j,k} =
(d_{j,k,1},\ldots,d_{j,k,M})\T \in \R^M$ collects all the pixels (\eg, in
lexicographic order) of the $j$-th sub-image in the $k$-th acquisition as
defined in Eq.~\eqref{eq:data-approx-model}. Similarly,
$\VIs^\Tag{det}_{j,k}(\VX) =
(\Is^\Tag{det}_{j,k,1},\ldots,\Is^\Tag{det}_{j,k,M})\T \in \R^M$ \LD{where the
terms $\Is^\Tag{det}_{j,k,m}$} are given by the model in
Eq.~\eqref{eq:direct-linear-model} applied to the Stokes parameters as a
function of the parameters of interest $\VX$, $S=S(\VX)$. For instance, if
$\VX=(\Iu, \Ip, \theta)$, then $S(\VX)$ is obtained by
Eq.~\eqref{eq:Jones-to-Stokes}. In the expression of the data fidelity term
given by Eq.~\eqref{eq:datafidelity}, $\M W_{j,k}$ is the precision matrix of
the data.  \LD{The precision matrix is} diagonal because pixels are considered as
mutually independent.  To account \LD{for the non-uniform noise variance and} for
invalid data (see \ref{sec:polarimetric-data}), we define the diagonal entries
of the precision matrix as:
\begin{equation}
  \forall \To{m}{1}{M}, \quad \bigl(\M{W}_{j,k}\bigr)_{m,m} = \begin{cases}
      \Sigma_{j,k,m}^ {-1}& \text{for valid data;} \\
      0 & \text{for invalid data,}\\
  \end{cases}
  \label{eq:weights}
\end{equation}
\LD{with $\Sigma_{j,k,m} = \Var(d_{j,k,m})$ the variance of a valid datum
$d_{j,k,m}$}. Invalid data include dead pixels, pixels incorrectly modeled by
our direct model because of saturation or of the coronograph, missing frames
for a given HWP angle, and unusable frames because of too strong atmospheric
turbulence or unproper coronograph centering.


Note that, in Eq.~\eqref{eq:datafidelity}, the Mahalanobis squared norms arise
from our Gaussian approximation of the statistics while the simple sum of these
squared norms for each sub-image in each frame is justified by the fact that
all frames and all sub-images are mutually independent.

\subsection{Regularization}

The problem of recovering the polarimetric parameters from the data is an
ill-conditioned inverse problem mainly due to the instrumental blur. 
Furthermore, the problem may also be ill-posed if there are too many invalid
data.  In the case of an ill-conditioned inverse problem, the maximum
likelihood estimator of the parameters of interest, that is the parameters
which minimize the data fidelity term $f_\Tag{data}(\VX)$ defined in
Eq.~\eqref{eq:datafidelity} alone, cannot be used because it is too heavily
corrupted by noise amplification. Explicitly requiring that the sought
parameters be somewhat regular is mandatory to avoid this
\citep{Titterington-1985-regularization, tarantola2005inverse}.  In practice,
this amounts to adding one or more regularization terms $f_\rho(\VX)$ to the
data-fidelity as assumed by the objective function defined in
Eq.~\eqref{eq:constrained-problem}.

\subsubsection{Edge-preserving smoothness}

We expect that the light distribution of circumstellar environments be mostly
smooth with some sharp edges, hence \emph{edge-preserving smoothness}
regularization \citep{charbonnier_deterministic_1997} appears to be the most suited choice
to this
kind of light distribution.  When considering the recovering of polarimetric
parameters, such a constraint can be directly imposed to the unpolarized
intensity $\Iu$, to the polarized intensity $\Ip$ or to the total intensity
$\Is$ by the following regularization terms:
\begin{align}
  f_{\Iu}^{}(\VX) &= \sum_n
    \sqrt{\Norm{\M{D}_{n} \VIu(\VX)}^2 + \mu_{\LD{\Iu}}^2} \,,
  \label{eq:regul-Iu} \\
  f_{\Ip}^{}(\VX) &= \sum_n
    \sqrt{\Norm{\M{D}_{n} \VIp(\VX)}^2 + \mu_{\LD{\Ip}}^2} \,,
  \label{eq:regul-Ip} \\
  f_{\Is}^{}(\VX) &= \sum_n
    \sqrt{\Norm{\M{D}_{n} \VIs(\VX)}^2 + \mu_{\LD{\Is}}^2} \,,
  \label{eq:regul-I}
\end{align}
where we denote in boldface sampled maps of polarimetric parameters, for
instance $\VIu \in \R^N$ the image of the unpolarized intensity or $\VIu(\VX)$
to make explicit that it is uniquely determined by the sought parameters $\VX$.
In the above expressions, \LD{$\mu_\rho > 0$ 
models the smoothing threshold} and
$\M{D}_{n}: \R^N \rightarrow \R^2$ is a linear operator which yields an
approximation of the 2D spatial gradient of its argument around the $n$-th pixel.
This operator is implemented by means of finite differences; more specifically
applying $\M{D}_{n}$ to a sampled map $\V{u}$ of a given parameter writes:
\begin{equation}
   \M{D}_{n}\,\V{u} = \begin{pmatrix}
    \V{u}_{n_{1}+1, n_{2}} - \V{u}_{n_{1}, n_{2}}\\
    \V{u}_{n_{1}, n_{2}+1} - \V{u}_{n_{1}, n_{2}}
  \end{pmatrix}\,.
  \label{eq:spatialgradient}
\end{equation}
where $(n_1,n_2)$ denotes the row and column indices of the $n$-th pixel in the
map.  At the edges of the support of the parameter maps, we simply assume
\emph{flat boundary conditions} and set the spatial gradient to zero there.

The regularizations in Eqs.~\eqref{eq:regul-Iu}--\eqref{eq:regul-I} implement
a hyperbolic version of a pseudo-norm of the spatial gradient of a given
component of the light distribution which behaves as an $L_2$-norm (\ie,
quadratically) for gradients much smaller than $\mu$ and as an $L_1$-norm
(\ie, linearly) for gradients much larger than $\mu$.  Hence imposing
smoothness for flat areas where the spatial gradient is small while 
avoiding strong penalization for larger spatial gradients at edges of
structures.

It has been shown \citep{lefkimmiatis2013hessian, chierchia_nonlocal_2014} that
grouping different sets of parameters in regularization terms that are
sub-$L_2$ norm of the gradient like the last one in
Eq.~\eqref{eq:regul-Iu}--\eqref{eq:regul-I} yields solutions in which strong
changes tend to occur at the same locations in the sets of parameters.  In
order to encourage sharp edges to occur at the same places in the Stokes
parameters $\Qs$ and $\Us$, we also consider using the following regularization
for these components:
\begin{align}
  f_{\Qs+\Us}^{}(\VX) &= \sum_n
    \sqrt{\Norm{\M{D}_{n} \VQs(\VX)}^2 + \Norm*{\M{D}_{n} \VUs(\VX)}^2 + 
    \mu_{\LD{\Qs+\Us}}^2} \,.
  \label{eq:regul-Q+U}
\end{align}

Many other regularizations implementing smoothness constraints can be found in
the literature from the simple quadratic one \citep{tikhonov1963regularization}
to the very popular \emph{total variation}
\citep[TV;][]{rudin_nonlinear_1992}\hide{ or the more recent Schatten norm
\CHANGE{(sic)} \citep{lefkimmiatis2013hessian}}.  Quadratic regularizations
\LD{tend} to yield strong ripples which, owing to the contrast of the
recovered maps, are an unacceptable nuisance while TV yields maps affected by a
so-called \emph{cartoon effect} (\ie, piecewise flat images) which is not
appropriate for realistic astronomical images.  We however note that the
hyperbolic edge-preserving regularization with a threshold \LD{$\mu_\rho$} set to a very
small level can be seen as a relaxed version of TV and has been widely used as a
differentiable approximation of this regularization.  Our choice of a
differentiable regularization is also motivated by the existence of efficient
numerical methods to minimize non-quadratic but differentiable objective
functions of many (millions or even billions) variables possibly with
additional strict constraints \LD{\citep{thiebaut2002optimization}}. See
\citet{denneulin2019gretsi, denneulin2020primal, denneulin2020theses} for a
comparison of possible advanced regularizers.


\subsubsection{Tuning of the hyperparameters}
\label{sec:tuning-regularization}

In the regularization function $f_\Tag{prior}(\VX)$, the terms defined in
Eq.~\eqref{eq:regul-Iu}--\eqref{eq:regul-I} and Eq.~\eqref{eq:regul-Q+U} can be
activated (or inhibited) by choosing the corresponding $\lambda_\rho > 0$ (or
$\lambda_\rho = 0$).  It is also required to tune the threshold level \LD{$\mu_\rho>0$}
in addition to the $\lambda_\rho$ multipliers.  All these hyperparameters have
an incidence on the recovered solution: the higher
$\lambda_\rho$ the
smoother the corresponding regularized component and 
\LD{a lowering of the threshold
$\mu_\rho$ allows us to capture sharper structures.} 
A number of practical methods
have been devised to automatically tune the hyperparameters:\LD{\emph{Stein's
Unbiased Risk Estimator} \citep[SURE: ][]{stein1981estimation,
eldar2008generalized, ramani_monte-carlo_2008, deledalle_stein_2014}, \emph{Generalized
Cross-Validation} \citep[GCV:][]{golub1979generalized},
the 
$L$-curve \citep{hansen1993use},}
\NewBeg
or \emph{hierarchical Bayesian strategies}
\citep{Molina-1994-hierarchical_Bayesian}. 

\begin{figure}
\centering
\begin{tikzpicture}
\draw (0,0) node[below right]{\includegraphics[width=0.45\textwidth]{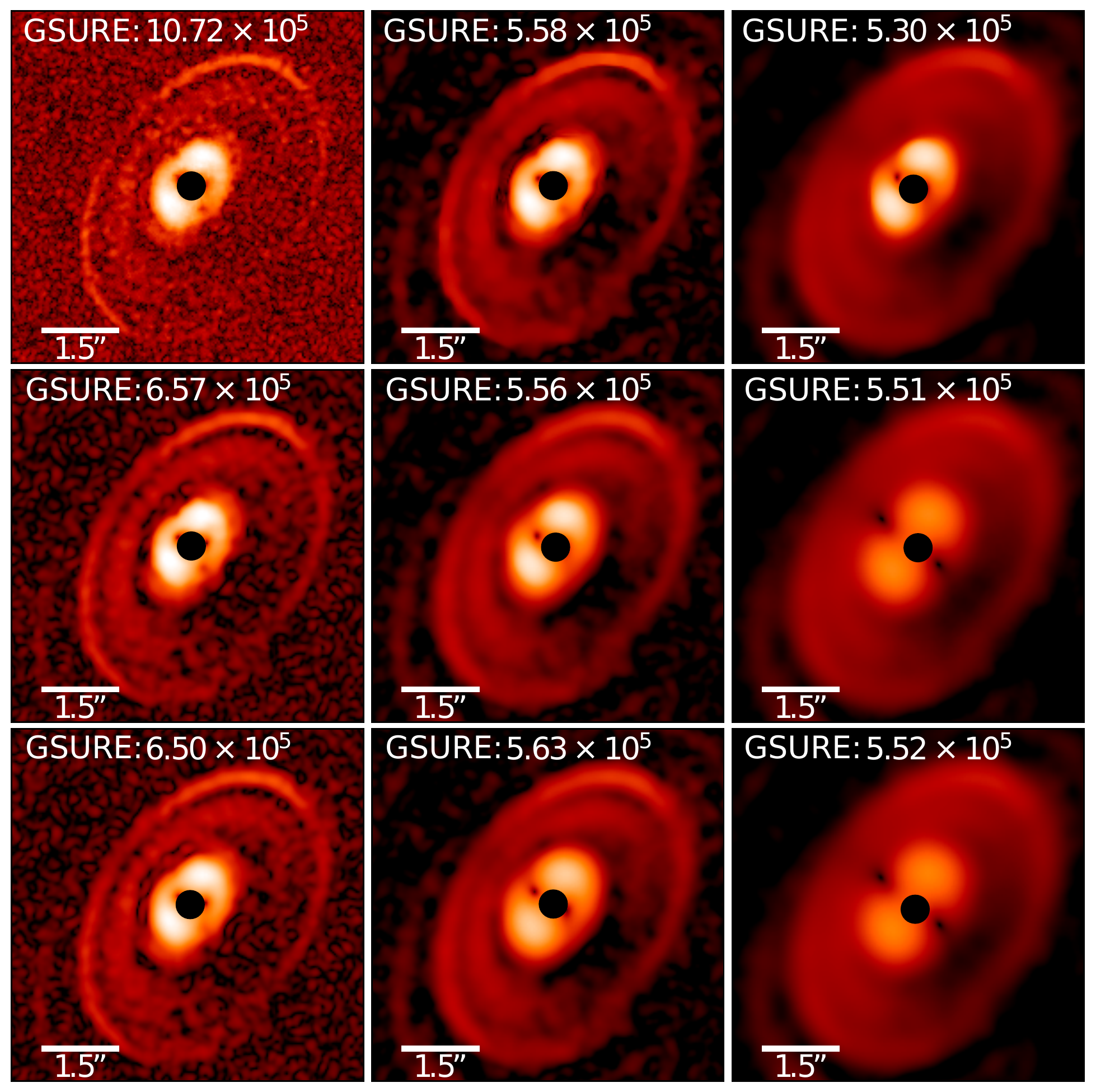}};
\draw (1.5, 0.25) node{$\mu_{Q+U}=10^{-3.5}$};
\draw (4.25, 0.25) node{$\mu_{Q+U}=10^{-2.5}$};
\draw (7, 0.25) node{$\mu_{Q+U}=10^{-1.5}$};

\draw (-0.25, -1.5) node[rotate=90]{$(\lambda/\mu)_{Q+U}=2$};
\draw (-0.25, -4.25) node[rotate=90]{$(\lambda/\mu)_{Q+U}=3$};
\draw (-0.25, -7) node[rotate=90]{$(\lambda/\mu)_{Q+U}=4$};

\end{tikzpicture}
\caption{\LD{Comparison of the non-linear RHAPSODIE reconstructions of RXJ~1615
\citep{avenhaus_disks_2018} for different values of $\lambda_{Q+U}$ and $\mu_{Q+U}$.
The value of the GSURE criterion is indicated for each reconstruction.}}
\label{fig:hyperpar_comp}
\end{figure}
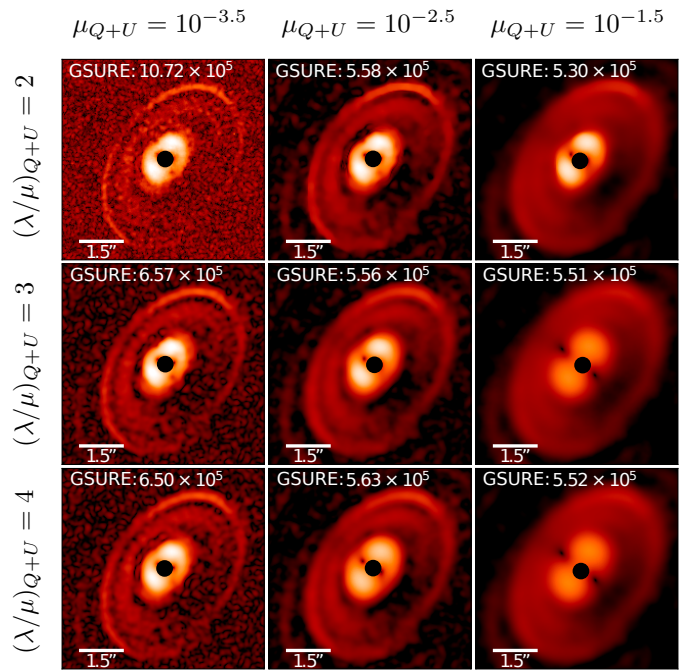

Unsupervised tuning of the hyperparameters with GSURE \citep[][]{
eldar2008generalized}, 
in a prediction error formulation, 
has been considered in this context.
Fig.~\ref{fig:hyperpar_comp} shows the value of GSURE for several reconstructions
of RXJ~1615 \citep{avenhaus_disks_2018}.
Since it is expected that there are relatively few sharp edges,
$\Norm{\M D_n\V X_\rho} \ll \mu_\rho$ should hold for most pixels $n$ in the
component $\V X_\rho$.  As a result, for most pixels $n$, the regularization
penalty behaves as a quadratic \citet{tikhonov1963regularization} smoothness weighted by
$\lambda_{\rho}/\mu_{\rho}$:
\begin{equation}
   \lambda_\rho\,\sqrt{\Norm{\M D_n\V X_\rho}^2 + \mu_\rho^2}
   \approx \lambda_\rho\,\mu_{\rho}
   + \frac{\lambda_\rho}{2\,\mu_\rho}\,\Norm{\M D_n\V X_\rho}^2\,.
\end{equation}
The strength of the blurring imposed by the regularization is therefore mostly
controlled by the value of $\lambda_\rho/\mu_\rho$ (\eg from top to bottom in Fig.~\ref{fig:hyperpar_comp}) while the sharp edges are
controlled by the threshold $\mu_\rho$ (\eg from left to right in Fig.~\ref{fig:hyperpar_comp}).  
\NewEnd


\LD{In Fig.~\ref{fig:hyperpar_comp}, we can observe that an automatic selection of the hyperparameters with GSURE would lead to an over regularized solution. In other high contrast contexts, we also observed this tendency of GSURE to over-smooth the result.  We think that devising a good unsupervised method for tuning the hyperparameters deserve further studies, and, for the results presented in this paper, we tuned the hyperparameters by hand by visually inspecting several reconstructions under different settings as presented in Fig.~\ref{fig:hyperpar_comp}.}



\LD{The polarized parameter $\Ip$ contribute to $\Is$, $\Qs$, and $\Us$ (see Eq.~\eqref{eq:Jones-to-Stokes}). To avoid a contamination of the extracted polarized parameters $\Ip$ by the unpolarized component $\Iu$, we regularize $\Iu$ and $\Ip$ separately.}
Hence, the regularization of the unpolarized component $\Iu$
should be done via $f_{\Iu}$ defined in Eq.~\eqref{eq:regul-Iu} rather than via
$f_{\Is}$ defined in Eq.~\eqref{eq:regul-I}.  For the polarized light, the
joint regularization of $\Qs$ and $\Us$ by $f_{\Qs+\Us}$ defined in
Eq.~\eqref{eq:regul-Q+U} is more effective than the regularization of $\Ip$
alone by $f_{\Ip}$ defined in Eq.~\eqref{eq:regul-Ip} which does not constrain
the angle $\theta$ of the linear polarization. In
Eq.~\eqref{eq:constrained-problem}, we therefore take the Set $1$ or Set $2$ of
hyperparameters presented in Table~\ref{tab:hyperpar}. The latter combination
is preferable as discussed previously.

\begin{table}[!ht]
  \centering
  \begin{tabular}{lrrrr}
    \hline
    \multicolumn{1}{c}{Hyperparameters} &
    \multicolumn{1}{c}{$\lambda_{\Ip}$} &
    \multicolumn{1}{c}{$\lambda_{\Iu}$} &
    \multicolumn{1}{c}{$\lambda_{\Is}$} &
    \multicolumn{1}{c}{$\lambda_{\Qs+\Us}$} \\
    \hline
    \hline
    Set $1$  & $=0$ & $\geq 0$ & $ =0$ & $\geq 0$ \\
    Set $2$  & $=0$ & $=0$ & $\geq 0$ & $\geq 0$ \\
    \hline
  \end{tabular}
  \caption{Set of hyperparameters used in the present work.}
  \label{tab:hyperpar}
\end{table}

\subsection{Imposing the positivity of the intensities}
\label{sec:positivity}

\LD{Imposing non-negativity constraints on the restored intensities has proven its efficiency} 
for astronomical imaging where large
parts of the images consist in background pixels whose value should be zero
\citep{biraud1969new}. Whatever the choice of the parametrization, the
intensities $\Is$, $\Iu$, and $\Ip$ should all be everywhere non-negative.

Since $\Is = \Iu + \Ip$, it is sufficient to require that $\Iu$ and $\Ip$ be
nonnegative.  Hence, for the set of parameters $\VX = (\VIu,\VIp,\V\theta)$,
the positivity constraint writes:
\begin{equation}
  \FeasibleSet = \Brace[\Big]{
    \Paren*{\VIu,\VIp,\V\theta} \in \R^{N \times 3} \SuchThat
    \forall n \in \IntRange{1,N},\ \Iu_n \geq 0,\ \Ip_n \geq 0
  } \, .
  \label{eq:feasible-set-Iu-Ip-theta}
\end{equation}

Expressed for the Stokes parameters $\VX = (\VIs,\VQs,\VUs)$, the positivity
\LD{yields} an epigraphical constraint:
\begin{equation}
  \FeasibleSet = \Brace[\Big]{
    \Paren*{\VIs,\VQs,\VUs} \in \R^{N \times 3} \SuchThat
    \forall n \in \IntRange{1,N},\ \Is_n \geq \sqrt{\Qs_{n}^2 + \Us_n^2}
  } \, .
  \label{eq:episet}
\end{equation}
Such a constraint can be found in \citet{birdi_sparse_2018}, but it has not yet
been implemented in high contrast polarimetric imaging.

Since $\Ip_n = \sqrt{\Qs_n^2 + \Us_n^2}$ (for all pixels $n$), the positivity
of the intensity $\Ip$ of the polarized light automatically holds if the
parameters $\VX = (\VIu,\VQs,\VUs)$ are considered.  It is then sufficient to
impose the positivity of the intensity $\Iu$ of the unpolarized light as
expressed by the following feasible set:
\begin{equation}
  \FeasibleSet = \Brace[\Big]{
    \Paren*{\VIu,\VQs,\VUs} \in \R^{N \times 3} \SuchThat
    \forall n \in \IntRange{1,N},\ \Iu_n \geq 0
  } \, .
  \label{eq:feasible-set-Iu-Q-U}
\end{equation}

\subsection{Choice of the polarimetric parameters}

Our method expresses the recovered parameters $\Es{\VX}$ as the solution of
a constrained optimization problem specified in
Eq.~\eqref{eq:constrained-problem}. As explained in
Sec.~\ref{sec:tuning-regularization}, the 
\LD{weights} of imposed regularization is
chosen via the values of the multipliers $\lambda_\rho$.

The constraints can be implemented by the feasible $\FeasibleSet$ for
different choices of the parameters $\VX$.  More specifically, $\VX =
(\VIu,\VIp,\V\theta)$, $\VX = (\VIs,\VQs,\VUs)$ or $\VX = (\VIu,\VQs,\VUs)$ can
be chosen.  Whatever the choice for $\VX$, the relations given in
Eq.~\eqref{eq:Jones-to-Stokes} and Eq.~\eqref{eq:Stokes-to-Jones} can be used
to estimate any parameter of interest given the recovered $\Es{\VX}$.  These
relations can also be used to compute the objective function $f(\VX)$ which
require the Stokes parameters needed by the direct model of the data
Eq.~\eqref{eq:direct-linear-model} and various polarimetric component depending
on the choice of the regularization.

With $\VX = (\VIs,\VQs,\VUs)$, the positivity constraints take the form of
an epigraphic \LD{constraint that is} more difficult to enforce
\LD{as being not} separable in the parameters \LD{space}. To solve the problem in
Eq.~\eqref{eq:constrained-problem} with such a constraint, an epigraphic
projection is required, leading to the use of a forward-backward scheme \citep{Combettes_P_2005_j-mms_sig_rpf}, reduced
in this context to a standard projected gradient descent. A description of the method for such a
minimization problem can be found in \cite{denneulin2020primal}. 

The choice
$\VX = (\VIs,\VQs,\VUs)$ may avoid some degeneracies because it ensures the
convexity of the problem in Eq.~\eqref{eq:constrained-problem}. 

With $\VX = (\VIu,\VIp,\V\theta)$ or $\VX = (\VIu,\VQs,\VUs)$, the positivity
constraints amounts to applying simple separable bound constraints on some
parameters. Since the objective is differentiable, a method such as VMLM-B
\citep{thiebaut2002optimization} can be used to solve the problem in
Eq.~\eqref{eq:constrained-problem}. \LD{The VMLM-B algorithm is a quasi-Newton method
with limited memory requirements and able to account for separable bound constraints.
VMLM-B is applicable to large size problems and only require to provide the bounds and a numerical
function to compute the objective function and its gradient.} 

In the following, we compare the performances of RHAPSODIE for the polarimetric
parameters $\VX = (\VIs,\VQs,\VUs)$ and $\VX = (\VIu,\VQs,\VUs)$ on simulated
synthetic datasets. \LD{We refer to \textit{linear RHAPSODIE} when we reconstruct $\VX = (\VIs,\VQs,\VUs)$ and to \textit{non-linear RHAPSODIE} when we reconstruct $\VX = (\VIu,\VQs,\VUs)$. Both parametrizations allow us to access the parameters of interest $\VIp$ and $\V\theta$ using 
Eq.~\eqref{eq:Stokes-to-Jones}.} The best choice of polarimetric parameters is then used to
process astrophysical datasets.
  

\section{Data calibrations}
\subsection{Detector calibration}
\label{subsec:detectorcalib}

Before the application of a reconstruction method, the calibration of the data
is essential to account for the noise and the artifacts linked to the
measurement. It allows for the estimation and correction of any pollution
induced by the sky background or the instrument as well as the detector
behavior in terms of errors on pixel values.

We use an inverse method to calibrate the raw data from these effects: the
quantity required for the calibration are jointly estimated from the likelihood
of the calibration data direct model \citep{denneulin2020theses}. In this method,
all calibration data are expressed as a function of the different contributions
(\ie, flux, sky background, instrumental background, gain, noise, and quantum
efficiency). All these quantities are then jointly estimated by the minimization
of the \LD{co-}log-likelihood of the data. Calibrated data are 
corrected 
\LD{for contribution of the bias and} the background and \LD{for non-uniform sensitivity and throughput. The calibration also provides associated weights computed according to the estimated variance, see Eq.~\eqref{eq:weights}.}
Finally,
\LDrevised{the} 
\LD{\LDrevised{defective} pixels are detected} by crossing several criteria, such
as their linearity, their covariance \LD{compared to that of} the other pixels, 
or the values of their likelihood \LD{in the calibration data}. 
This calibration method produces calibrated data outputs
$(d_k)_{\To{k}{1}{K}}$ and their weights $(\M{W}_k)_{\To{k}{1}{K}}$.

\subsection{Instrumental calibration}

\begin{figure*}
\centering \includegraphics[width=\textwidth]{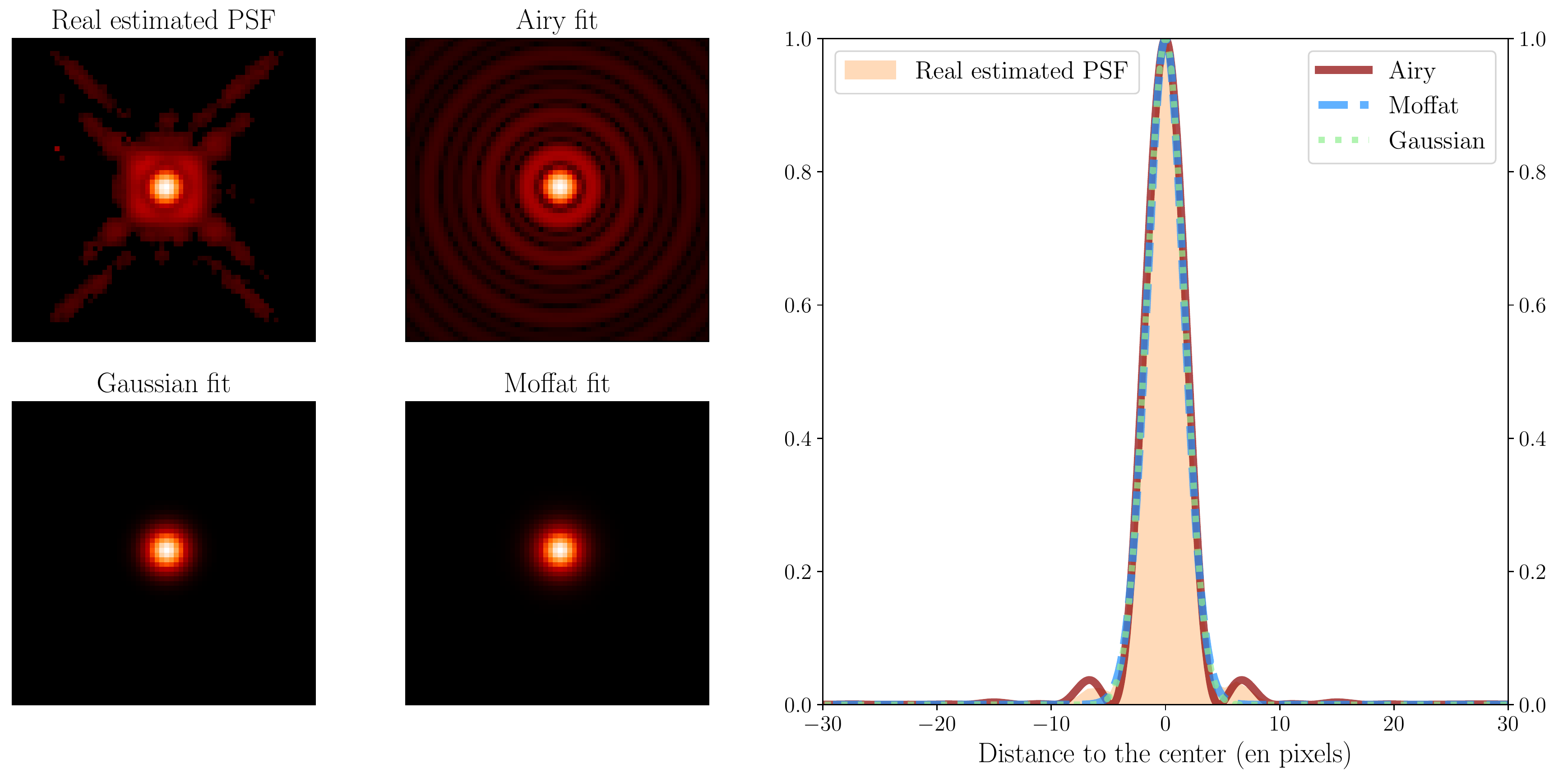}
\caption{\LD{Example of the fitting of the true PSF on the empirical SPHERE/IRDIS data of the target HD~106906 reduced with RHAPSODIE. Observed PSF in H band (upper left) and
different parametrization (Airy, Gaussian, and Moffat). Intensities are in log-scale to enhance the faint diffraction patterns.}} \label{fig:PSF_FIT}
\end{figure*}

The instrumental calibration 
\LD{is a required step} for the estimation, from dedicated data, of
the instrumental PSF and of the star centers on each side of the detector.
Since the star is placed behind the coronagraphic mask, simultaneous PSF
estimation is not possible. To estimate the PSF, we use a dedicated flux
calibration (\texttt{STAR-FLUX}) that is acquired just before and after the
science exposure by offsetting the telescope to about $0.5$ arcsec with respect to
the coronagraphic mask by using the SPHERE tip/tilt mirror
\citep{beuzit_sphere:_2019}. \LD{Consequently, the PSF is recorded with similar atmospheric conditions (listed in Table~\ref{tab:DataSetInfo}) as the science observations.} When performing this calibration, suitable neutral
density filters are inserted to avoid detector saturation. It has been shown in
\citep{beuzit_sphere:_2019} that these neutral density filters do not affect
the PSF shape and thus its calibration. This instrumental calibration leads to
the estimation of the PSF 
\LD{modeled through the operators $\M{A}_k$.  
The PSF model does not include the spiders to remain rotation invariant
(see Fig.~\ref{fig:PSF_FIT}). }

\LD{In the case of synthetic observations, the assumed 2D PSF is extracted from the real data RY~Lup (\eg similar to the top left image on Fig.~\ref{fig:PSF_FIT}). For astrophysical observations, we fit a 2D PSF model on pre-reduced PSF data for each observed target (\eg Gaussian, Airy, and Moffat fits on Fig.~\ref{fig:PSF_FIT}).} 

During the coronographic observing sequence, the star point spread function
peak is hidden by the coronagraphic mask and its position was determined using
a special calibration (\texttt{STAR-CENTER}) where four faint replicas of the
star image are created by giving a bi-dimensional sinusoidal profile to the
deformable mirror \citep[see ][]{beuzit_sphere:_2019}. The \texttt{STAR-CENTER}
calibration was repeated before and after each science observation, and
resulting center estimations were averaged. In addition we use the derotator
position and the true north calibration from \cite{maire2016sphere} to extract
the angle of rotation of the north axis. These instrumental calibration steps
lead to the estimation of the transformation operators $\M{T}_{j,k}$ which
rotate and translate the maps of interest to make the centers and the
north axes coincide with those in the data.

\subsection{Polarization calibration}

When the light is reflected by the optical devices in the instrument, some
instrumental polarization is introduced, resulting in a loss of polarized
intensity and cross-talk contamination between \LD{Stokes parameters} $Q$ and $U$.

The classical method to compensate for this instrumental polarization is to
employs the azimuthal \LD{$Q_{\phi}$ and $U_{\phi}$} parameters to reduce the noise floor
in the image \citep{avenhaus_structures_2014}. Because this method is limited
to face-on disks, we instead use the method developed by
\cite{van_holstein_polarimetric_2020} which rely on the pre-computed
calibration of the instrumentation polarization as a function of the
observational configurations. The pipeline IRDAP
\citep{van_holstein_polarimetric_2020} yields the possibility to determine and
correct the instrumental polarization  in the signal reconstruction, yet this
reconstruction requires to estimate first the $\Qs$ and $\Us$ parameters, to
perform the instrumental polarization correction. After computing the double
difference, IRDAP uses a Mueller matrix model of the instrument carefully
calibrated using real on sky data to correct for the polarized intensity $\Ip$
(created upstream of the HWP) and crosstalk of the telescope and instrument to
compute the model-corrected $\Qs$  and $\Us$ images.

Using IRDAP, we estimate the instrumental transmission parameters
$(\nu_{j,k,l})_{j\in \{1,\;2\}, \To{k}{1}{K}, \To{\ell}{1}{L}}$ from the
Mueller matrix to calibrate the instrumental polarization in our datasets.

IRDAP also determines the corresponding uncertainty by measuring the stellar
polarization for each HWP cycle individually and by computing the standard error
of the mean over the measurements. Finally, IRDAP creates an additional set of
$\Qs$ and $\Us$ images by subtracting the measured stellar polarization from the
model-corrected $\Qs$  and $\Us$ images. We use a similar method to correct for the
stellar polarization which is responsible for strong light pollution at small
separation in particular for faint disks such as debris disks. The contribution
of this stellar light is estimated as different factors of $\Iu$ in both $\Qs$
and $\Us$, respectively $\varepsilon_\Qs$ and $\varepsilon_\Us$. We estimate
this stellar contribution by using a pixel annulus $\Omega$ located at a
separation where there is no disk signal (either near the edge of the
coronograph or at the separation corresponding to the adaptive optics cut-off
frequency). Both corrections factors $\varepsilon_\Qs$ and $\varepsilon_\Us$ as
follow:
\begin{equation}
    \begin{cases}
        \varepsilon_{\Qs}= \big(\sum_{n \in \Omega} \Qs_n/\Iu_n\big)/N_\Omega\\
        \varepsilon_{\Us}= \big(\sum_{n \in \Omega} \Us_n/\Iu_n\big)/N_\Omega,
    \end{cases}
\end{equation} 
where $N_\Omega$ is the number of pixels of the ring $\Omega$. We then compute
$\Qs^\Tag{cor}=\Qs - \varepsilon_\Qs \Iu$ and $\Us^\Tag{cor}=\Us -
\varepsilon_\Us \Iu$ in order to create an additional set of $\Qs^\Tag{cor}$
and $\Us^\Tag{cor}$ images by subtracting the measured stellar polarization
from the model-corrected Q  and U images.

\section{Applications on high contrast polarimetric data}
\label{sec:application}

\LD{In 
this section, we will compare two RHAPSODIE configurations.
The first configuration defined as \textit{without deconvolution}, means that $\Is^\Tag{det}_{j,k,m}$ in the data fidelity term \eqref{eq:datafidelity} does not include the convolution, leading to the simplification of the equations \eqref{eq:direct-linear-model} and \eqref{eq:PSF-factorization}: \begin{align}
  \Is^\Tag{det}_{j,k,m}
  &= \sum_{\ell=1}^{3}\,\nu_{j,k,\ell}\,\sum_{n=1}^N \, (\M T_{j,k})_{m,n}\,
  S_{\ell,n} \, ,
  \label{eq:direct-linear-model-nodec}
\end{align}
 (\ie, $\M A_k$ is the identity). Such a configuration of RHAPSODIE aims to be comparable to the state-of-the-art methods (\ie, Double Difference and IRDAP). These methods do not include any deconvolution and we aim to show that RHAPSODIE also performs well in such a case. }
 
 \LD{The second configuration defined as \textit{with deconvolution} states for the full RHAPSODIE capabilities. Such a configuration of RHAPSODIE aims to show the benefits of the global model compared to an \textit{a posteriori} deconvolution to improve the angular resolution.}

\subsection{Application on synthetic data}

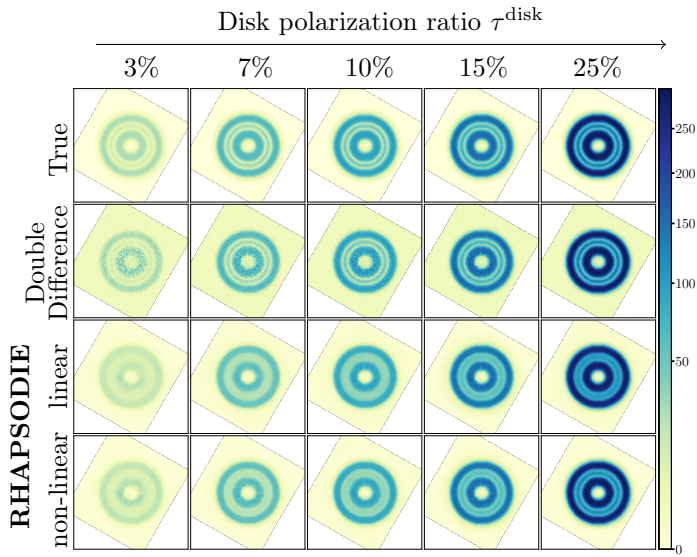
\begin{figure}[!b]\vspace{-0.2cm}
    \centering
    \begin{tikzpicture}
        \begin{scope}
            \draw (0,0) node[below right]{\includegraphics[width=0.45\textwidth]{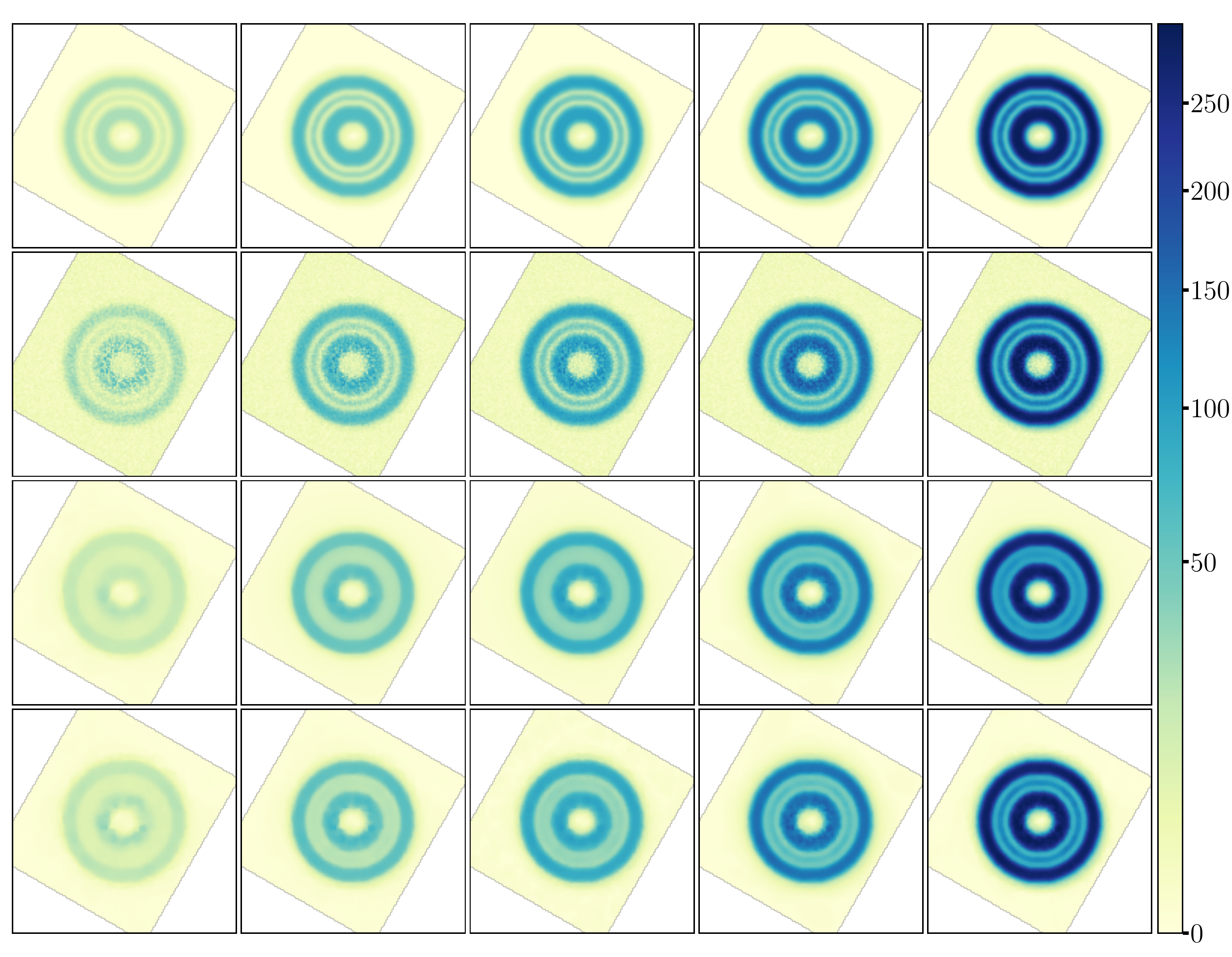}};
            \draw (-0.5,-4.9) node[rotate=90]{\bf RHAPSODIE};
            \draw (0, -5.6) node[rotate=90] {non-linear};
            \draw (0, -4.1) node[rotate=90] {linear};
            \draw (-0.3, -2.6) node[rotate=90] {Double};
            \draw (0, -2.6) node[rotate=90] {Difference};
            \draw (0, -1.1) node[rotate=90] {True};
            \draw[->] (0.5,0.3) -> (8,0.3) node[midway, above]{Disk polarization ratio $\tau^\Tag{disk}$};
            \draw (1.1,0) node{$3\%$};
            \draw (2.6,0) node{$7\%$};
            \draw (4.1,0) node{$10\%$};
            \draw (5.6,0) node{$15\%$};
            \draw (7.1,0) node{$25\%$};
        \end{scope}
        \end{tikzpicture}\vspace{-0.4cm}
    \caption{Visual comparison of the reconstructed polarized intensity $\Ip$ with the state-of-the-art Double Difference and the RHAPSODIE methods without deconvolution.}
    \label{fig:compvisunodeconv}
\end{figure}

\begin{figure}[!b]\vspace{-0.2cm}
\centering
    \begin{tikzpicture}\vspace{-0.2cm}
        \begin{scope}
            \draw (0,0) node[below right]{\includegraphics[width=0.45\textwidth]{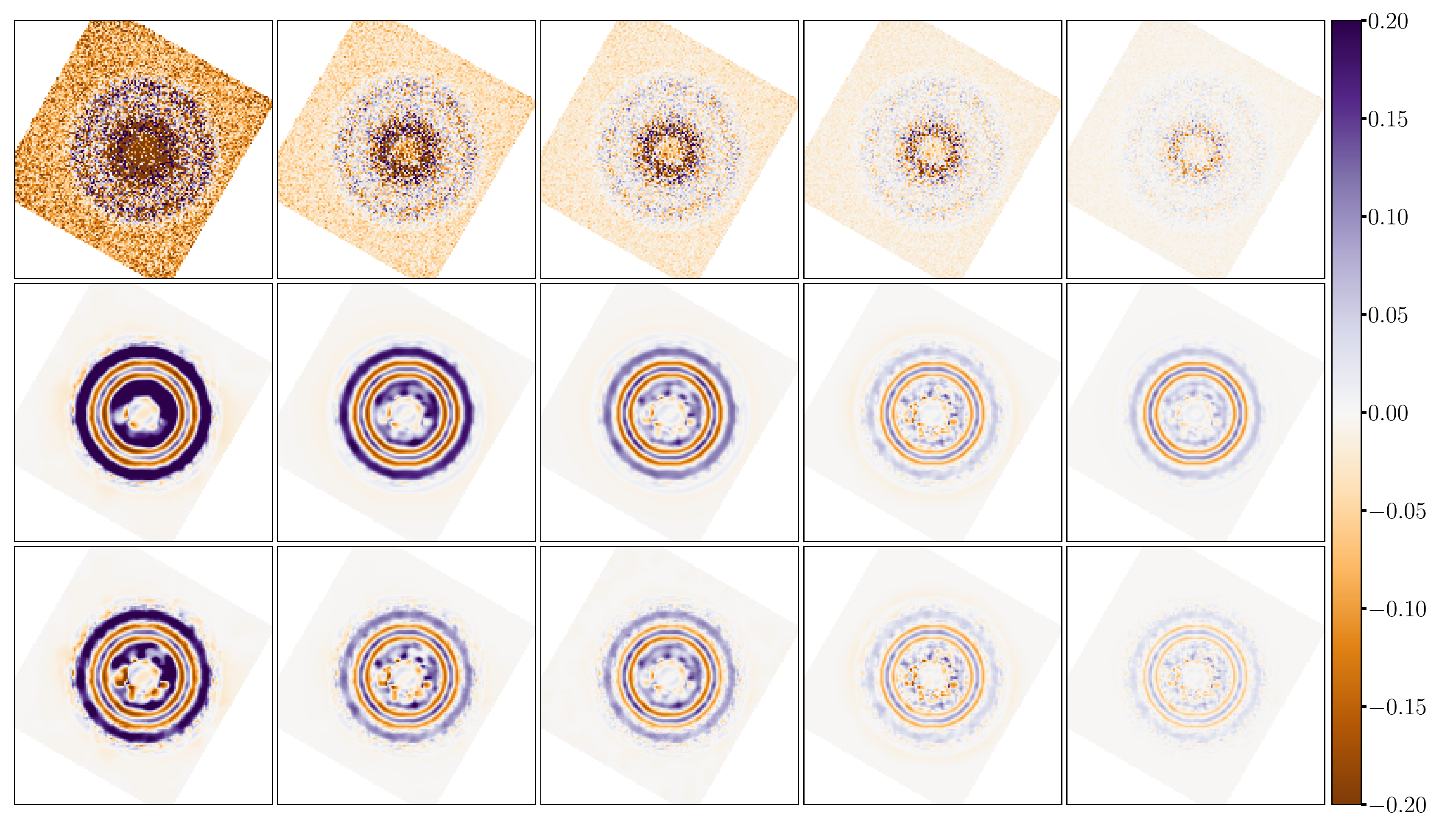}};
            \draw (-0.5,-3.2) node[rotate=90]{\bf RHAPSODIE};
            \draw (0, -3.9) node[rotate=90] {non-linear};
            \draw (0, -2.5) node[rotate=90] {linear};
            \draw (-0.3, -1.1) node[rotate=90] {Double};
            \draw (0, -1.1) node[rotate=90] {Difference};
            \draw[->] (0.5,0.3) -> (8,0.3) node[midway, above]{Disk polarization ratio $\tau^\Tag{disk}$};
            \draw (1.1,0) node{$3\%$};
            \draw (2.6,0) node{$7\%$};
            \draw (4.1,0) node{$10\%$};
            \draw (5.6,0) node{$15\%$};
            \draw (7.1,0) node{$25\%$};
        \end{scope}
        \end{tikzpicture}\vspace{-0.4cm}
    \caption{Maps of \LD{errors} of the reconstructions displayed on Fig.~\ref{fig:compvisunodeconv}. These \LD{errors} are obtained as the difference between the true and the reconstructed images.}
    \label{fig:compvisunodeconvres}
\end{figure}

\begin{figure}[!b]
\centering
\includegraphics[width=0.48\textwidth]{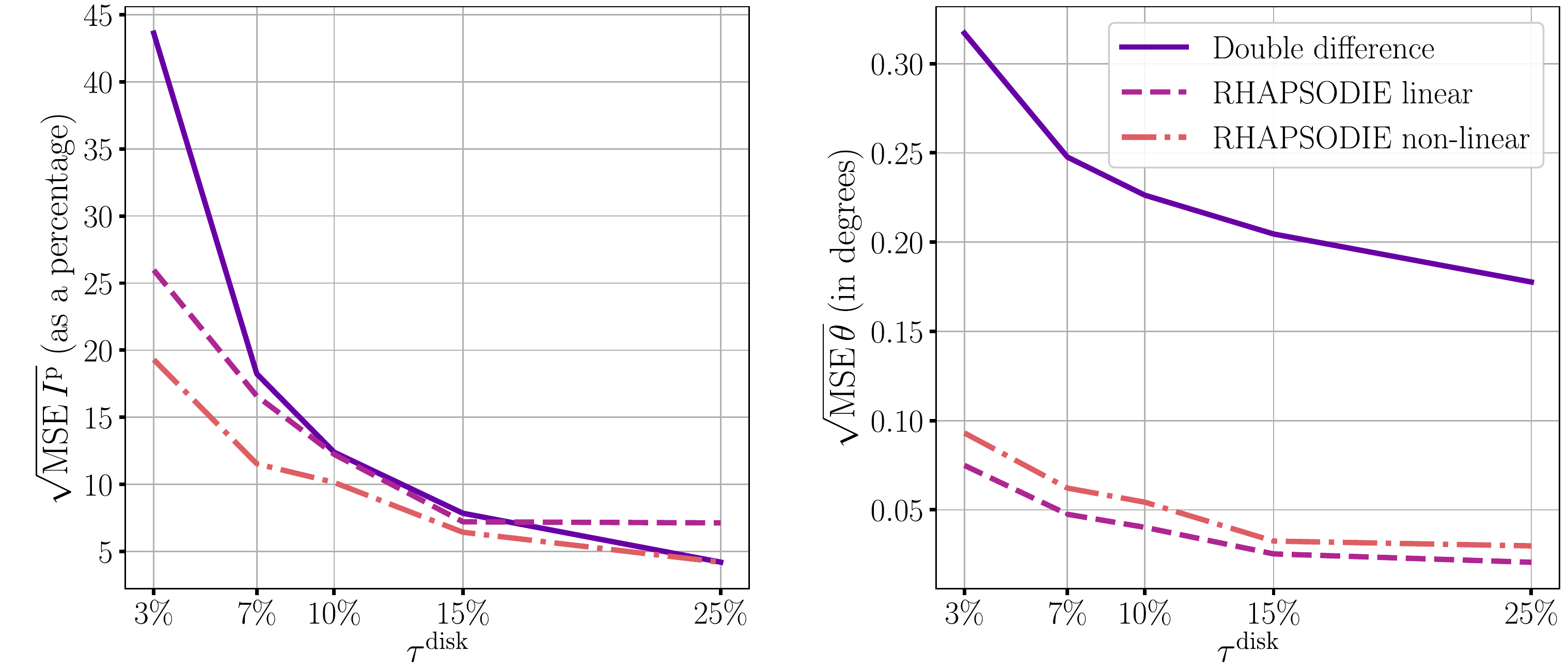}\vspace{-0.4cm}
\caption{Comparison of the MSE between the true map and the estimated map of the polarized intensity $\Ip$ and of the angle of polarization $\theta$ for the Double Difference and the linear and non-linear RHAPSODIE methods without deconvolution.}
\label{fig:msenodec}
\end{figure}

\begin{figure}[!b]\vspace{-0.08cm}
    \centering
    \begin{tikzpicture}
        \begin{scope}
            \draw (0,0) node[below right]{\includegraphics[width=0.45\textwidth]{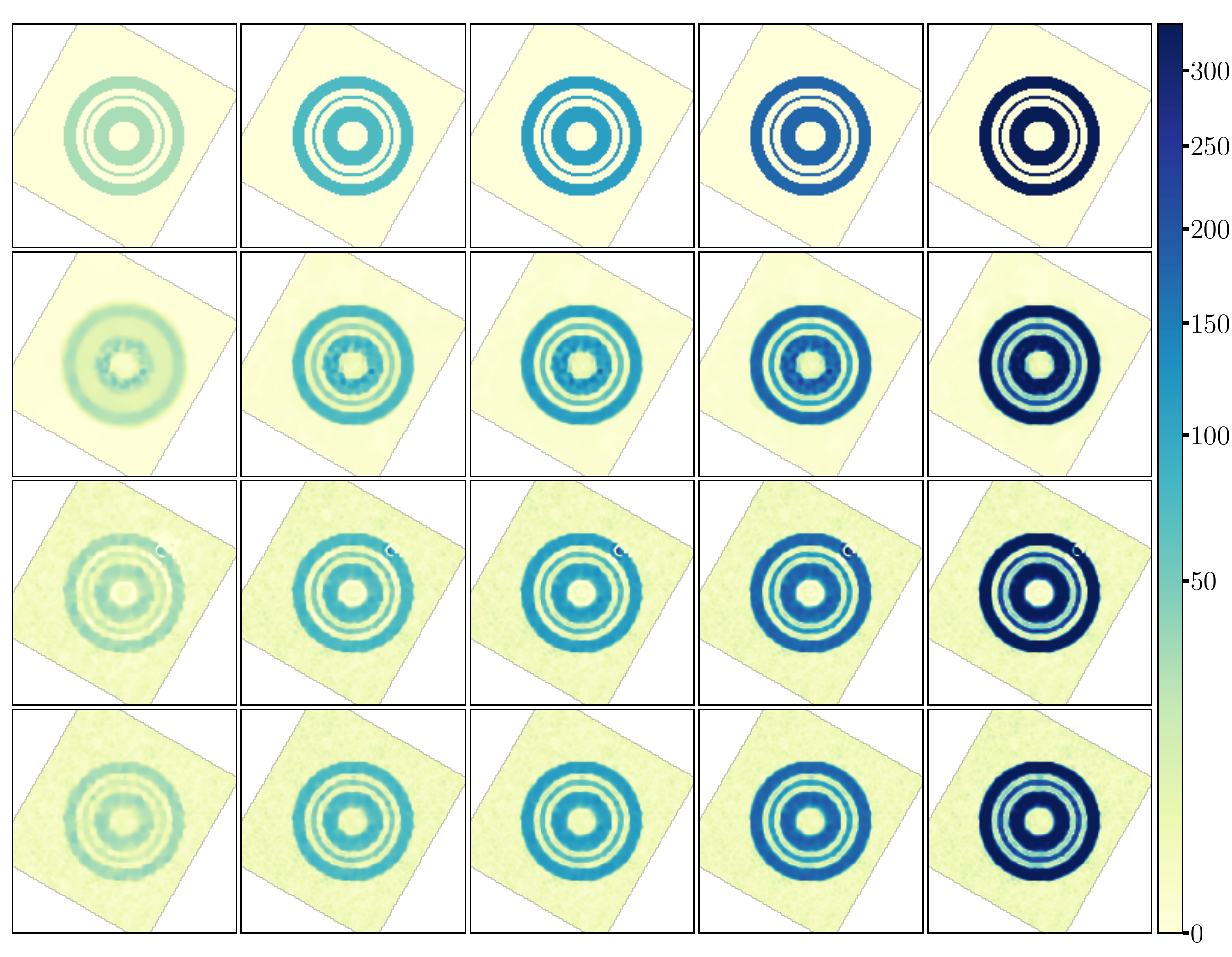}};
            \draw (-0.5,-4.9) node[rotate=90]{\bf RHAPSODIE};
            \draw (0, -5.6) node[rotate=90] {non-linear};
            \draw (0, -4.1) node[rotate=90] {linear};
            \draw (-0.3, -2.6) node[rotate=90] {Double};
            \draw (0, -2.6) node[rotate=90] {Difference};
            \draw (0, -1.1) node[rotate=90] {True};
            \draw[->] (0.5,0.3) -> (8,0.3) node[midway, above]{Disk polarization ratio $\tau^\Tag{disk}$};
            \draw (1.1,0) node{$3\%$};
            \draw (2.6,0) node{$7\%$};
            \draw (4.1,0) node{$10\%$};
            \draw (5.6,0) node{$15\%$};
            \draw (7.1,0) node{$25\%$};
        \end{scope}
        \end{tikzpicture}\vspace{-0.4cm}
    \caption{Visual comparison of the reconstructed polarized intensity $\Ip$ with the state-of-the-art Double Difference and the RHAPSODIE methods \LD{with} deconvolution. }
    \label{fig:compvisudeconv}
\end{figure}

\begin{figure}[!b]\vspace{-0.2cm}
    \centering
    \begin{tikzpicture}\vspace{-0.2cm}
        \begin{scope}
            \draw (0,0) node[below right]{\includegraphics[width=0.45\textwidth]{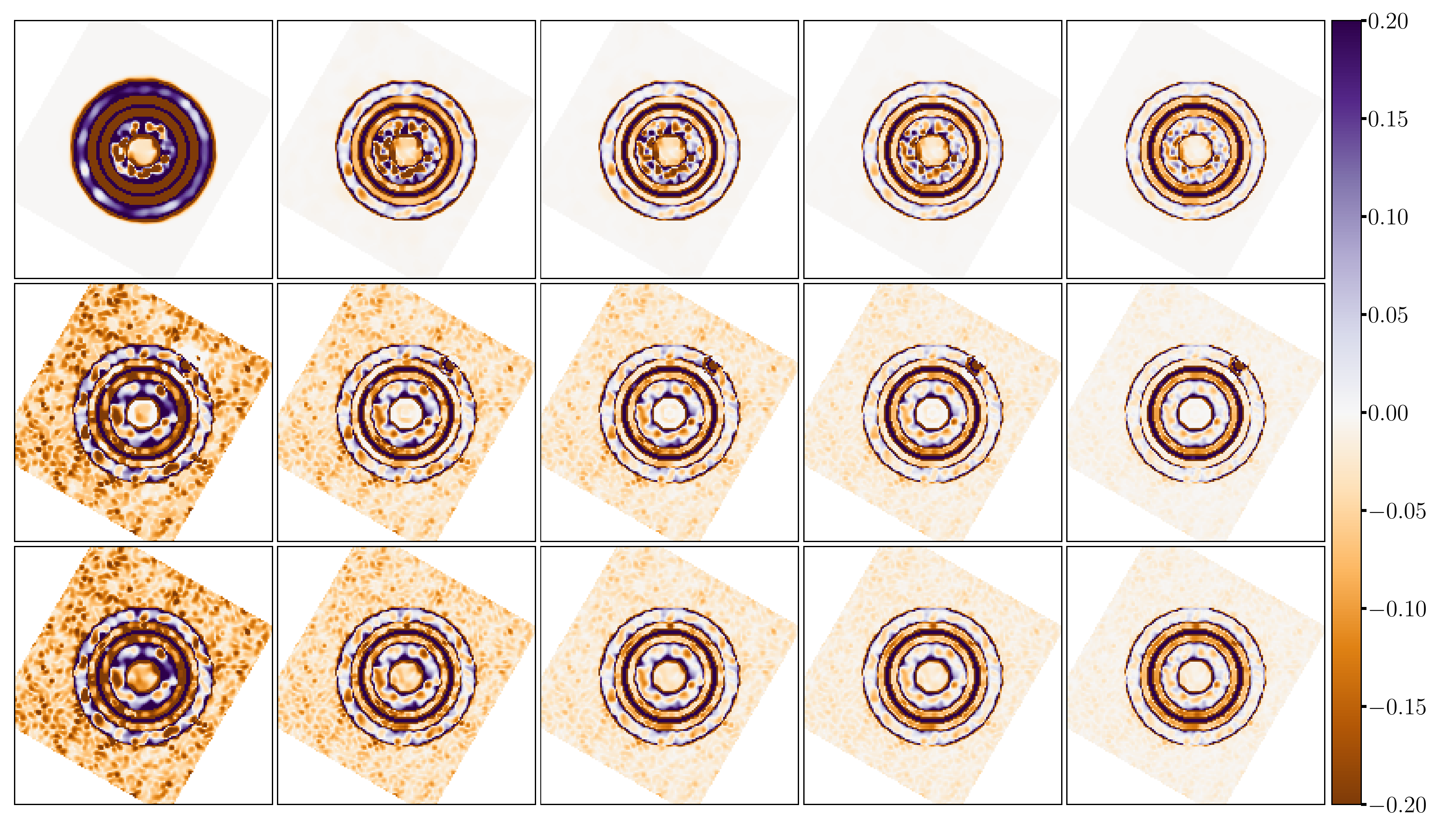}};
            \draw (-0.5,-3.2) node[rotate=90]{\bf RHAPSODIE};
            \draw (0, -3.9) node[rotate=90] {non-linear};
            \draw (0, -2.5) node[rotate=90] {linear};
            \draw (-0.3, -1.1) node[rotate=90] {Double};
            \draw (0, -1.1) node[rotate=90] {Difference};
            \draw[->] (0.5,0.3) -> (8,0.3) node[midway, above]{Disk polarization ratio $\tau^\Tag{disk}$};
            \draw (1.1,0) node{$3\%$};
            \draw (2.6,0) node{$7\%$};
            \draw (4.1,0) node{$10\%$};
            \draw (5.6,0) node{$15\%$};
            \draw (7.1,0) node{$25\%$};
        \end{scope}
        \end{tikzpicture}\vspace{-0.4cm}
    \caption{Maps of \LD{errors} of the reconstructions displayed on
    Fig.~\ref{fig:compvisudeconv}. These \LD{errors} are obtained as the
    difference between the true and the reconstructed images.}
    \label{fig:compvisudeconvres}
\end{figure}

\begin{figure}[!b]
\centering \includegraphics[width=0.48\textwidth]{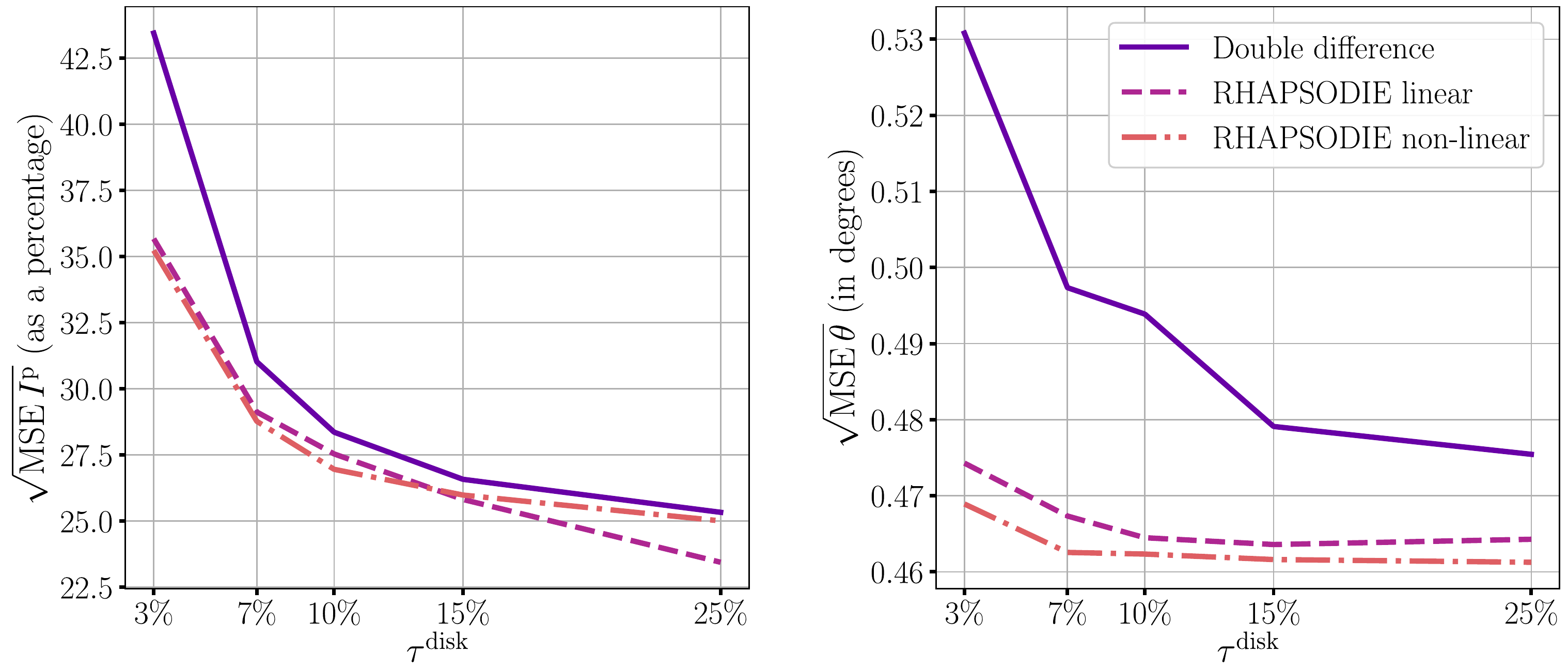}\vspace{-0.4cm}
\caption{Comparison of the MSE between the true map and the estimated map of
the polarized intensity $\Ip$ and of the angle of polarization $\theta$ for the
Double Difference and the linear and non-linear RHAPSODIE methods with
deconvolution.} \label{fig:msedec}
\end{figure}

\LD{The performance of the linear and non-linear methods are first evaluated on synthetic}
datasets, without (resp. with) the deconvolution displayed on
Fig.~\ref{fig:compvisunodeconv}, Fig.~\ref{fig:compvisunodeconvres} and
Fig.~\ref{fig:msenodec} (resp. Fig.~\ref{fig:compvisudeconv},
Fig.~\ref{fig:compvisudeconvres} and Fig.~\ref{fig:msedec}).  These datasets are
composed of unpolarized residual stellar flux, mixed with unpolarized and
polarized disk flux. We produce several synthetic datasets following steps
given in Appendix~\ref{sec:datasimu}, for different ratios of the polarization
of the disk, called $\tau^\Tag{disk}$ \LD{and defined in the equation \eqref{eq:taudisk}}.

The results of the RHAPSODIE methods are compared to the results obtained with
the classical Double Difference method \citep{tinbergen_astronomical_2005, avenhaus_structures_2014}. The Double Difference is applied on
recentered and rotated datasets with the bad pixels interpolated. For the
comparison with deconvolution, the results of the Double Difference are
deconvolved after the reduction. \LD{In order to 
provide fair comparisons with RHAPSODIE, we propose to use an inverse approach rather than applying a high-pass spatial filtering. 
The deconvolved Double Difference reconstructions are obtained by solving the following problem:}
\begin{align}
    (\Es{\Qs}, \Es{\Us}) \in & \argmin_{(\Qs, \Us) \in \R^N \times \R^N} \Bigg[\Vert \Es{\Qs}^\Tag{D. D.} - \M{A} \Qs \Vert^2 + \Vert \Es{\Us}^\Tag{D. D.} - \M{A} \Us \Vert^2 \notag \\
    & \quad + \LD{ \lambda_{(\Qs +\Us)^\Tag{D. D.}} f_{(\Qs +\Us)^\Tag{D. D.}}(\Qs,\Us)} \Bigg]
\end{align}
\LD{where $\Es{\Qs}^\Tag{D. D.}$ and $\Es{\Us}^\Tag{D. D.}$ are the Stokes parameters reconstructed with the Double Difference and $\M A$ represents the convolution by the PSF.}
This deconvolution method \LD{performs the deconvolution of Stokes parameters $\Qs$ and $\Us$
jointly in order to} sharpen the polarized intensity image, but does
not recover the polarization signal lost by averaging over close-by
polarization signals with opposite sign and may even introduce artificial
structures that are not present in the original source.

\LD{For the reconstruction with RHAPSODIE, the hyperparameters of regularization $\lambda_\rho$, and $\mu_\rho$ are
chosen in order to minimize the total Mean Square Error (MSE), \ie the sum of the MSE
between each estimated parameter $\Es{\Iu}$, $\Es{\Ip}$ and $\Es{\theta}$, obtained from $\Es{\VX}$ from Eq.~\eqref{eq:Stokes-to-Jones}, and the ground truth $\Tr{\Iu}$, $\Tr{\Ip}$ and $\Tr{\theta}$. The total MSE is given by:
\begin{align}
    \textrm{MSE}^\Tag{tot}  &= \sum_{n=1}^{N_{\Iu}} \Expect\Big[ \big(\Es{\Iu} - \Tr{\Iu}\big)^2 \Big] \notag \\
                            & \quad + \sum_{n=1}^{N_{\Ip}}  \Expect\Big[ \big(\Es{\Ip} - \Tr{\Ip} \big)^2 \Big] \notag \\
                            & \quad + \sum_{n=1}^{N_{\theta}}  \Expect\Big[\textrm{angle} \big(\erm^{2 \irm ( \Es{\theta} - \Tr{\theta})}\big)^2 \Big].
\label{eq:EQMparametermulticomp}
\end{align}
where,  $N_{\Iu}$, $N_{\Ip}$, and $N_{\theta}$ are the number of pixels with
signal of interest in the respective $\Iu$, $\Ip$, and $\theta$ maps.}

\LD{For the deconvolution of the Double Difference results, $\lambda_{(\Qs +\Us)^\Tag{D. D.}}$ and $\mu_{(\Qs +\Us)^\Tag{D. D.}}$ are chosen to minimize only the sum of the MSE on $\Iu$ and $\theta$.}


\LD{For the reconstructions without the deconvolution, Fig.~\ref{fig:compvisunodeconv} and 
Fig.~\ref{fig:compvisunodeconvres} show that the reconstructions are less noisy with RHAPSODIE.
The inner circle is always better reconstructed. Moreover, the non-linear reconstruction 
(\ie, minimization on $\Iu$, $\Qs$ and $\Us$) is better than the linear when $\tau^\Tag{disk}$ grows. 
In both configurations of RHAPSODIE (non-linear and linear), the thin ring is not as well reconstructed 
as with the Double Difference.} It is possible to recover
such sharper structure with RHAPSODIE, by reducing the regularization
weight (\ie, reducing the hyperparameter \LD{$\lambda_{\Qs + \Us}$}). \LD{However}, if \LD{$\lambda_{\Qs + \Us}$}
is too small, the data will be overfitted and the noise in the \LD{reconstructed images}
will be amplified. \LD{It is thus necessary to keep a good trade-off between a
smooth solution and a solution close to the data. Classically, minimizing the
MSE is a good trade-off between underfitting and overfitting. 
The MSE curves in Fig.~\ref{fig:msenodec} are coherent with the observations. When $\tau^\Tag{disk}=25\%$,
the non-linear RHAPSODIE MSE and Double Difference MSE are equivalent, but on the reconstructions, we can see that
if the thin circle is better reconstructed with the Double Difference, the inner circle is better reconstructed with RHAPSODIE.
Moreover, we can see that the angle error is more than three-time smaller with RHAPSODIE compared to the Double Difference error.}\\

\LD{For the reconstructions with the deconvolution, the RHAPSODIE methods better 
reconstructions of the polarized intensity as shown on the
reconstructions in Fig.~\ref{fig:compvisudeconv}, mostly for the inner ring, even if
the background is less biased with the proposed deconvolved Double Difference (as shown on the 
error maps in Fig.~\ref{fig:compvisudeconvres}). In
fact, as shown on the MSE displayed in Fig.~\ref{fig:msedec}, for $\tau^\Tag{disk}=3\%$, RHAPSODIE 
delivers a more accurate polarized intensity estimation. The errors 
of the reconstructions are larger for the Double Difference
than for the RHAPSODIE methods. The RHAPSODIE methods also allow us to achieve a better reconstruction of the angle of
polarization, mostly with the non-linear estimation. 
The non-linear reconstruction appears to be more efficient as being not polluted by the artifacts of deconvolution of the unpolarized point source companion as
seen on Fig.~\ref{fig:compvisudeconvres}. Even if according to the MSE in Fig.~\ref{fig:msedec},
for $\tau^\Tag{disk}=25\%$, the MSE is smaller for the linear RAPSODIE. in fact, the space between 
the outer ring and the thin ring is better reconstructed with such a configuration 
(\LDrevised{see Fig.~\ref{fig:compvisudeconv} and Fig.~\ref{fig:compvisudeconvres}).}}

According to these results, the RHAPSODIE methods are better than the
state-of-the-art methods, in particular the non-linear RHAPSODIE method. 
\LD{The benefits of our contribution is clearly visible in the case of faint disks
and structures which are the most common on astrophysical data.}  This
is why in the following section dedicated to the astrophysical data, we select
the non-linear RHAPSODIE method with manual selection of the hyperparameters.


\subsection{Astrophysical data}

RHAPSODIE was applied to several IRDIS datasets dedicated to protoplanetary,
transition and debris disks, to test the efficiency of our method and to
compare it to the state-of-the-art method. For each reconstruction, we present
the map of the projected intensities, by using the standard azimuthal Stokes
parameters $\Qs_\phi$ and $\Us_\phi$ estimated from:
\begin{equation}
\begin{cases}
\Qs_\phi=\Qs_n \cos \left( 2 \psi_n \right)+ \Us_n \sin \left( 2 \psi_n \right)\\
\Us_\phi=\Us_n \cos \left( 2 \psi_n \right)-  \Qs_n \sin \left( 2 \psi_n \right),
\end{cases}
\label{eq:porthopperp}
\end{equation}
with $\psi_n = \arctan \left( \frac{n_1-n^\Tag{center}_1}{n_2-n^\Tag{center}_2}\right)$ where $\left(n_1,n_2\right)$ denotes the row and the column indices of the $n$-th pixel in the map and $\left( n_1^\Tag{center}, n_2^\Tag{center}\right)$ those of the pixel center. 
We also present the maps of polarized intensities $\Ip$ and the angles of polarization $\theta$ estimated by the different methods.

\begin{table*}[!t]
  \centering
  \begin{tabular}{lllrrrrrrrrrrrrr}
    \hline
    \multicolumn{1}{c}{Target} &
    \multicolumn{1}{c}{Date} &
    \multicolumn{1}{c}{Filter} &
    \multicolumn{1}{c}{$\Delta_t$ \LDrevised{(s)}}&
    \multicolumn{1}{c}{\LD{HWP} \LDrevised{cycles}} &
    \multicolumn{1}{c}{$K^\Tag{miss}$} &
    \multicolumn{1}{c}{$\Delta_t^\Tag{tot}$ \LDrevised{(s)}}&
    \multicolumn{1}{c}{\LD{seeing}} &
    \multicolumn{1}{c}{\LD{$\tau_0$ (ms)}}\\
    \hline
    \hline
    TW~Hydrae & $2015$-$04$-$01$ & H &$16$ & $22$ & $13$ & $5424$& 1.15 & 2.2\\
    IM~Lupus & $2016$-$03$-$14$ & H & $64$ & $6$ & $7$ & $2624$ &1.40 & 2.7\\
    MY~Lupus & $2016$-$03$-$16$ & H & $64$ & $7$ & $18$ & $2432$& 1.63 & 1.5\\
    RY~Lupus & $2016$-$05$-$27$ & H & $32$ & $8$ & $0$ & $4096$ & 0.65 & 3.3\\
    T~Chae & $2016$-$02$-$20$ & H &  $32$ & $30$ &$0$ & $3840$ & 1.8 & 2.2\\
    RXJ~1615 & $2016$-$03$-$15$ & H &  $64$ &$11$ & $7$ & $5184$ & 0.7 & 4.1\\
    HD~106906 & $2019$-$01$-$17/18/20$ & H & $32$ & $42$ & $0$ & $5376$ & 0.54 & 14.0\\
    HD~61005 & $2015$-$05$-$02$ & H & $16$ & $12$ & $16$ & $2816$ & 1.5 & 3.2\\
    AU~MIC & $2017$-$06$-$20$ & J & $16$ & $23$ & $0$ & $11776$ & 1.4 & 2.2\\
    \hline
  \end{tabular}
  \caption{\LD{Information of the datasets used for the target reconstruction: the name of the target, the date of the observation, the filter used, the exposition time $\Delta_t$ for one acquisition, the number of cycles of HWP, the number of frame missing or removed $K^\Tag{miss}$, the total exposition time of the observation $\Delta_t^\Tag{tot}$ considering $K^\Tag{miss}$, the seeing, and observation conditions $\tau_0$. The total number of frames in each dataset is given by $K=\Delta_t^\Tag{tot} \, / \,\Delta_t + K^\Tag{miss}$. }}
  \label{tab:DataSetInfo}
\end{table*}

\begin{table*}[!t]
  \centering
  \begin{tabular}{lllrrrrrrrrrrr}
    \hline
    \multicolumn{1}{c}{Target} &
    \multicolumn{1}{c}{$\lambda_{\Iu}^\Tag{no-dec}$} &
    \multicolumn{1}{c}{$\lambda_{\Qs + \Us}^\Tag{no-dec}$} &
    \multicolumn{1}{c}{$\mu_{\Iu}^\Tag{no-dec}$}&
    \multicolumn{1}{c}{$\mu_{\Qs + \Us}^\Tag{no-dec}$}&
    \multicolumn{1}{c}{$\lambda_{\Iu}^\Tag{dec}$} &
    \multicolumn{1}{c}{$\lambda_{\Qs + \Us}^\Tag{dec}$} &
    \multicolumn{1}{c}{$\mu_{\Iu}^\Tag{dec}$} &
    \multicolumn{1}{c}{$\mu_{\Qs + \Us}^\Tag{dec}$} &
    \multicolumn{1}{c}{PSF}\\
    \hline
    \hline
    TW~Hydrae  & $10^{4.0}$ & $10^{4.0}$ & $10^{-4.0}$ & $10^{-4.0}$ & $10^{4.0}$ & $10^{4.0}$ & $10^{-5.0}$ & $10^{-5.0}$ & Moffat \\
    IM~Lupus &$10^{3.0}$ & $10^{4.0}$ & $10^{-0.8}$ & $10^{-4.0}$ & $10^{5.0}$ & $10^{5.5}$ & $10^{-3.0}$& $10^{-3.0}$ & Airy\\
    MY~Lupus & $10^{2.0}$ & $10^{0.4}$ & $10^{-0.7}$ &$10^{-3.6}$ & $10^{1.0}$ & $10^{0.5}$  & $10^{-3.0}$& $10^{-3.0}$ & Moffat\\
    RY~Lupus &  $10^{4.0}$ & $10^{5.5}$ & $10^{-3.0}$ & $10^{-3.0}$ & $10^{5.0}$ & $10^{3.0}$  & $10^{-3.0}$ & $10^{-3.0}$ & Moffat\\
    T~Chae &  $10^{0.0}$ & $10^{-1.0}$ & $10^{-3.0}$  & $10^{-3.0}$ & $10^{0.5}$ & $10^{-1.0}$ & $10^{-2.0}$ & $10^{-2.0}$ & Moffat \\
    RXJ~1615 & $10^{0.0}$ & $10^{0.0}$ & $10^{-4.0}$ & $10^{1.0}$ & $10^{1.0}$ & $10^{-1.5}$ & $10^{-3.0}$ & $10^{-3.0}$ & Moffat\\
    HD~106906 &  $10^{2.5}$ & $10^{1.5}$ & $10^{-4.0}$ & $10^{-4.0}$ & $10^{5.0}$ & $10^{1.7}$ & $10^{2.2}$ & $10^{-1.8}$ & Airy \\
    HD~61005 &  $10^{5.0}$ & $10^{4.1}$ & $10^{-0.7}$ & $10^{-3.4}$ & $10^{5.0}$ & $10^{4.1}$ & $10^{-0.7}$ & $10^{-3.9}$ & Moffat \\
    AU~MIC &  $10^{5.0}$ & $10^{4.3}$ & $10^{-4.2}$& $10^{-3.2}$ & $10^{5.0}$ & $10^{4.3}$& $10^{-1.2}$ & $10^{-4.2}$ & Moffat \\
    \hline
  \end{tabular}
  \caption{\LD{Informations of the parameters used for the target reconstruction presented in this section: the name of the target and the values of the hyperparameters  $\lambda_{\Iu}^\Tag{no-dec}$, $\lambda_{\Qs + \Us}^\Tag{no-dec}$, $\mu^\Tag{no-dec}$ (resp. $\lambda_{\Iu}^\Tag{dec}$, $\lambda_{\Qs + \Us}^\Tag{dec}$, $\mu^\Tag{dec}$) used for the reconstruction with RHAPSODIE without deconvolution (resp. with deconvolution), and the parametrization model of the PSF used.}}
  \label{tab:RHAPSODIEInfo}
\end{table*}

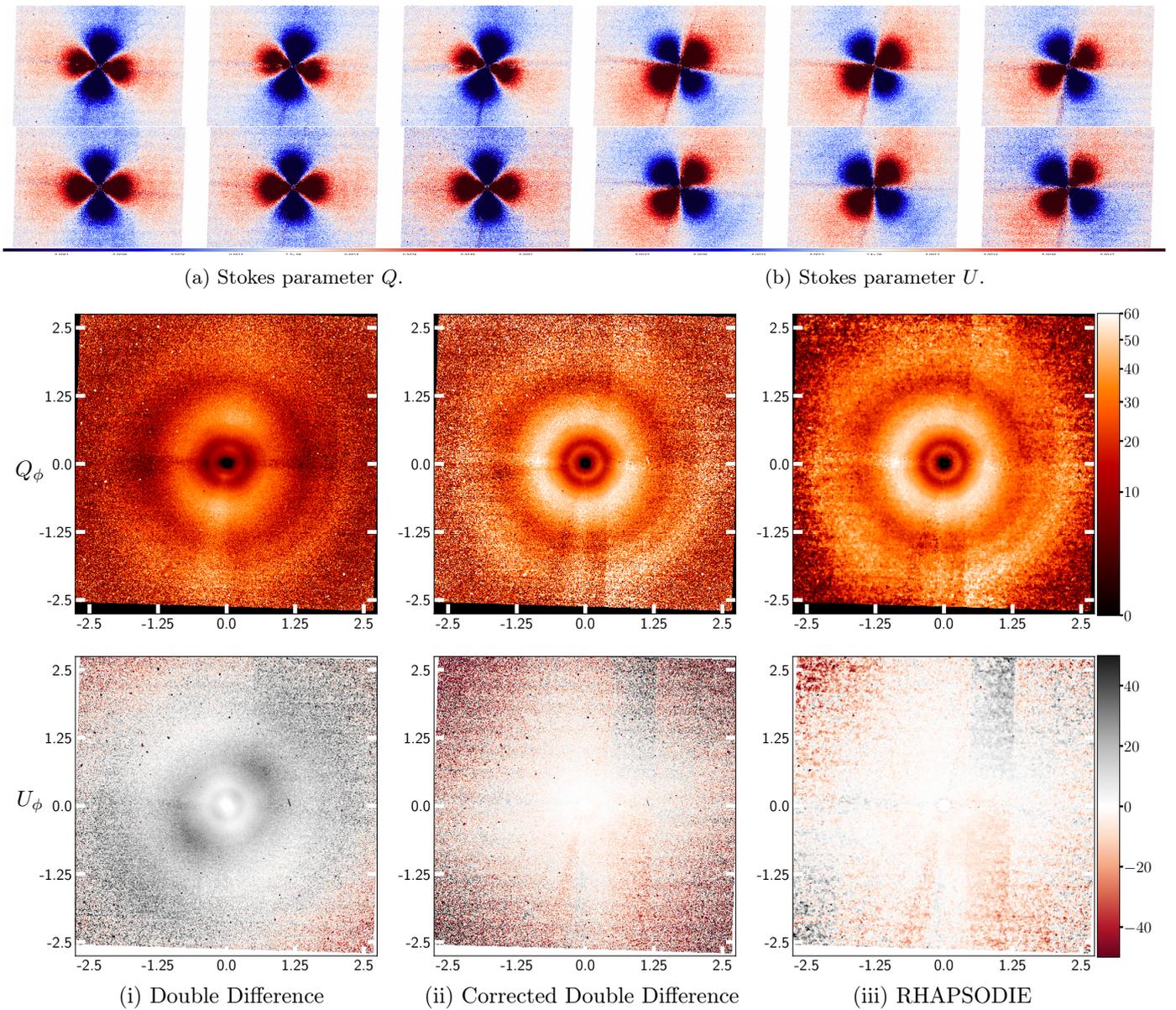
\begin{figure*}
    \centering
    \begin{subfigure}[b]{0.495\textwidth}
        \includegraphics[width=\textwidth]{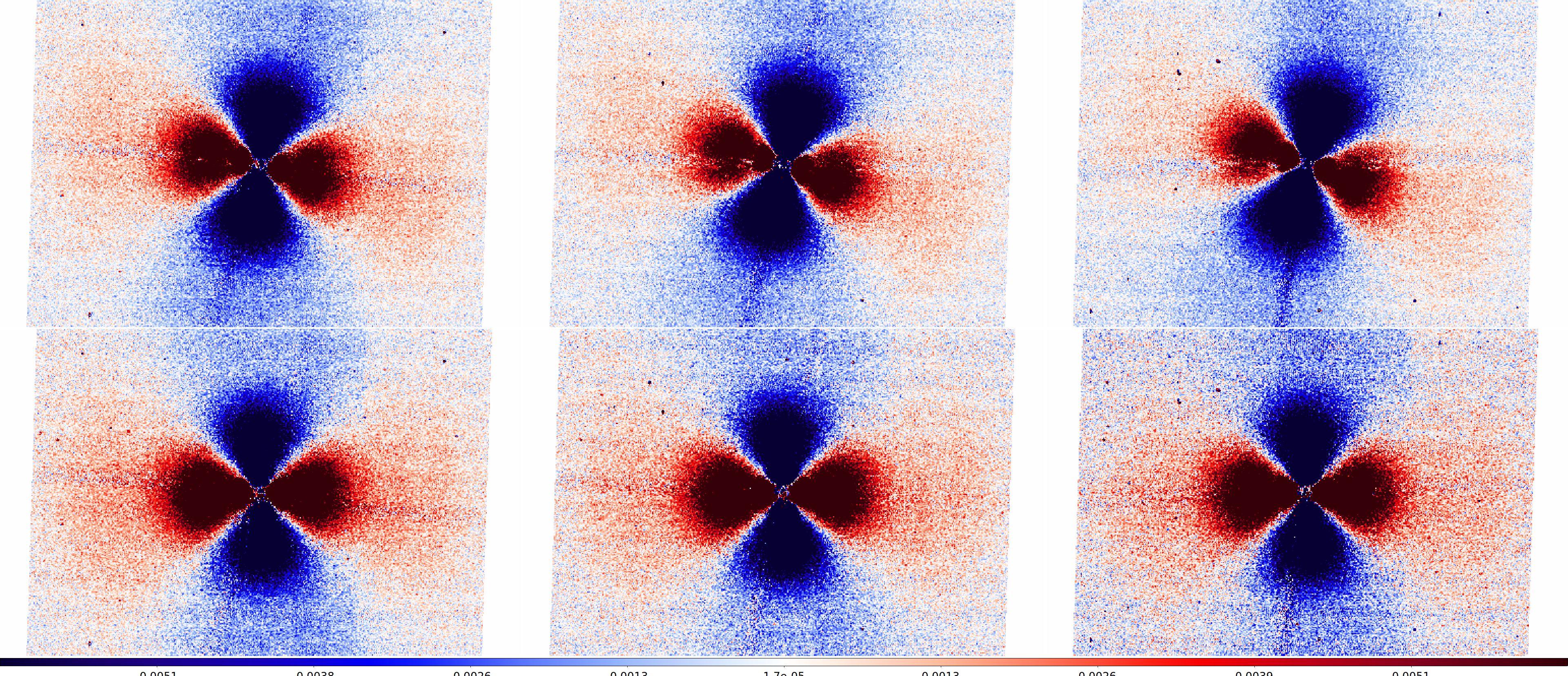}
        \caption{Stokes parameter $\Qs$.}
        \label{fig:ResultscompTWhydraeQ}
        \end{subfigure}\begin{subfigure}[b]{0.495\textwidth}
        \includegraphics[width=\textwidth]{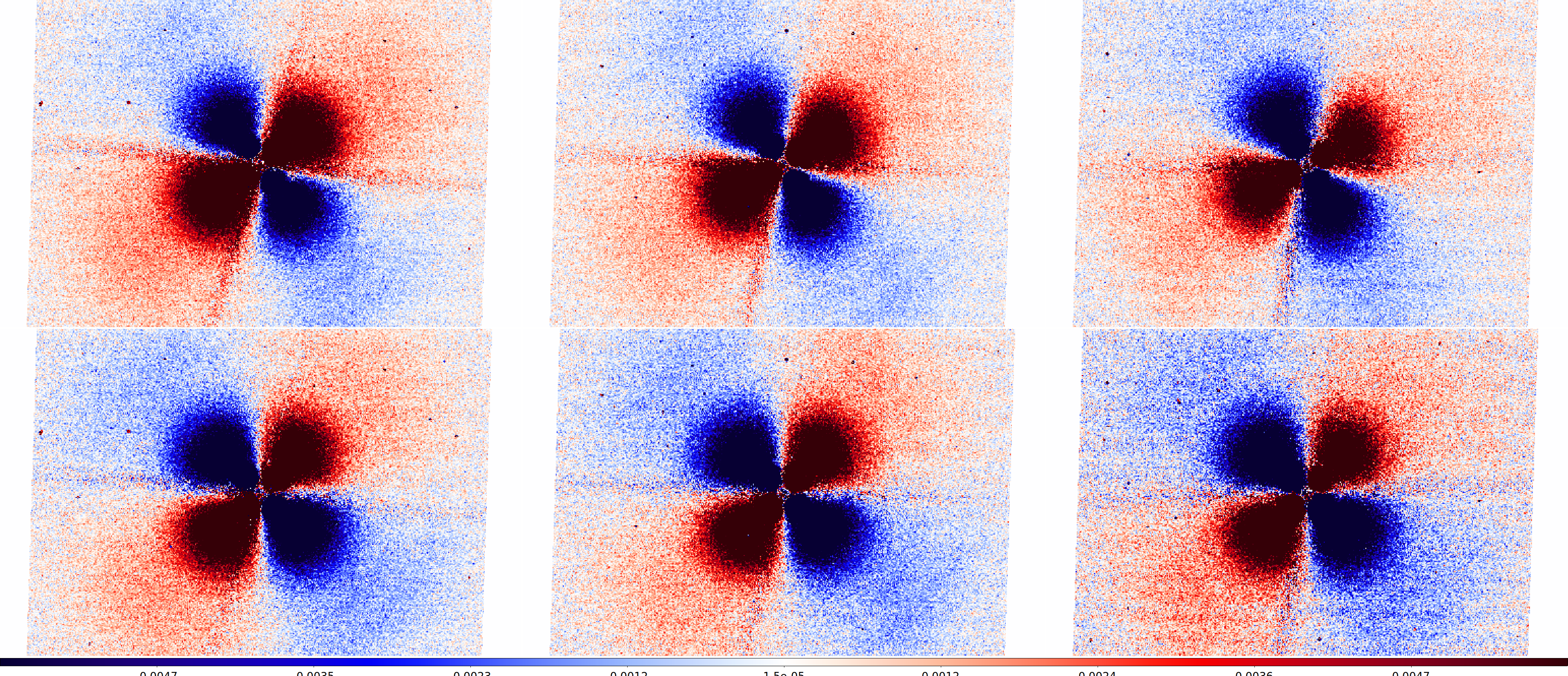}
        \caption{Stokes parameter $\Us$.}
        \label{fig:ResultscompTWhydraeU}
        \end{subfigure}
    \begin{subfigure}[b]{\textwidth}
    \centering
    \begin{tikzpicture}
        \draw (0,0) node[above right]{\includegraphics[width=0.95\textwidth]{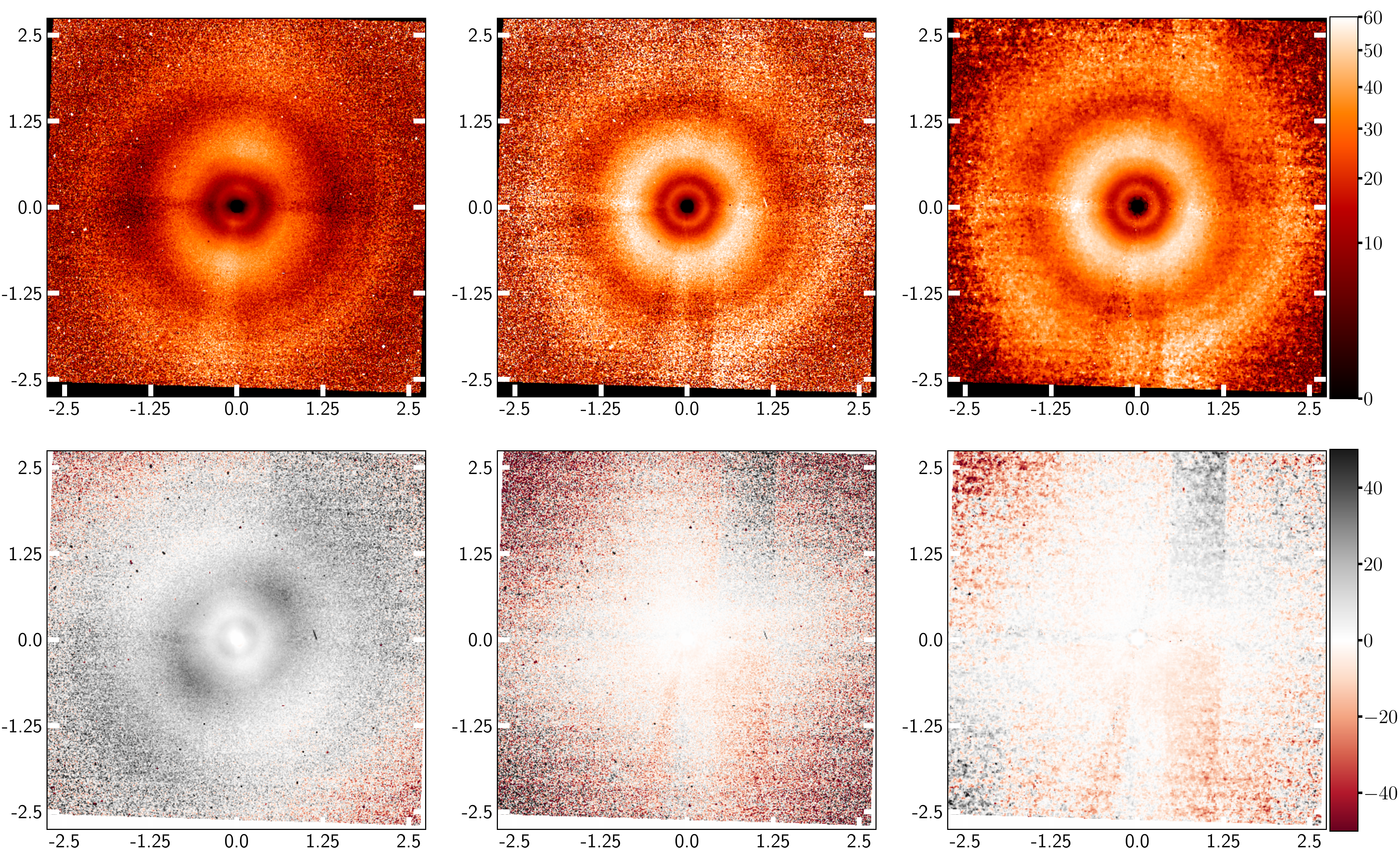}};
        \draw(0,8.) node{$Q_\phi$};
        \draw(0,2.8) node{$U_\phi$};
        \draw(3,0) node[below]{(i) Double Difference};
        \draw(8.65,0) node[below]{(ii) Corrected Double Difference};
        \draw(14.3,0) node[below]{(iii) RHAPSODIE};
    \end{tikzpicture}
        \caption{Azimutal Stokes parameters displayed in arcseconds : (i) Double Difference, (ii) Double Difference with the instrumental polarization correction, (iii) RHAPSODIE which includes by default the instrumental polarization correction. The intensities are multiplied in each pixel by the distance to the star $r^2$.}
        \label{fig:ResultscompTWhydraeQphi}
    \end{subfigure}
    \caption{Reconstruction of the $\Qs$ (a) and $\Us$ (b) parameters \LD{of the target TW~Hydrae}, for the first three cycle of HWP rotation, without (upper row) and with (lower row) the correction of the polarization. Without the correction, both $\Qs$ and $\Us$ are rotated and attenuated. The $\Qs_{\phi}$ and $\Us_{\phi}$ images reconstructed from the entire dataset are presented in (c). All the reconstructions are done without deconvolution to demonstrate mainly the efficiency of the instrumental polarization correction and the benefits of RHAPSODIE.}
    \label{fig:ResultscompTWhydrae}
\end{figure*}

First, the reconstructions of the target TW~Hydrae with the Double Difference
and RHAPSODIE without deconvolution are compared in the
Fig.\ref{fig:ResultscompTWhydrae}, without and with the correction of the
instrumental polarization. The images are scaled  by the square of the
separation to account for the drop-off of stellar illumination with distance.

The instrumental polarization in this dataset introduces a polarization
rotation and an attenuation (loss of polarization signal) of the intensity with
varying time during the observations which can be monitored easily because the
disk is face-on. If uncorrected, the combination of data from multiple
polarimetric cycles will result in very poorly constrained polarimetric
intensity measurements. When we correct the instrumental polarization (see
Fig.~\ref{fig:ResultscompTWhydraeQphi}) by using the IRDAP method, these
effects are compensated and the disk reconstructed is more accurate
(Fig.~\ref{fig:ResultscompTWhydraeQphi} (iii). As a result, the contamination
of the \LD{$\Us_\phi$} signal from cross-talk is decreased and becomes negligible as seen
in Fig.~\ref{fig:ResultscompTWhydraeQphi} \LD{(iii)}.

The comparison with \cite{2017Boekel,de_boer_polarimetric_2020} shows that our
method is less impacted by the bad pixels and improves the disk SNR in the area
where the signal is low. On the other hand, it is slightly more sensitive to
detector flat calibration accuracy. It is worth noticing that IRDAP has a
dedicated correction of the detector response nonuniformity (flat field
variation between the detector column) for the various amplifiers that is not
implemented in RHAPSODIE. However, the flat calibration of observations more
recent than 2016 have improved and do allow better calibration and as a
consequence do not impact our method efficiency anymore.

\begin{figure*}
    \centering
    \begin{tikzpicture}
        \begin{scope}
            \draw (0,0) 
            node[right]{\includegraphics[width=0.98\textwidth]{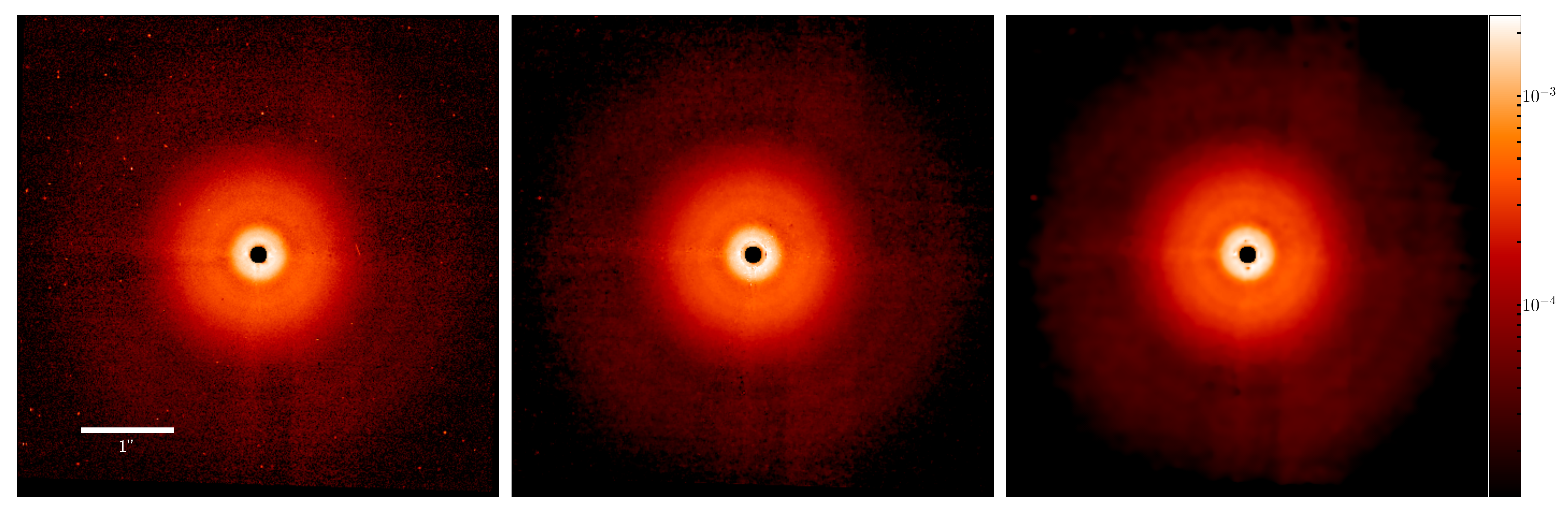}};
            \draw (0,0) node[rotate=90]{TW~Hydrae};
            \draw (3,3) node{Double Difference};
            \draw (8.75,3) node{RHAPSODIE (without dec.)};
            \draw (14.5,3) node{RHAPSODIE (with dec.)};
        \end{scope}

        \begin{scope}[yshift=-6cm]
            \draw (0,0) 
            node[right]{\includegraphics[width=0.98\textwidth]{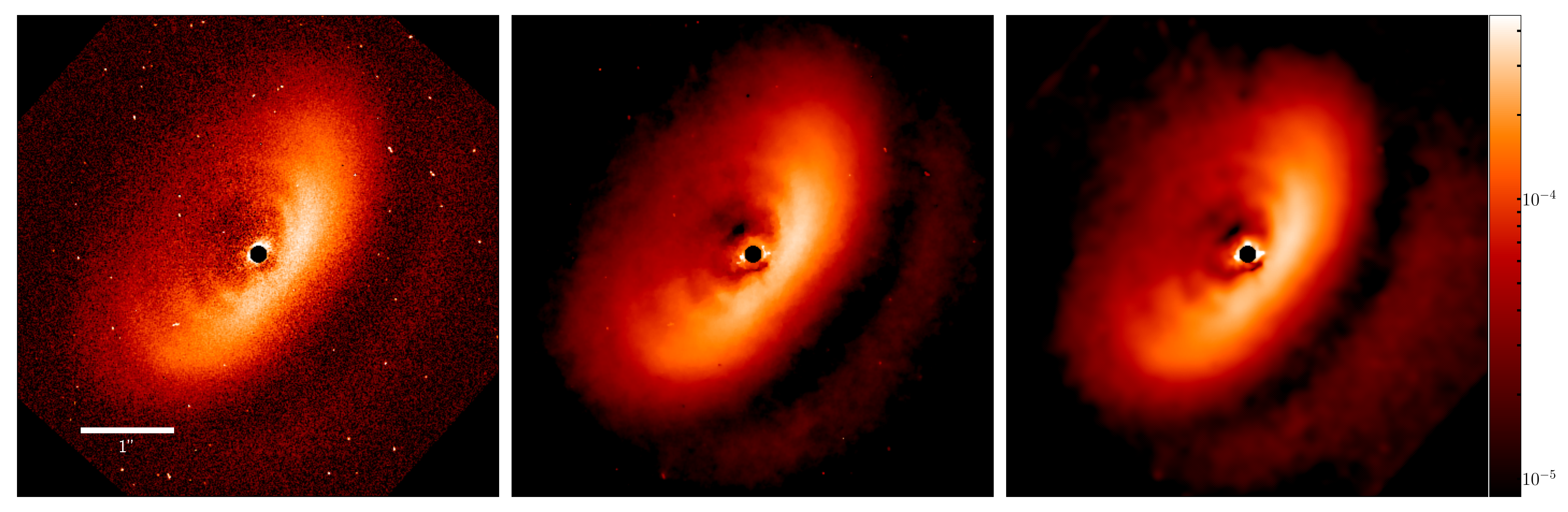}};
            \draw (0,0) node[rotate=90]{IM~Lupus};
        \end{scope}
        \begin{scope}[yshift=-12cm]
            \draw (0,0) 
            node[right]{\includegraphics[width=0.98\textwidth]{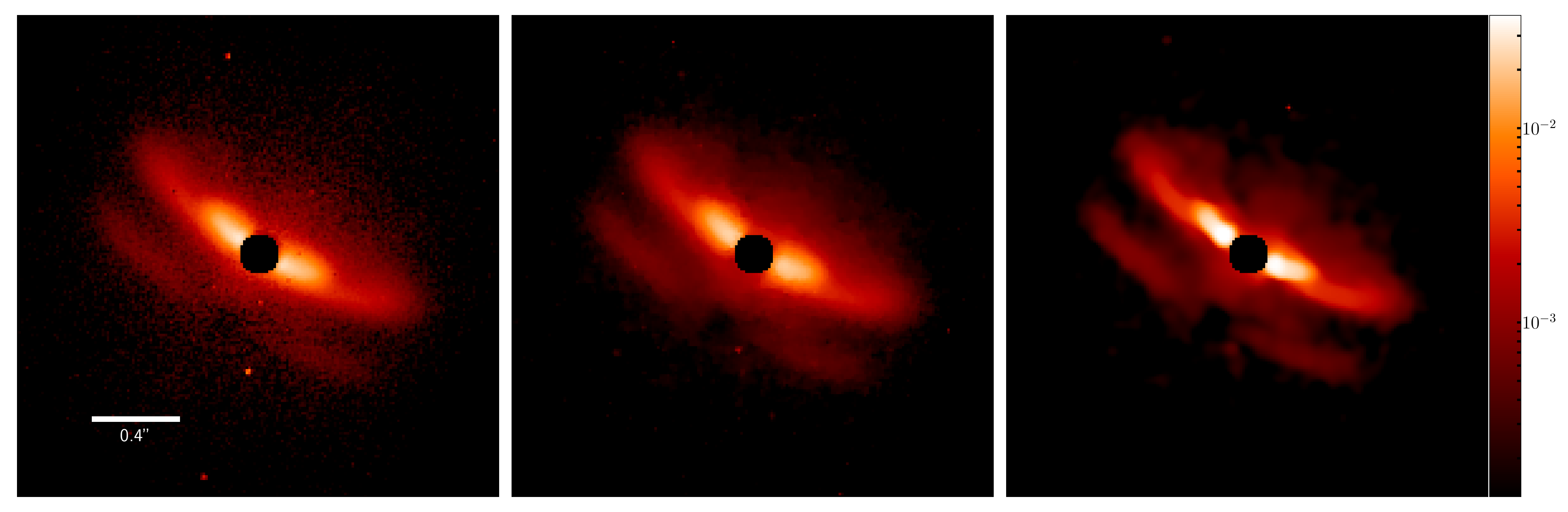}};
            \draw (0,0) node[rotate=90]{MY~Lupus};
        \end{scope}
        \end{tikzpicture}
    \caption{\LD{Reconstructions of the polarized intensity $\Ip$ of the protoplanetary disks TW~Hydrae, IM~Lupus, and MY~Lupus. From the left to the right, the reconstructions have been obtained with the Double Difference, RHAPSODIE \textit{without deconvolution} and RHAPSODIE \textit{with deconvolution}. The maps are displayed in logarithmic scale and normalized in contrast to the unpolarized stellar flux. The pixels lying underneath the coronograph are masked in black. 
    North is up and East is to the left in all frames.}}
    \label{fig:proto_ip}
\end{figure*}

\begin{figure*}
    \centering
    \begin{tikzpicture}
        \begin{scope}
            \draw (0,0) 
            node[right]{\includegraphics[width=0.98\textwidth]{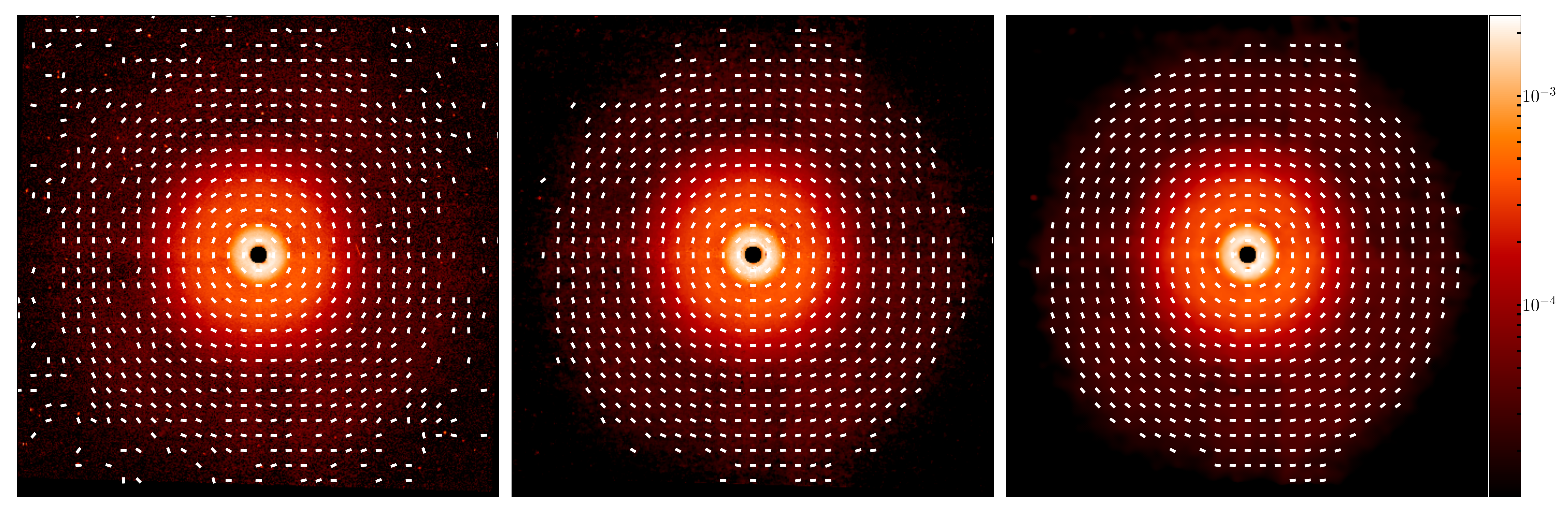}};
            \draw (0,0) node[rotate=90]{TW~Hydrae};
            \draw (3,3) node{Double Difference};
            \draw (8.75,3) node{RHAPSODIE (without dec.)};
            \draw (14.5,3) node{RHAPSODIE (with dec.)};
        \end{scope}

        \begin{scope}[yshift=-6cm]
            \draw (0,0) 
            node[right]{\includegraphics[width=0.98\textwidth]{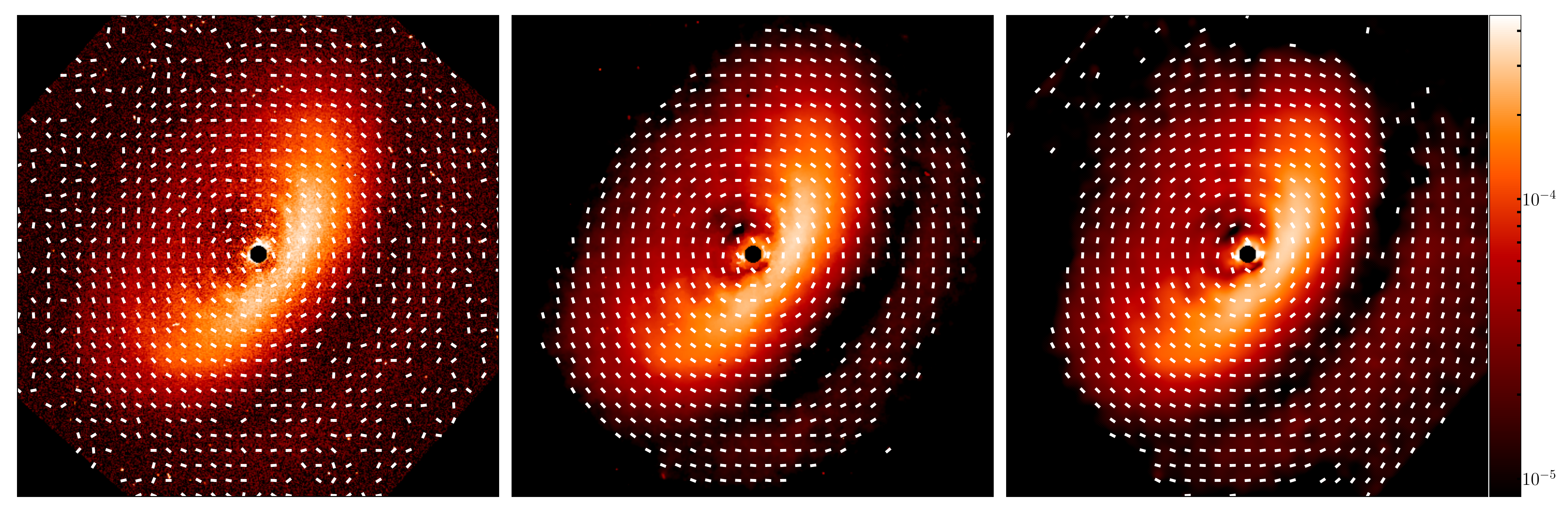}};
            \draw (0,0) node[rotate=90]{IM~Lupus};
        \end{scope}
        \begin{scope}[yshift=-12cm]
            \draw (0,0) 
            node[right]{\includegraphics[width=0.98\textwidth]{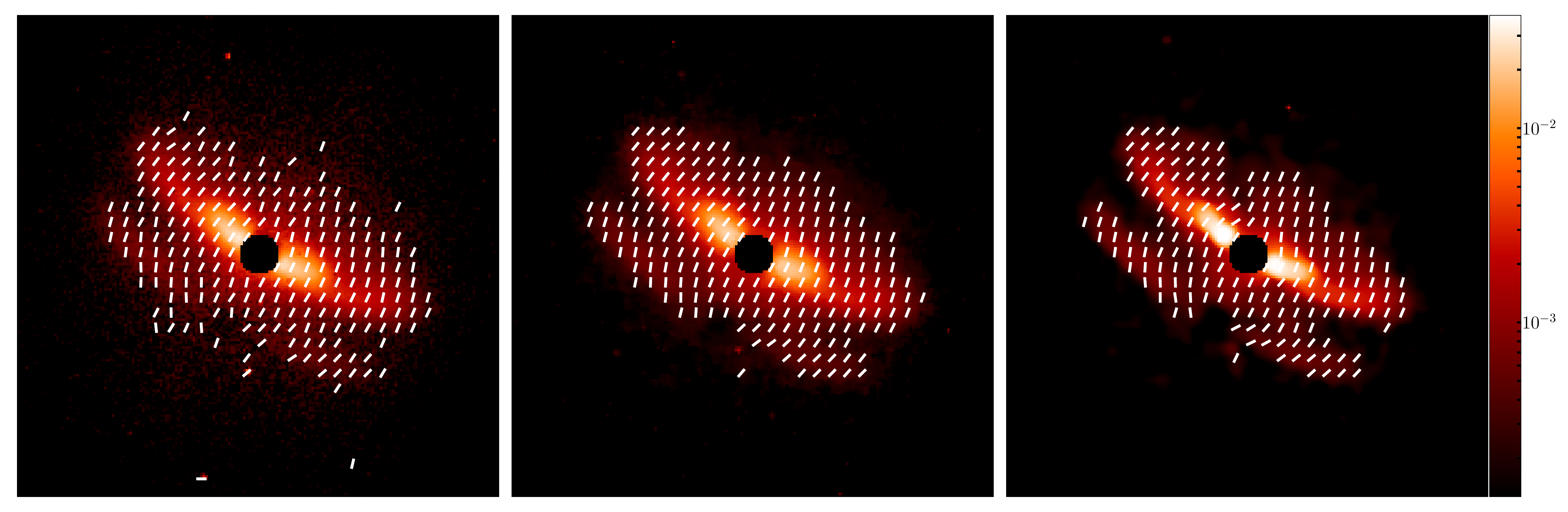}};
            \draw (0,0) node[rotate=90]{MY~Lupus};
        \end{scope}
        \end{tikzpicture}
    \caption{\LD{Reconstructions of the polarization angle $\theta$ of the protoplanetary disks TW~Hydrae, IM~Lupus, and MY~Lupus. From the left to the right, the reconstructions have been obtained with the Double Difference, RHAPSODIE \textit{without deconvolution} and RHAPSODIE \textit{with deconvolution}. The maps are displayed in  logarithmic scale and normalized in contrast to the unpolarized stellar flux. The pixels lying underneath the coronograph are masked in black. 
    North is up and East is to the left in all frames.}}
    \label{fig:proto_theta}
\end{figure*}

\begin{figure*}
    \centering
    \begin{tikzpicture}
        \begin{scope}
            \draw (0,0) 
            node[right]{\includegraphics[width=0.98\textwidth]{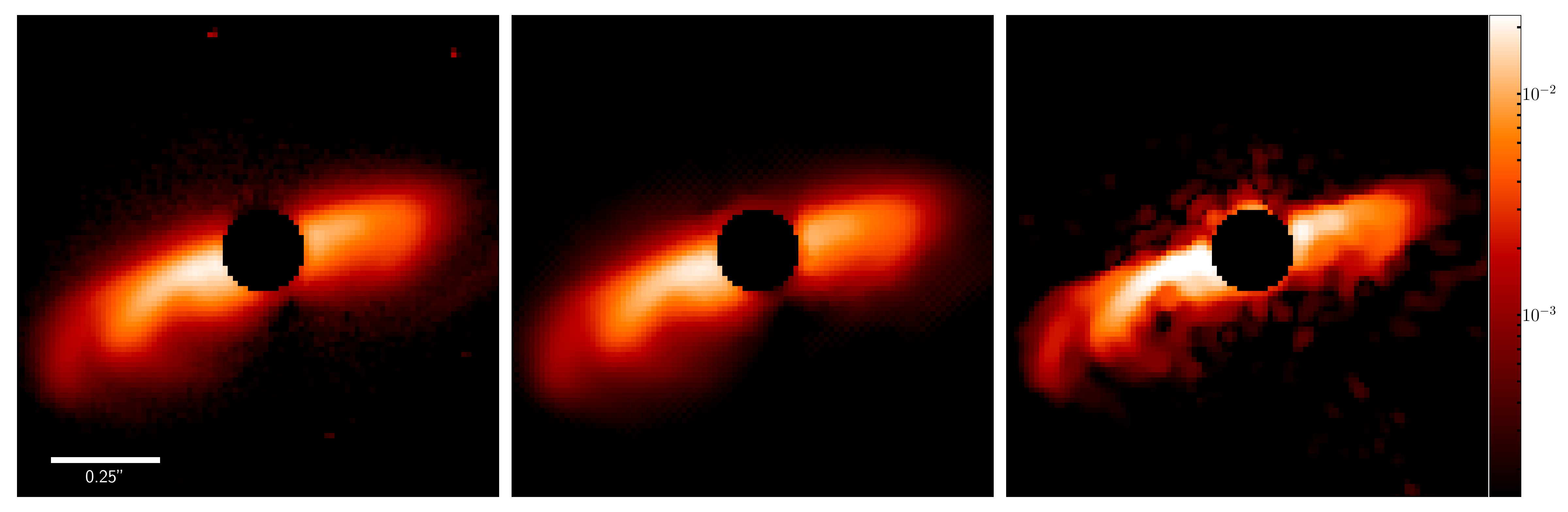}};
            \draw (0,0) node[rotate=90]{RY~Lupus};
            \draw (3,3) node{Double Difference};
            \draw (8.75,3) node{RHAPSODIE (without dec.)};
            \draw (14.5,3) node{RHAPSODIE (with dec.)};

        \end{scope}
        \begin{scope}[yshift=-6cm]
            \draw (0,0) 
            node[right]{\includegraphics[width=0.98\textwidth]{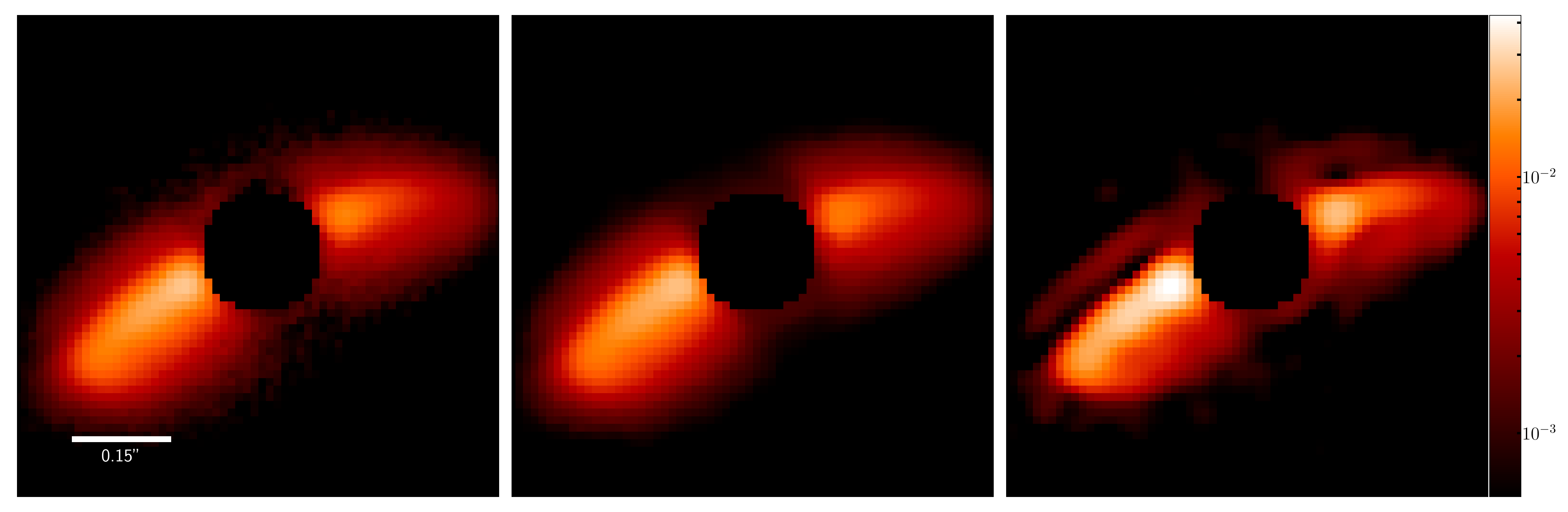}};
            \draw (0,0) node[rotate=90]{T~Chae};

        \end{scope}
        \begin{scope}[yshift=-12cm]
            \draw (0,0) 
            node[right]{\includegraphics[width=0.98\textwidth]{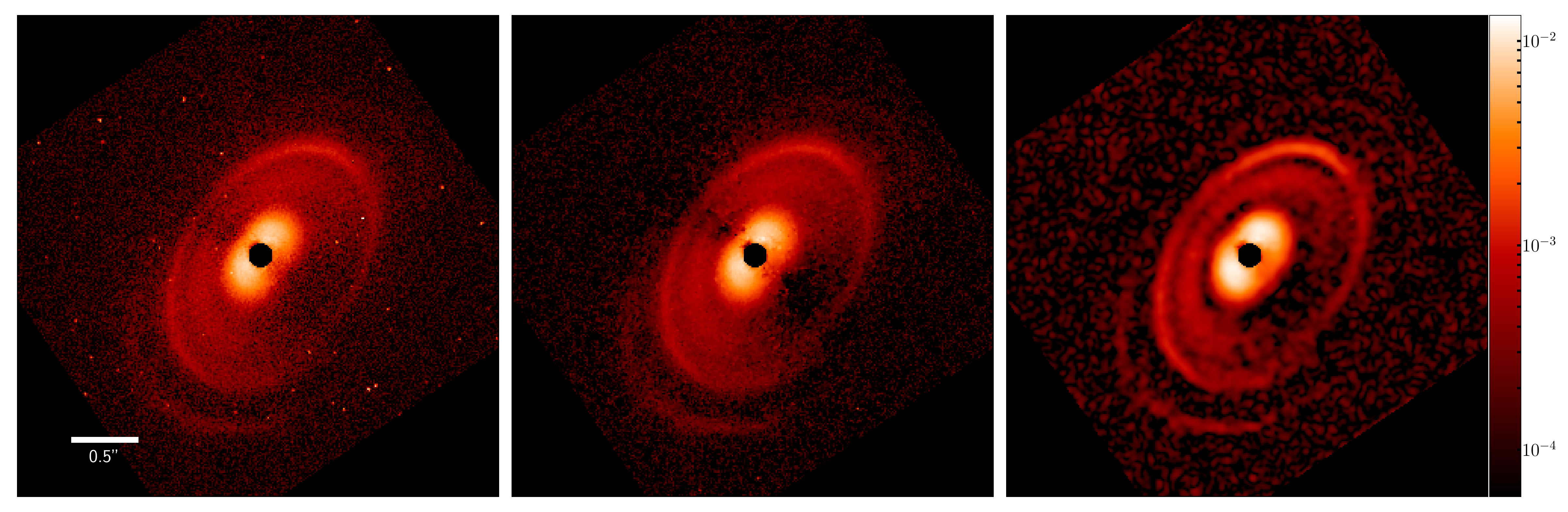}};
            \draw (0,0) node[rotate=90]{RXJ 16 15};

        \end{scope}
        \end{tikzpicture}
    \caption{\LD{Reconstructions of the polarized intensity $\Ip$ of the transition disks RY~Lupus, T~Chae, and RXJ~1615. From the left to the right, the reconstructions have been obtained with the Double Difference, RHAPSODIE \textit{without deconvolution} and RHAPSODIE \textit{with deconvolution}. The maps are displayed in logarithmic scale and normalized in contrast to the unpolarized stellar flux. The pixels lying underneath the coronograph are masked in black. 
    North is up and East is to the left in all frames.}}
    \label{fig:transition_ip}
\end{figure*}

\begin{figure*}
    \centering
    \begin{tikzpicture}
        \begin{scope}
            \draw (0,0) 
            node[right]{\includegraphics[width=0.98\textwidth]{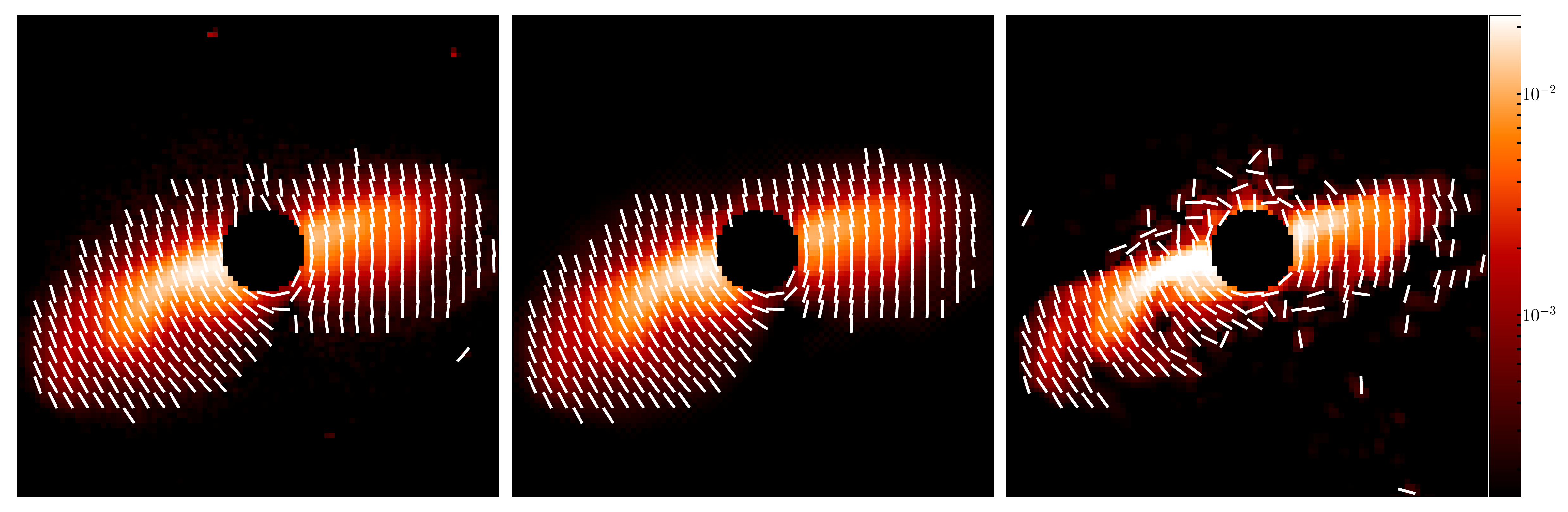}};
            \draw (0,0) node[rotate=90]{RY~Lupus};
            \draw (3,3) node{Double Difference};
            \draw (8.75,3) node{RHAPSODIE (without dec.)};
            \draw (14.5,3) node{RHAPSODIE (with dec.)};

        \end{scope}
        \begin{scope}[yshift=-6cm]
            \draw (0,0) 
            node[right]{\includegraphics[width=0.98\textwidth]{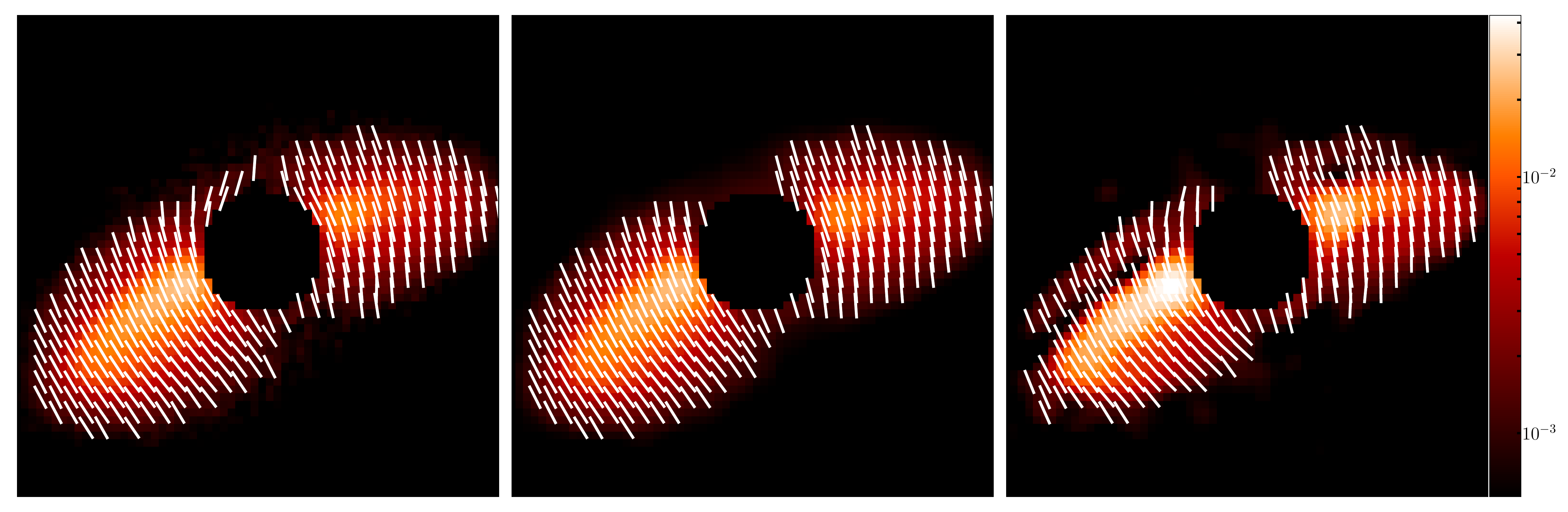}};
            \draw (0,0) node[rotate=90]{T~Chae};

        \end{scope}
        \begin{scope}[yshift=-12cm]
            \draw (0,0) 
            node[right]{\includegraphics[width=0.98\textwidth]{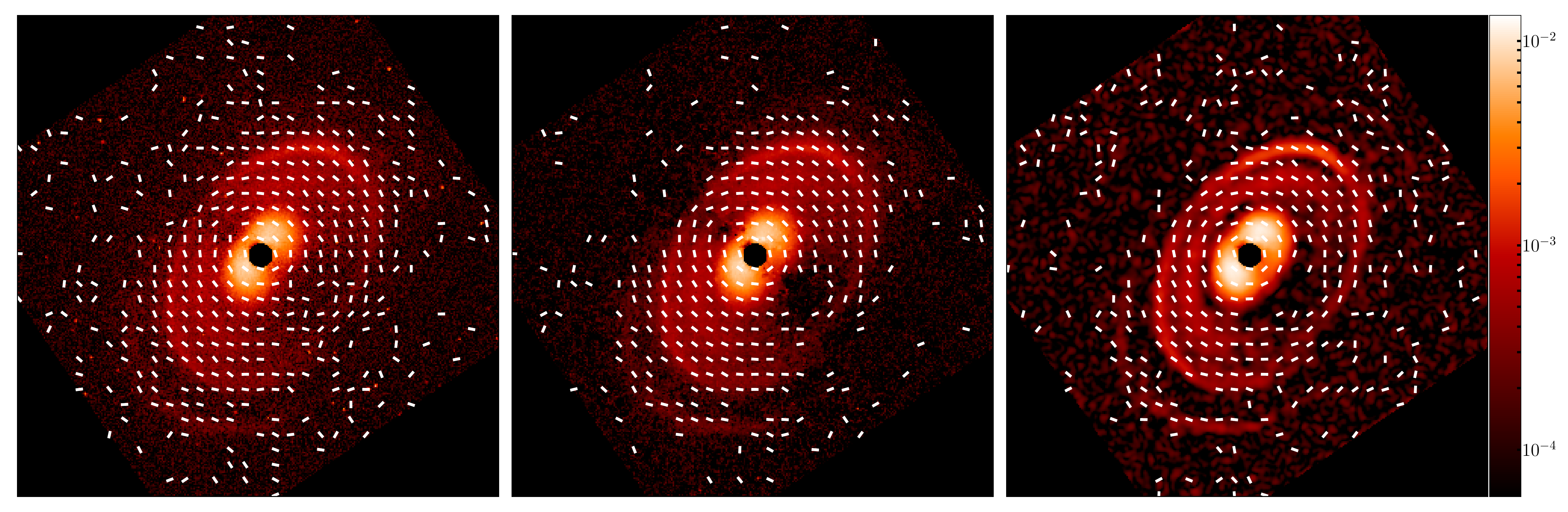}};
            \draw (0,0) node[rotate=90]{RXJ 16 15};

        \end{scope}
        \end{tikzpicture}
    \caption{\LD{Reconstructions of the polarization angle $\theta$ of the transition disks RY~Lupus, T~Chae, and RXJ~1615. From the left to the right, the reconstructions have been obtained with the Double Difference, RHAPSODIE \textit{without deconvolution} and RHAPSODIE \textit{with deconvolution}. The maps are displayed in logarithmic scale and normalized in contrast to the unpolarized stellar flux. The pixels lying underneath the coronograph are masked in black. 
    North is up and East is to the left in all frames.}}
    \label{fig:transition_theta}
\end{figure*}

\begin{figure*}

    \centering
    \begin{tikzpicture}
        \begin{scope}
            \draw (0,0) 
            node[right]{\includegraphics[width=0.98\textwidth]{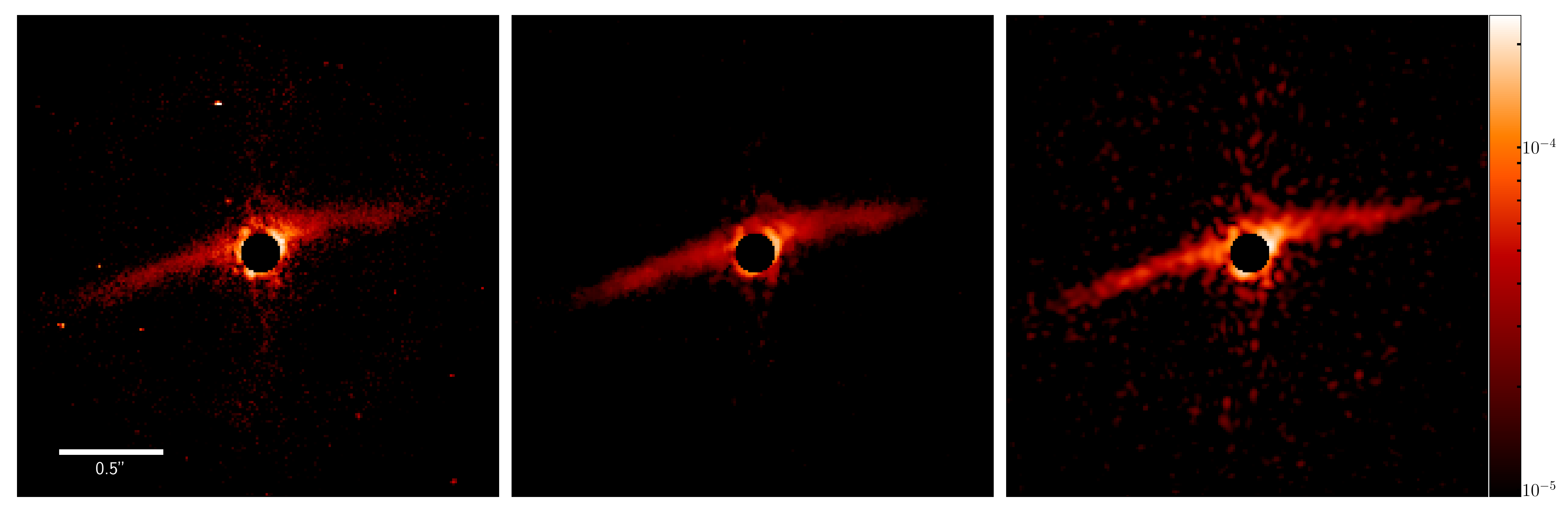}};
            \draw (0,0) node[rotate=90, above]{HD~106906};
            \draw (3,3) node{Double Difference};
            \draw (8.75,3) node{RHAPSODIE (without dec.)};
            \draw (14.5,3) node{RHAPSODIE (with dec.)};

        \end{scope}
        \begin{scope}[yshift=-6cm]
            \draw (0,0) 
            node[right]{\includegraphics[width=0.98\textwidth]{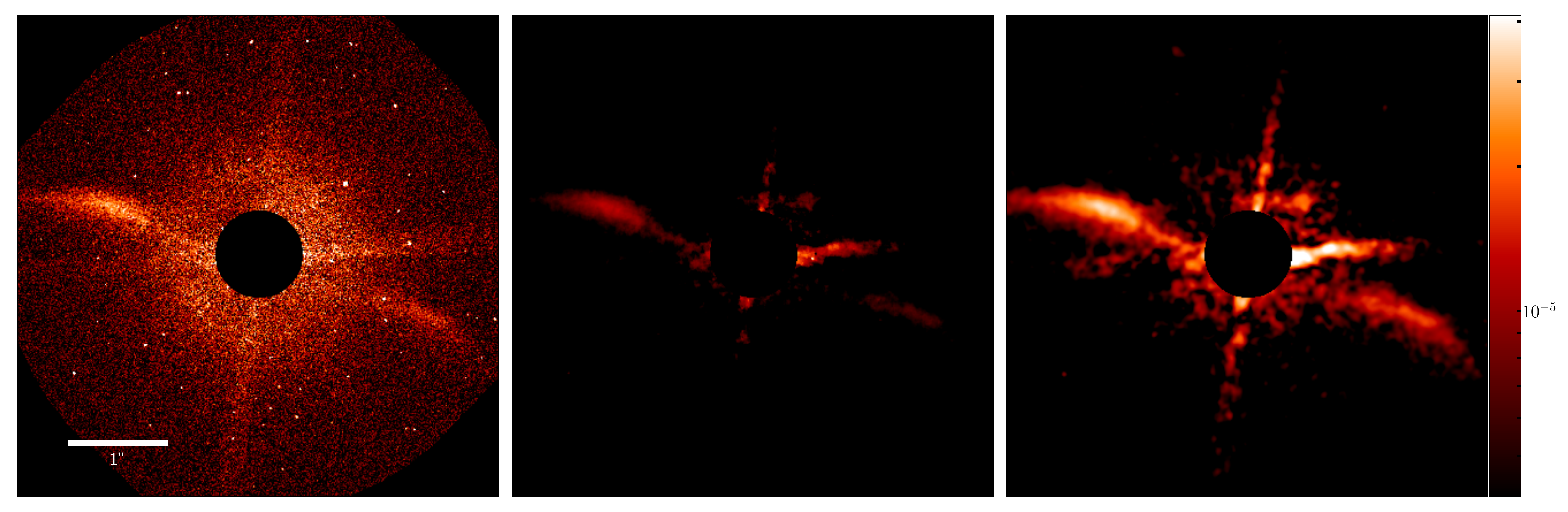}};
            \fill[white] (0,0) rectangle (0.3,2.8);
            \draw (0,0) node[rotate=90, above]{HD~61005};

                 \end{scope}
        \begin{scope}[yshift=-12cm]
            \draw (0,0) 
            node[right]{\includegraphics[width=0.98\textwidth]{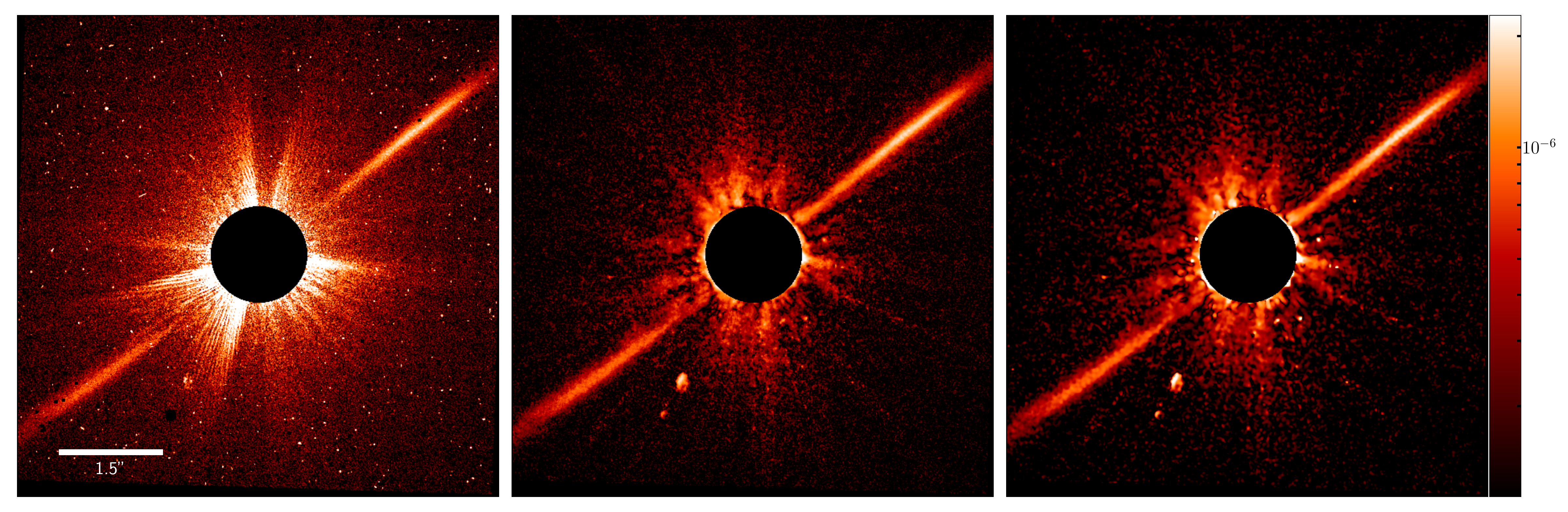}};
            \draw (0,0) node[rotate=90, above]{AU~MIC};

        \end{scope}
    \end{tikzpicture}
    \caption{\LD{Reconstructions of the polarized intensity $\Ip$ of the debris disks HD~106906, HD~61005, and AU~MIC. From the left to the right, the reconstructions have been obtained with the Double Difference, RHAPSODIE \textit{without deconvolution} and RHAPSODIE \textit{with deconvolution}. The maps are displayed in  logarithmic scale and normalized in contrast to the unpolarized stellar flux. The pixels lying underneath the coronograph are masked in black. 
    North is up and East is to the left in all frames.}}
    \label{fig:debris_ip}
\end{figure*}

\begin{figure*}

    \centering
    \begin{tikzpicture}
        \begin{scope}
            \draw (0,0) 
            node[right]{\includegraphics[width=0.98\textwidth]{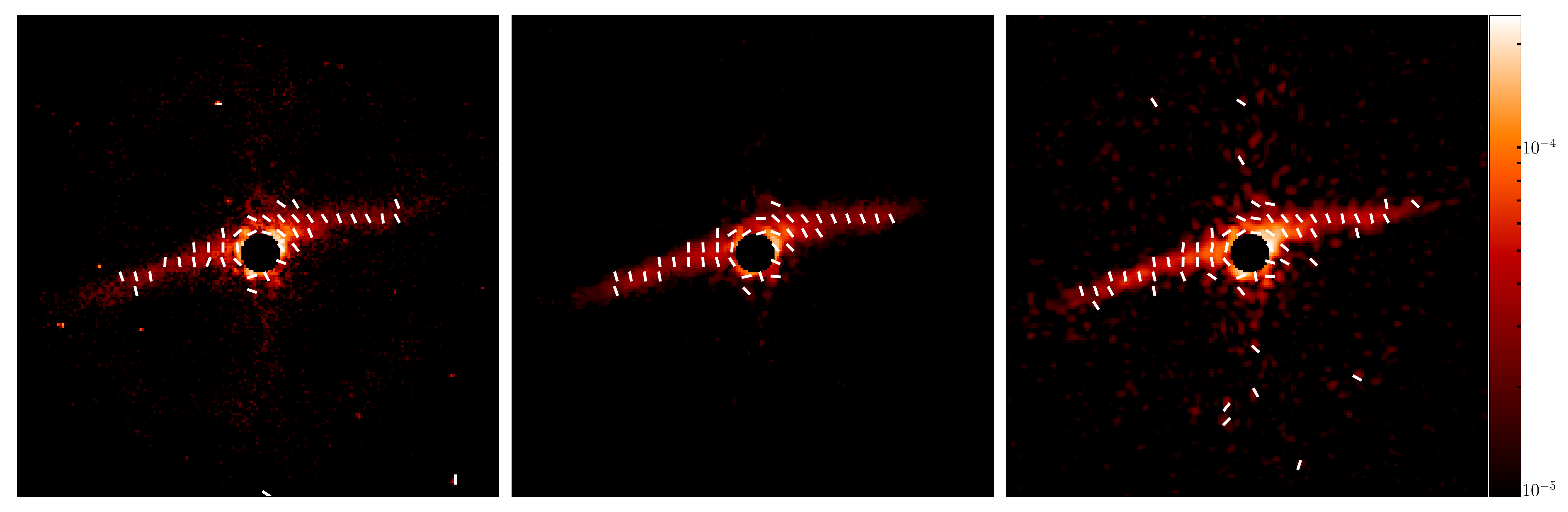}};
            \draw (0,0) node[rotate=90, above]{HD~106906};
            \draw (3,3) node{Double Difference};
            \draw (8.75,3) node{RHAPSODIE (without dec.)};
            \draw (14.5,3) node{RHAPSODIE (with dec.)};

        \end{scope}
        \begin{scope}[yshift=-6cm]
            \draw (0,0) 
            node[right]{\includegraphics[width=0.98\textwidth]{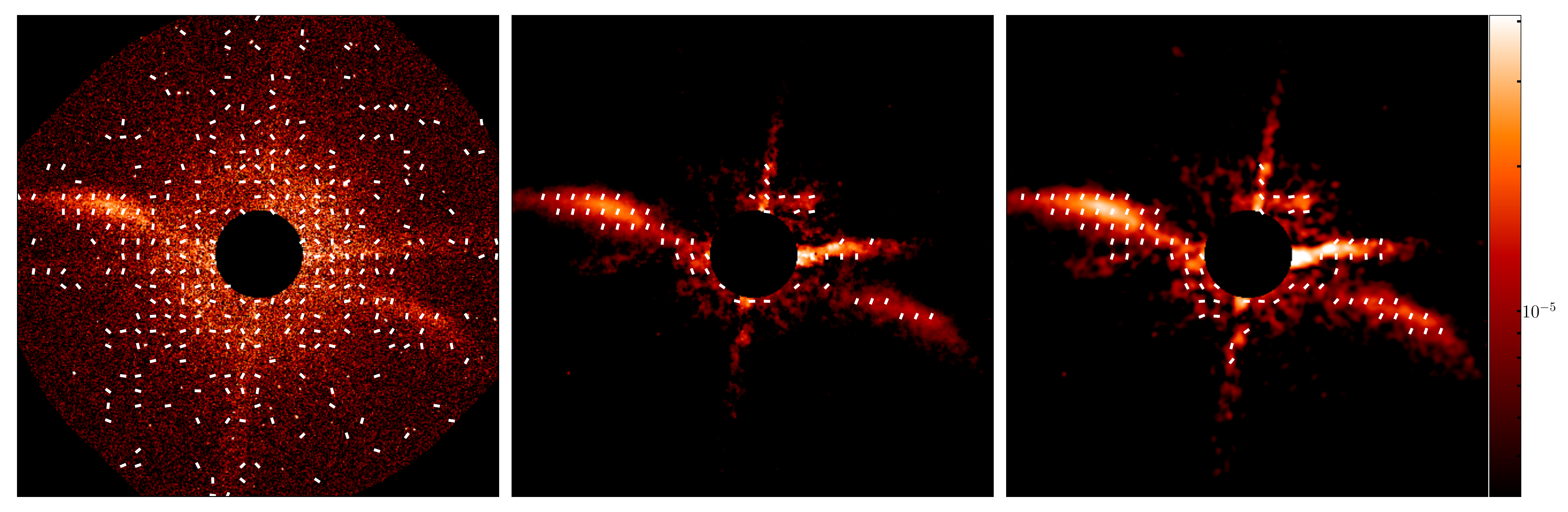}};
            \fill[white] (0,0) rectangle (0.3,2.8);
            \draw (0,0) node[rotate=90, above]{HD~61005};

                 \end{scope}
        \begin{scope}[yshift=-12cm]
            \draw (0,0) 
            node[right]{\includegraphics[width=0.98\textwidth]{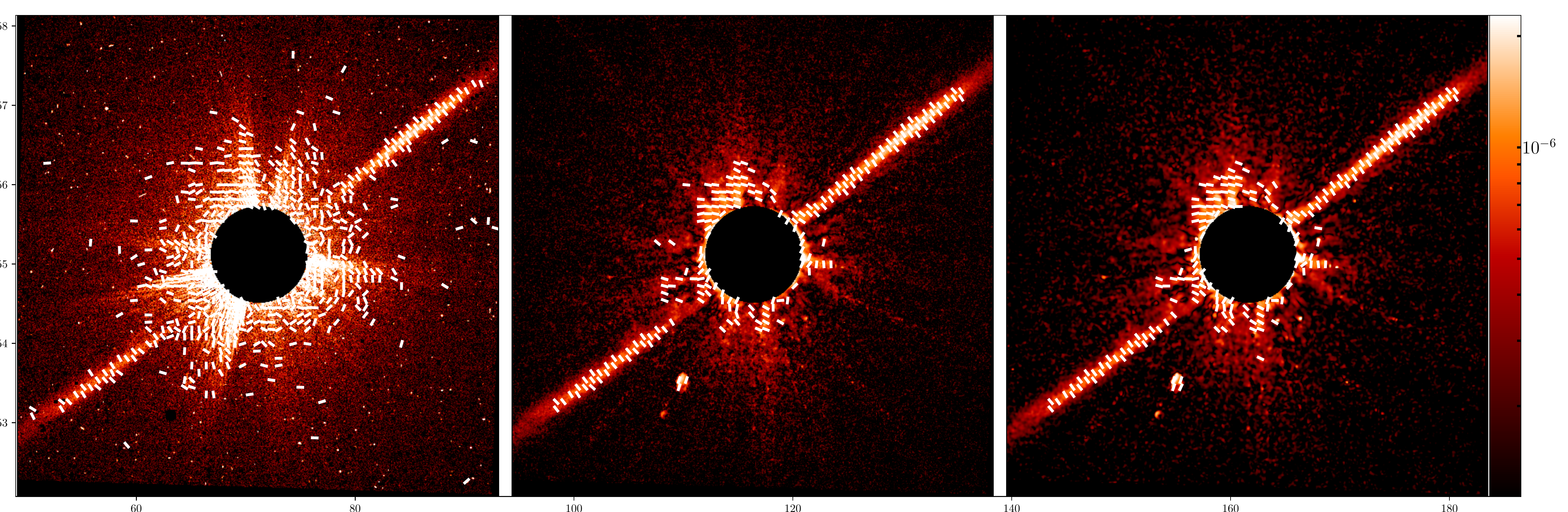}};
            \draw (0,0) node[rotate=90, above]{AU~MIC};

        \end{scope}
    \end{tikzpicture}
    \caption{\LD{Reconstructions of the polarized intensity $\Ip$ of the debris disks HD~106906, HD~61005, and AU~MIC. From the left to the right, the reconstructions have been obtained with the Double Difference, RHAPSODIE \textit{without deconvolution} and RHAPSODIE \textit{with deconvolution}. The maps are displayed in logarithmic scale and normalized in contrast to the unpolarized stellar flux. The pixels lying underneath the coronograph are masked in black. 
    North is up and East is to the left in all frames.}}
    \label{fig:debris_theta}
\end{figure*}

\begin{figure*}[!b]
\centering
 \begin{tikzpicture}
        \begin{scope}
            \draw (0,0) 
            node[right]{\includegraphics[width=0.96\textwidth]{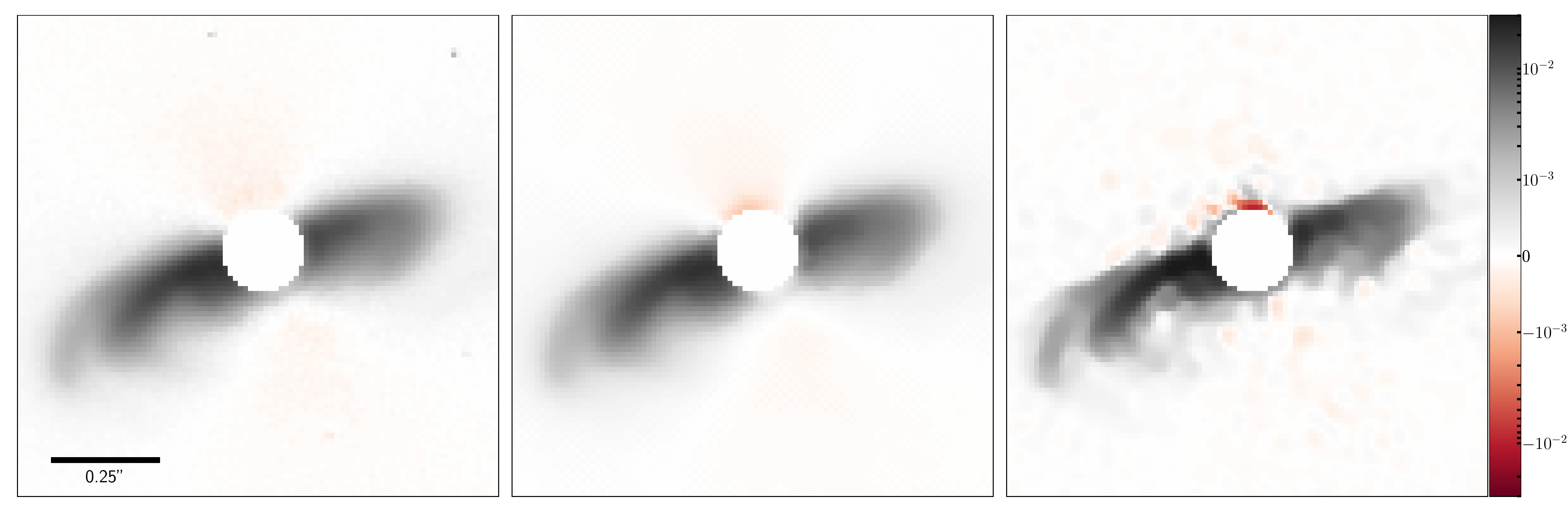}};
            \draw (0,0) node[rotate=90, above]{RY~Lupus $Q_\phi$};
            \draw (3,3) node{Double Difference};
            \draw (8.75,3) node{RHAPSODIE (without dec.)};
            \draw (14.5,3) node{RHAPSODIE (with dec.)};

        \end{scope}
        \begin{scope}[yshift=-5.5cm]
            \draw (0,0) 
            node[right]{\includegraphics[width=0.96\textwidth]{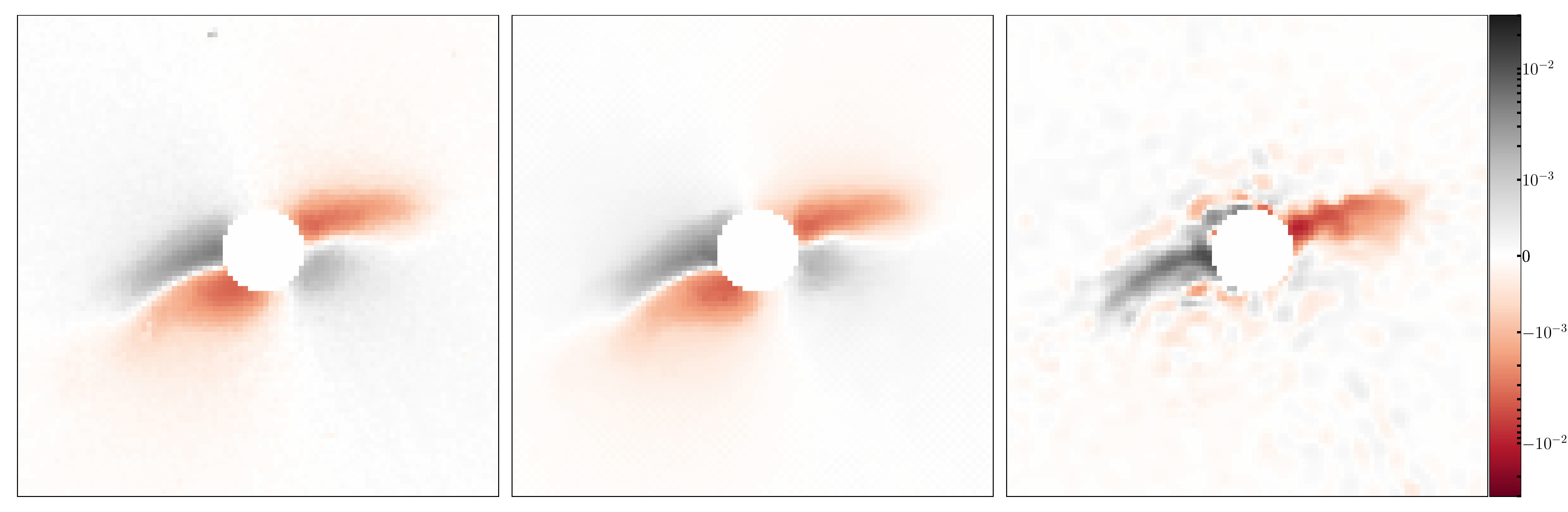}};
            \fill[white] (0,0) rectangle (0.3,2.8);
            \draw (0,0) node[rotate=90, above]{RY~Lupus $U_\phi$};

                 \end{scope}
        \begin{scope}[yshift=-11.75cm]
            \draw (0,0) 
            node[right]{\includegraphics[width=0.96\textwidth]{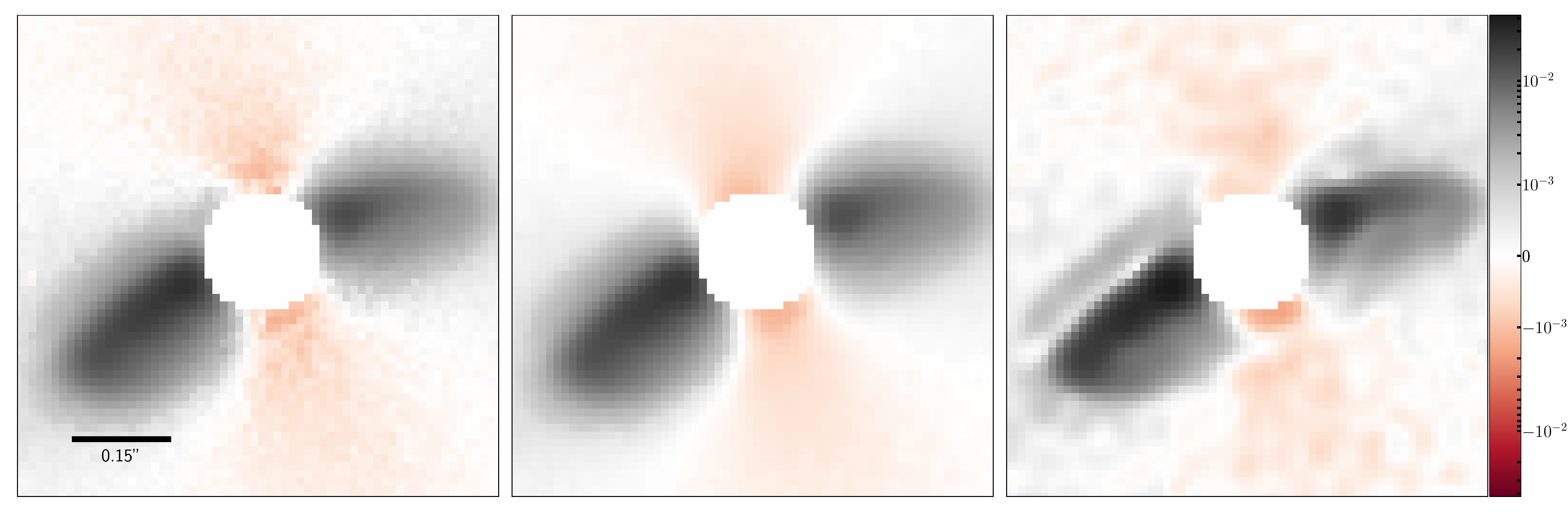}};
            \draw (0,0) node[rotate=90, above]{T~Chae $Q_\phi$};

        \end{scope}

        \begin{scope}[yshift=-17.25cm]
            \draw (0,0) 
            node[right]{\includegraphics[width=0.96\textwidth]{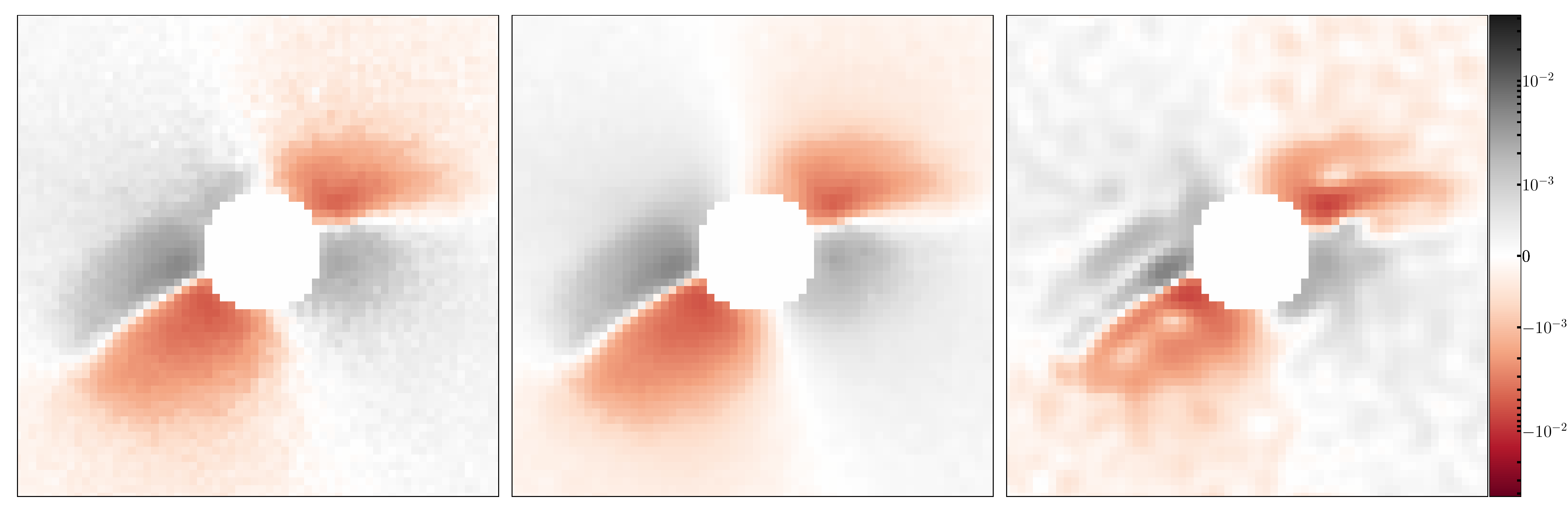}};
            \draw (0,0) node[rotate=90, above]{T~Chae $U_\phi$};

        \end{scope}
    \end{tikzpicture}
\caption{\LD{$Q_\phi$ and $U_\phi$ projections of the reconstructions of RY~Lupus and T~Chae. The maps are displayed in symmetrical logarithmic scale and normalized in contrast to the unpolarized stellar flux. The coronograph is masked in white.}}
\label{fig:qphi_uphi}
\end{figure*}
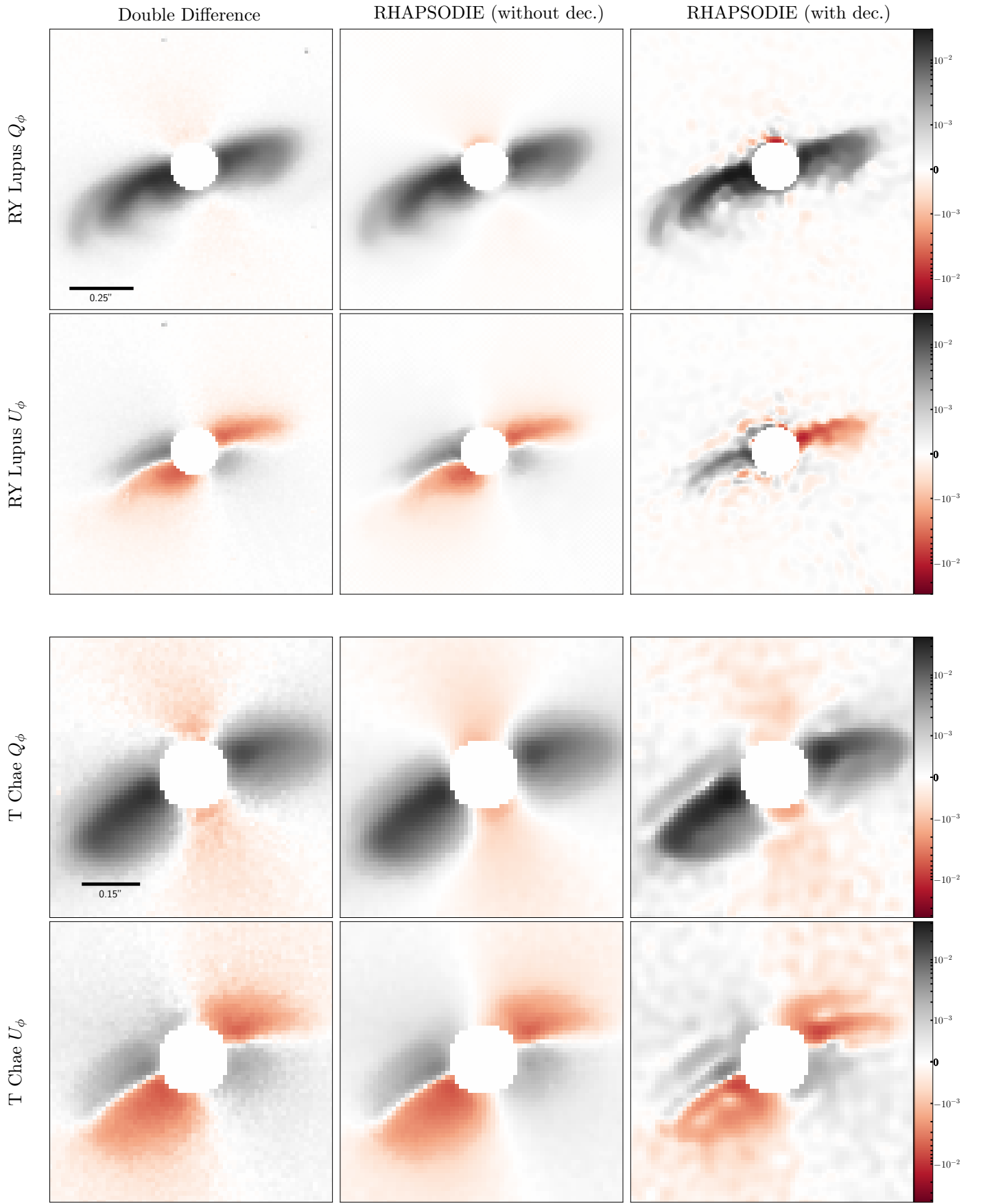

The efficiency of the method we have developed is demonstrated on the \LD{Figures
\ref{fig:proto_ip}, \ref{fig:proto_theta}, \ref{fig:transition_ip}, \ref{fig:transition_theta}, \ref{fig:debris_ip}, 
\ref{fig:debris_theta}, \LDrevised{and Fig.~\ref{fig:qphi_uphi}}. Fig.~\ref{fig:proto_ip} and Fig.~\ref{fig:proto_theta} 
present the Double Difference and RHAPSODIE reconstructions of the protoplanery
disks TW~Hydrae \citep{2017Boekel}, IM~Lupus \citep{avenhaus_disks_2018} and MY
Lupus \citep{avenhaus_disks_2018}. Fig.~\ref{fig:transition_ip}, \ref{fig:transition_theta}, \LDrevised{and Fig.~\ref{fig:qphi_uphi}} present RHAPSODIE reconstructions of transition disks
RY~Lupus \citep{langlois_first_2018} and T Chamaeleontis \citep{pohl_new_2017}. \LDrevised{Fig.~\ref{fig:transition_ip}, \ref{fig:transition_theta} also present RHAPSODIE reconstructions of transition disk}
RXJ~1615 \citep{2016deBoer}. Fig.~\ref{fig:debris_ip} and Fig.~\ref{fig:debris_theta} present the RHAPSODIE reconstructions of the debris disks
HD~106906 \citep{2015Kalas,2016Lagrange}, HD~61005 \citep{2016Olofsson}, and AU
Mic \citep{2018Boccaletti}.
These reconstructions are all normalized in contrast to the unpolarized stellar flux (estimated when registering the PSF off-centered from the coronograph) by taking into account the  transmission of the neutral-density filter used when registering the PSF to prevent saturation.}

\LD{The
contrast between the selected disks polarized light and their host stars unpolarized light responsible for the level of stellar pollution are also very different (ranging from $3\cdot10^{-2}$ to $8\cdot10^{-6}$). The achievable contrast is more favorable for highly inclined disk such as MY~Lup \LD{on Fig.~\ref{fig:proto_ip}} because in such a case,} the
star likely shines partially through the disk, which is dimming the starlight
and thus decreasing the contrast between the star and the disk.

In all cases, \LD{our} method produces high-quality \LD{reconstructions} of the disk polarized
signal and minimizes the artifacts from bad pixels. 
The comparison of these
reductions with the state-of-the-art confirms the benefits of our method but it
is more difficult to quantify the gain on real data not knowing the truth and
should be \LD{strengthened} by using numerical models of these objects. In addition,
our method \LD{can produce deblurred results which clearly enhances the
angular resolution and is beneficial for interpreting the disk morphology
and for studying its} physical properties. \LD{In all cases, the deconvolution sharpens the polarized intensity
image, helps to recover the polarization signal lost by averaging over
close-by polarization signals with opposite sign, and does not introduce
artificial structures that are not present in the original source (see
Fig.~\ref{fig:transition_ip} (RY~Lup)). As mentioned before the
hyperparameters could be further tuned to adjust the smoothing of the
deconvolved images according to the required noise/angular resolution trade-off.}

For several of these disks (IM~Lup, MY~Lup on Fig.\ref{fig:proto_ip}), the outer edge of the disk and
thus the lower disk surface have been detected in \cite{avenhaus_disks_2018}
and are confirmed by our method. \LD{The deconvolution of these datasets allows to
further highlight these fainter features, to enhance or reveal the midplane gaps in the case of T~Cha (on Fig.~\ref{fig:transition_ip}) which was not identified by
\cite{pohl_new_2017}. The reason is that without deconvolution, the PSF smears
light from the disk upper and lower sides into the midplane gap.} 

\LD{Except for Au~Mic, the debris disk presented on Fig.~\ref{fig:debris_ip} are much fainter in contrast than the protoplanetary or the
transition disks in polarized intensity.} As a consequence these datasets have
required careful stellar polarization compensation. The HD~106906 debris disk
is viewed close to edge-on in polarized light as reported in
\cite{van_holstein_polarimetric_2020,2020Esposito}. The image clearly shows the
known East-West brightness asymmetry of the disk, which was detected in total
intensity \citep{2015Kalas, 2016Lagrange}. Thanks to the deep dataset and good
reconstruction, we also detect the backward-scattering far side of the disk to
the west of the star, just south of the brighter near side of the disk. This
feature is further highlighted by the deconvolution.

\LD{The reader may remark that the RHAPSODIE deblurred reconstructions
of RXJ~1615 (on Fig.~\ref{fig:transition_ip}) and \LDrevised{HD~106906} (on
Fig.~\ref{fig:debris_ip}) seems noisier than
without deconvolution. This is due to hyperparameters purposely chosen so as
to achieve the best angular resolution with the side effect of 
slight noise amplification. Thus
the noise is not negligible in the reconstruction. 
As shown in Fig.~\ref{fig:hyperpar_comp}, increasing the regularization weight to smooth the background would lead to a loss of the thin structures.}

\LD{We also analyzed two datasets (HD~61005 and Au~Mic) taken under very bad conditions. For these two datasets, our
method also proves to be efficient in using incomplete polarization cycles to recover the disk polarized signal, and
to deconvolve this signal despite strong artifacts produced in addition by the rotating
spiders} (\ie, unmasked by the Lyot stop)  when observing in field stabilized.
The difference in the spider position during the polarimetric cycle results in
an artificial polarimetric signal when using standard data reduction techniques.
It is worth noticing that the strength of our method to deal with these
artifacts comes from its ability to use weighted maps to account for them.
For instance, the spiders are weighted by their variance during their rotation
frame by frame in addition to the inclusion of a static bad pixel ponderation.
As a result, the contribution of the spiders to the polarimetric signal when using our
method is decreased compared to the other methods. The noise which is created
by these spiders remains, as seen on Fig.~\ref{fig:debris_ip}, and can
generate artifacts in the deconvolution as seen for HD~61005. Counteracting the
\LD{artifacts caused by} the spiders can be efficiently done by performing DPI observation in
pupil tracking \LD{mode} as proposed in \cite{van_holstein_combining_2017}. \LD{We have also validated the efficiency of RHAPSODIE in such observations which allows a further gain in the reduction of the instrumental artifacts from the telescope spiders.}

Another advantage of our method is its ability to use incomplete polarimetric
cycles which are discarded by the classical methods. This capability of our
method leads to an increase in SNR which was quantified more precisely using
our model of the data in \cite{denneulin2020theses}. To benefit from
this improvement, the instrumental polarization has to remain small because
incomplete polarimetric cycles do not benefit from the instrumental
polarization compensation performed by estimating both couples $Q$ and $-Q$ (and $U$
and $-U$, respectively).

\section{Conclusion}



We developed a new method to extract the polarimetric signal using an inverse
\LD{problems approach} that \LD{exploits a model of the measured} signal
formation process. 
The method includes \LD{a} weighted data fidelity term \LD{which} takes into account \LD{the blur and the polarization due to the instrument}, and effectively disentangles polarized signal of interest from stellar contamination.
In order to avoid noise amplification in the minimization of the data fidelity term, an edge preserving smoothing penalization is added allowing to favor smooth estimates almost everywhere. The associated minimization problem is solved by standard optimization techniques. Our method enables to accurately measure the polarized intensity and angle of
linear polarization of circumstellar disks by taking into account the noise
propagation and the observed objects convolution. It has the capability to use
incomplete polarimetry cycles (when the instrumental polarization is small)
which enhances the sensitivity of these observations. It also takes proper
account for bad pixels by using weighted maps instead of interpolating them.
These bad pixels can cause systematic errors of several tenths of percent in
the polarization measurements as shown by \citep[][]{hostein2020prep}
In addition, the effect of bad pixel interpolation could also have some impact
when reaching $0.1$\% polarimetric accuracy.

We have validated the method on both simulated and archive data from SPHERE/IRDIS and compared its performances with the state-of-the-art methods. We have implemented the method in an end-to-end data-analysis package called RHAPSODIE. The method we developed improves the overall performances in particular at low SNR/small polarized flux compared to standard methods. 

By increasing the sensitivity and including deconvolution, this method will allow for more accurate studies of the orientation and morphology of the disks, especially in the innermost regions. It also will enable more accurate
measurements of the angle of linear polarization at low SNR, which would allow for more in-depth studies of dust properties. Finally, the method will enable more accurate measurements of the polarized intensity which is critical to construct scattering phase functions.

\LD{RHAPSODIE is the first regularized inverse approach implemented for high contrast polarimetric imaging.}
\LD{It demonstrates the benefits of advanced signal processing methods in this domain.}

\section{Acknowledgements}

We acknowledge Rob Van Holstein for his help with the instrumental polarization
correction based on careful instrumental calibrations he performed which are
implemented in the IRDAP tool he developed. \LD{We also thank the anonymous referee for her/his careful reading of the manuscript as well as her/his insightful comments and suggestions.}This work has made use of the
SPHERE Data Centre, jointly operated by OSUG/IPAG (Grenoble), PYTHEAS/LAM/CeSAM
(Marseille), OCA/Lagrange (Nice), Observatoire de Paris/LESIA (Paris), and
Observatoire de Lyon (OSUL/CRAL). This work is supported by the French National
Programms PNP and the Action Spécifique Haute Résolution Angulaire (ASHRA) of
CNRS/INSU co-funded by CNES. The authors are grateful to the LABEX Lyon
Institute of Origins (ANR-10-LABX-0066) of the Université de Lyon for its
financial support within the program ``Investissements d'Avenir''
(ANR-11-IDEX-0007) of the French government operated by the National Research
Agency (ANR). This paper is based on observations made with ESO Telescopes at
the La Paranal Observatory under programme ID: 598.C-0359, 095.C-0273,
0102.C-0916, 096.C-0523, 097.C-0523, 096.C-0248. 


\bibliographystyle{aa}
\bibliography{manuscrit}

\begin{appendix}
\section{Synthetic dataset simulation}
\label{sec:datasimu}

In order to evaluate the performance of the RHAPSODIE method, synthetic data have been created. These synthetic data are designed to reproduce astrophysical cases.  First, the truth $N=128 \times 128$ maps $\Tr{\Iu}$, $\Tr{\Ip}$ et $\Tr{\theta}$ are created. Such a value of $N$ pixels fits the main Region Of Interest (ROI) size. 

\begin{figure*}[b]
    \begin{center}\resizebox{0.99\textwidth}{!}{
        \begin{tikzpicture}
            \begin{scope}[scale=1]
                \draw (0,0) node[left]{\includegraphics[width=17cm]{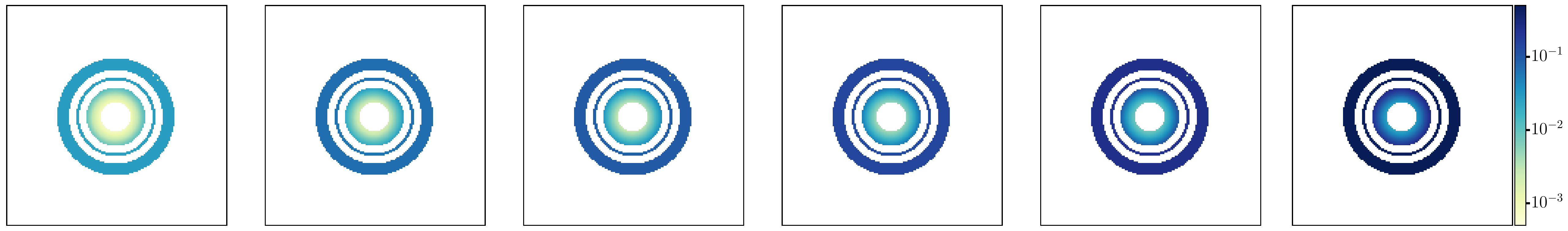} };
                \draw (-17.3,0) node[rotate=90]{$\tau^\Tag{total}$};
                \draw (-15.75,1.5) node{$\tau^\Tag{disk}=3\%$};
                \draw (-13,1.5) node{$\tau^\Tag{disk}=7\%$};
                \draw (-10.25,1.5) node{$\tau^\Tag{disk}=10\%$};
                \draw (-7.5,1.5) node{$\tau^\Tag{disk}=15\%$};
                \draw (-4.75,1.5) node{$\tau^\Tag{disk}=25\%$};
                \draw (-2,1.5) node{$\tau^\Tag{disk}=50\%$};
                
                \draw (0,-3)  node[left]{\includegraphics[width=17cm]{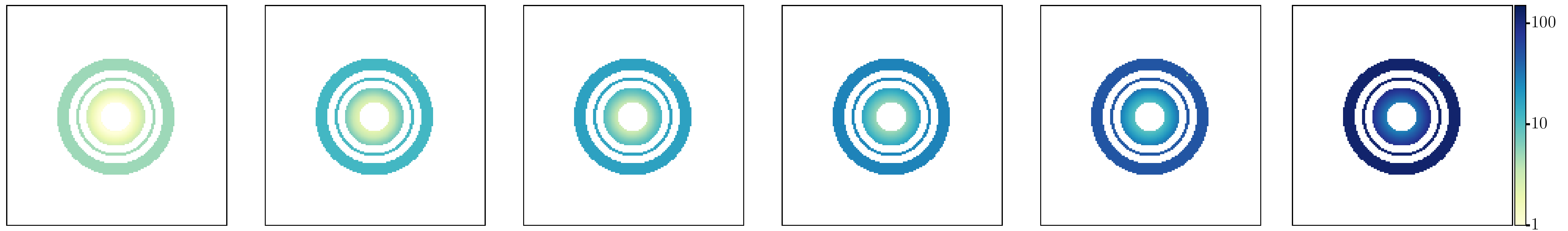} };
                \draw (-17.25,-3) node[rotate=90]{SNR};
                
            \end{scope}
        \end{tikzpicture}}
    \end{center}
    \caption{\LD{Total ratios of polarization $\tau^\Tag{total}$ maps, with respect to the total intensities, and SNR maps for the different values of disk polarized fractions $\tau^\Tag{disk}$.}}
    \label{fig:SNR_TAU}
\end{figure*}
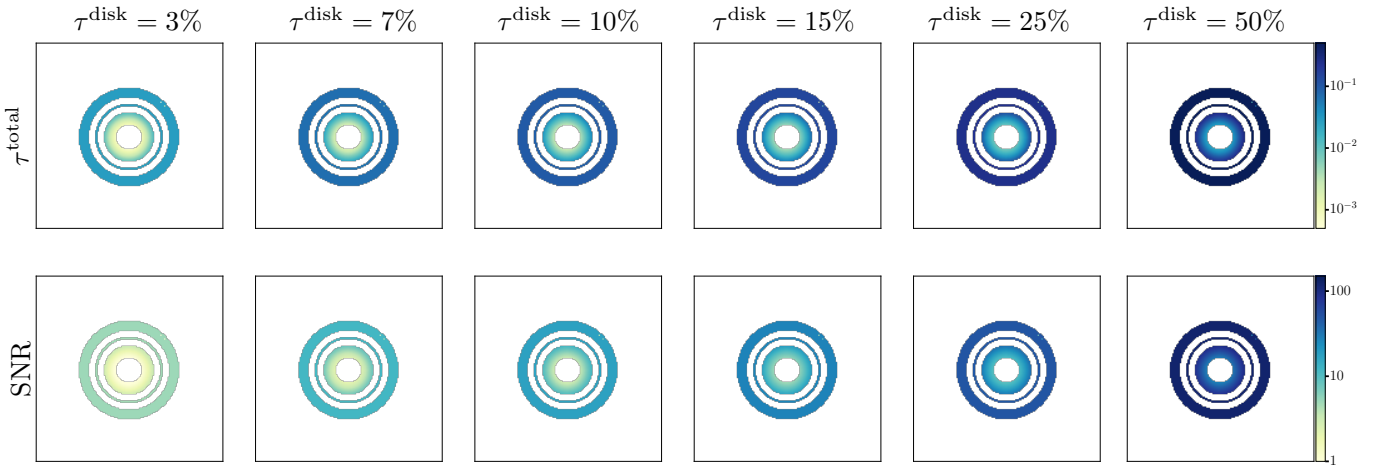

The hardest disk structures to reconstruct are faint, small and lightly
polarized structures, and consequently have high contrast with the unpolarized
stellar intensity and a low SNR. This is why the synthetic environment we generated a disk
with three rings of equal brightness but with a different contrast with the
unpolarized stellar intensity. This disk is partially polarized with a linearly
polarization $\Ip$ and a polarization angle $\theta \in ]-\pi, \pi]$ and an
unpolarized component ${\Iu}^\Tag{disk}$. The disk polarization ratio between
both component is given by:
\begin{equation}
    \tau^\Tag{disk} =\frac{\Ip}{{\Iu}^\Tag{disk}+\Ip}.
\label{eq:taudisk}    
\end{equation}
These synthetic images are then combined as maps of Stokes parameters $\Is$,
$\Qs$ and $\Us$.

\begin{figure*}[!b]
    \resizebox{0.98\textwidth}{!}{
        \begin{tikzpicture}
            \begin{scope}[scale=1]
                \draw (0,20) node[right]{\includegraphics[height=5cm]{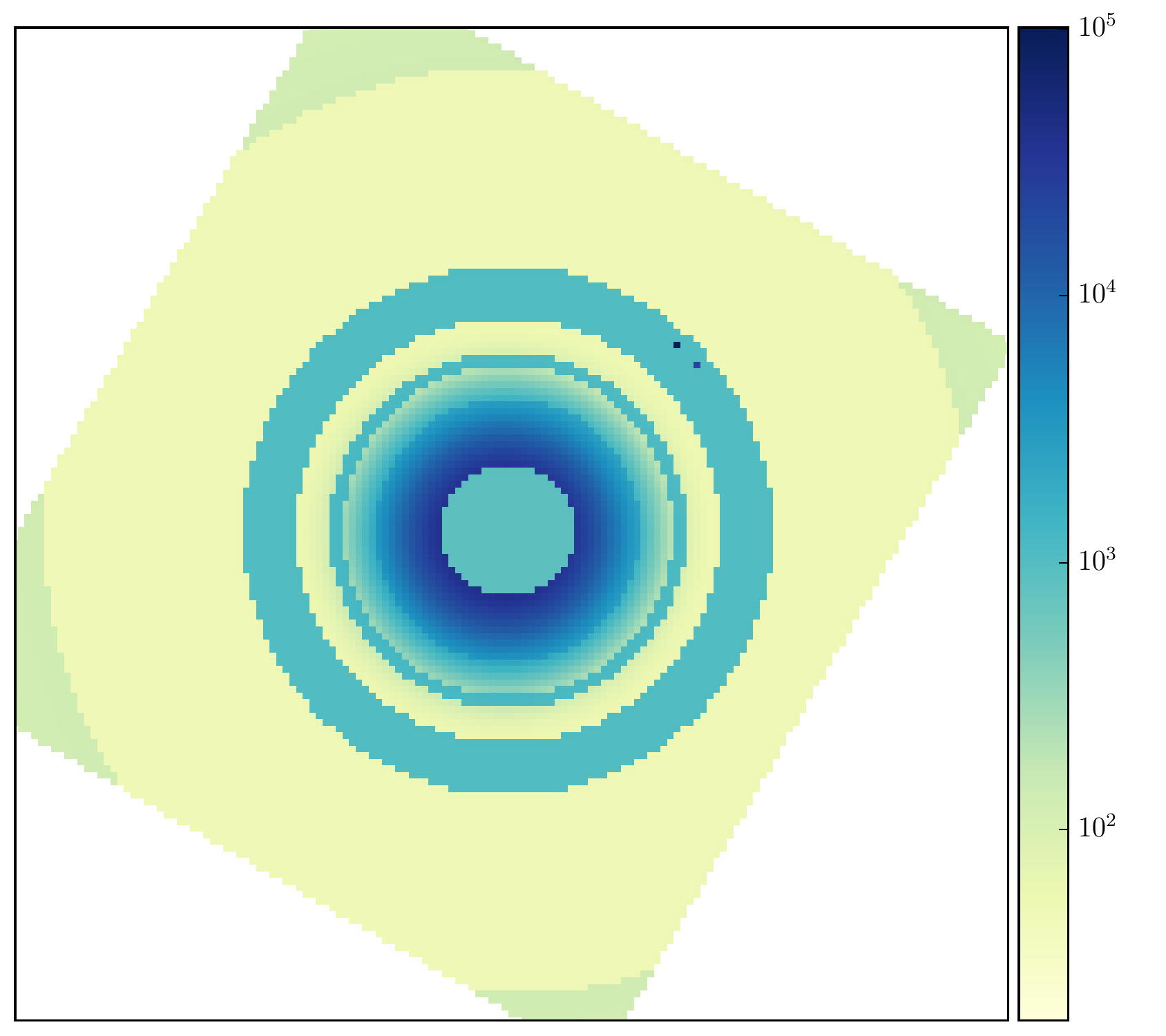} };
                \draw (6,20) node[right]{\includegraphics[height=5cm]{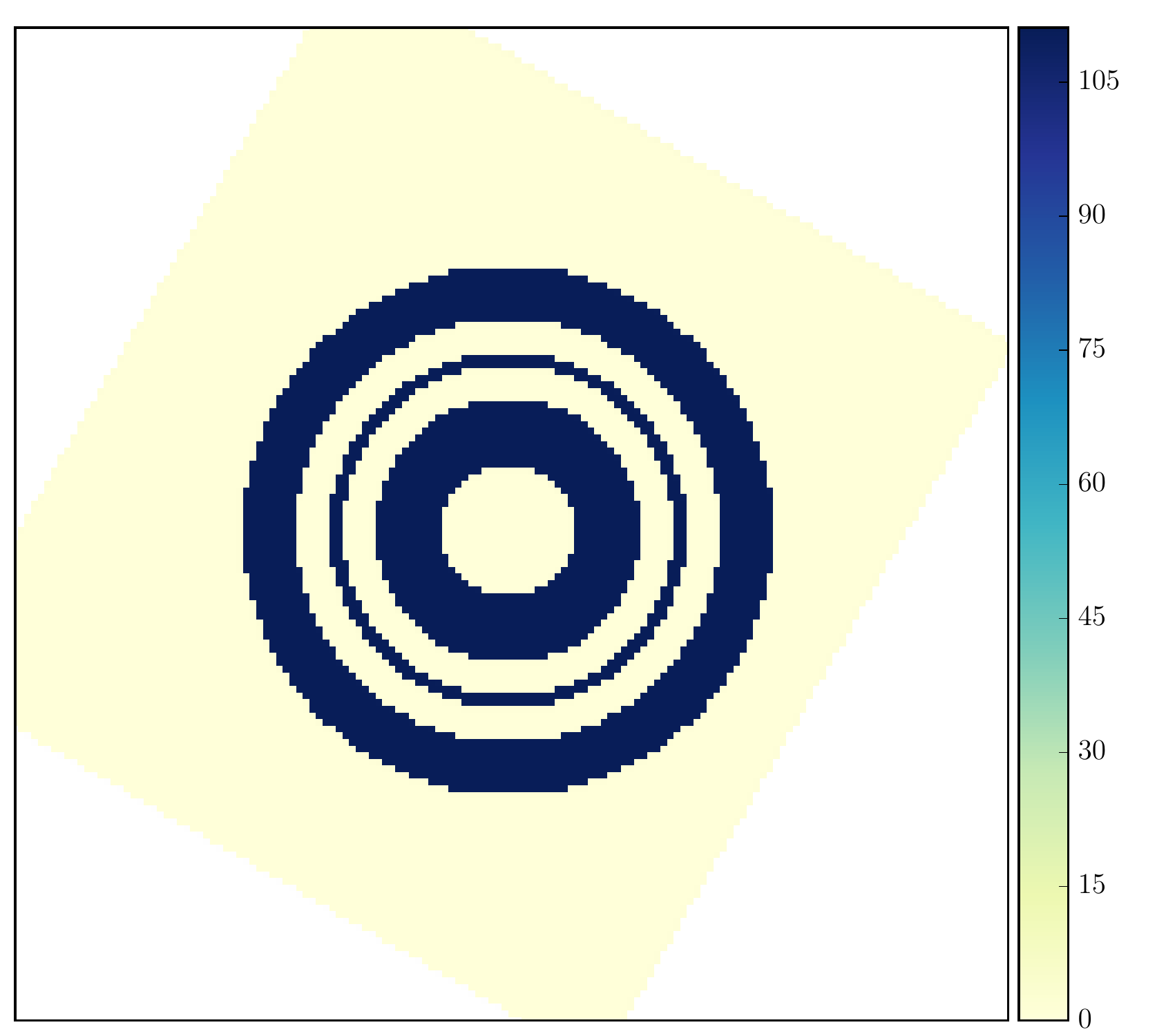} };
                \draw (12,20) node[right]{\includegraphics[height=5cm]{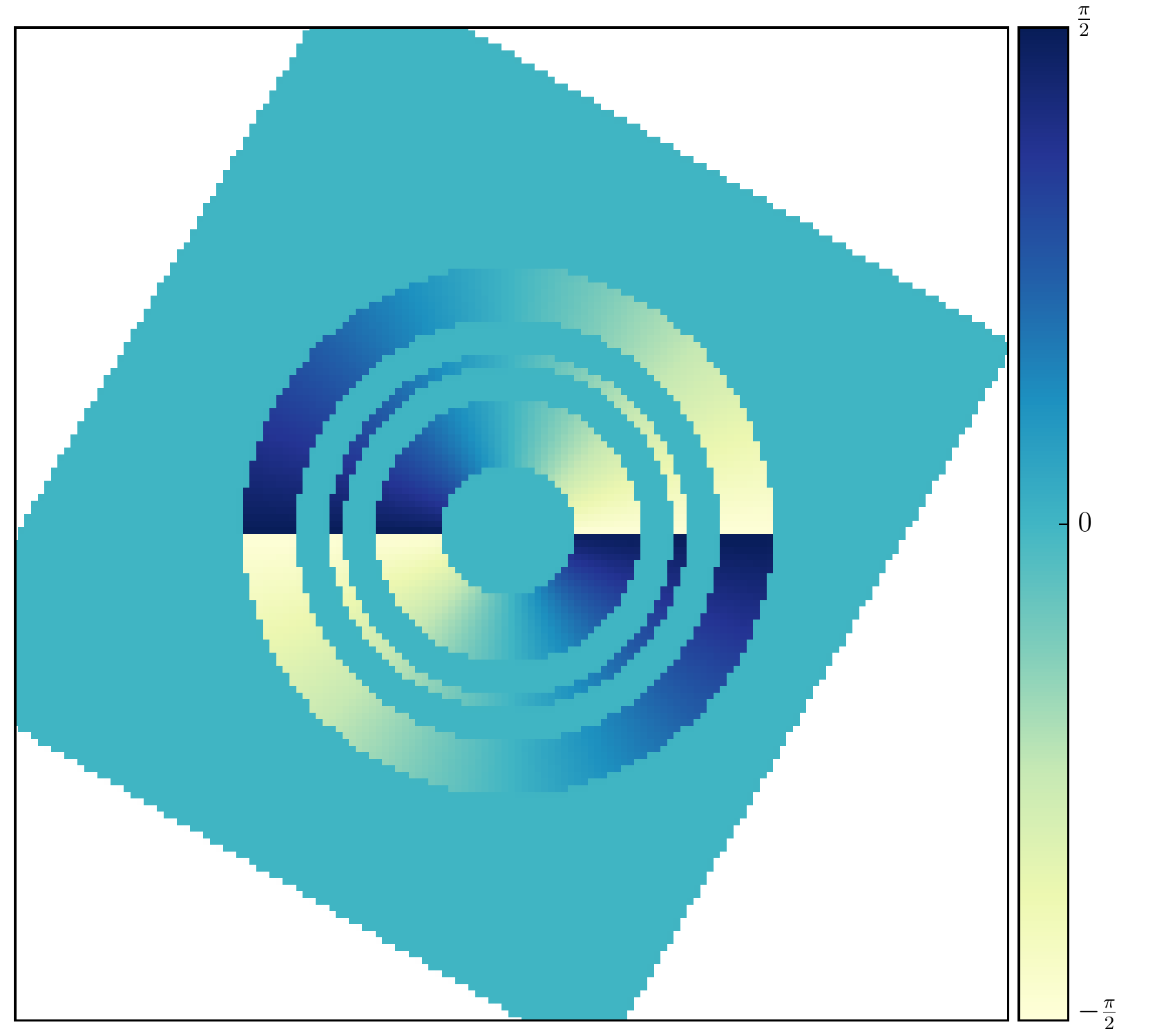} };
                
                \draw (2.5, 17) node{$\Tr{\Iu}$};
                \draw (8.5, 17) node{$\Tr{\Ip}$};
                \draw (14.5, 17) node{$\Tr{\theta}$};
                
                \draw (8.5, 17) node[left, rotate=90]{\resizebox{0.5cm}{12cm}{$\{$ }};
                
                \draw[->, thick] (2.5,16.3) -- (2.5, 16.9);
                \draw (2.5, 16) node[rotate=-90]{\resizebox{0.5cm}{3cm}{$\{$ }};
                \draw (1.5,15) node{\includegraphics[height=1.4cm]{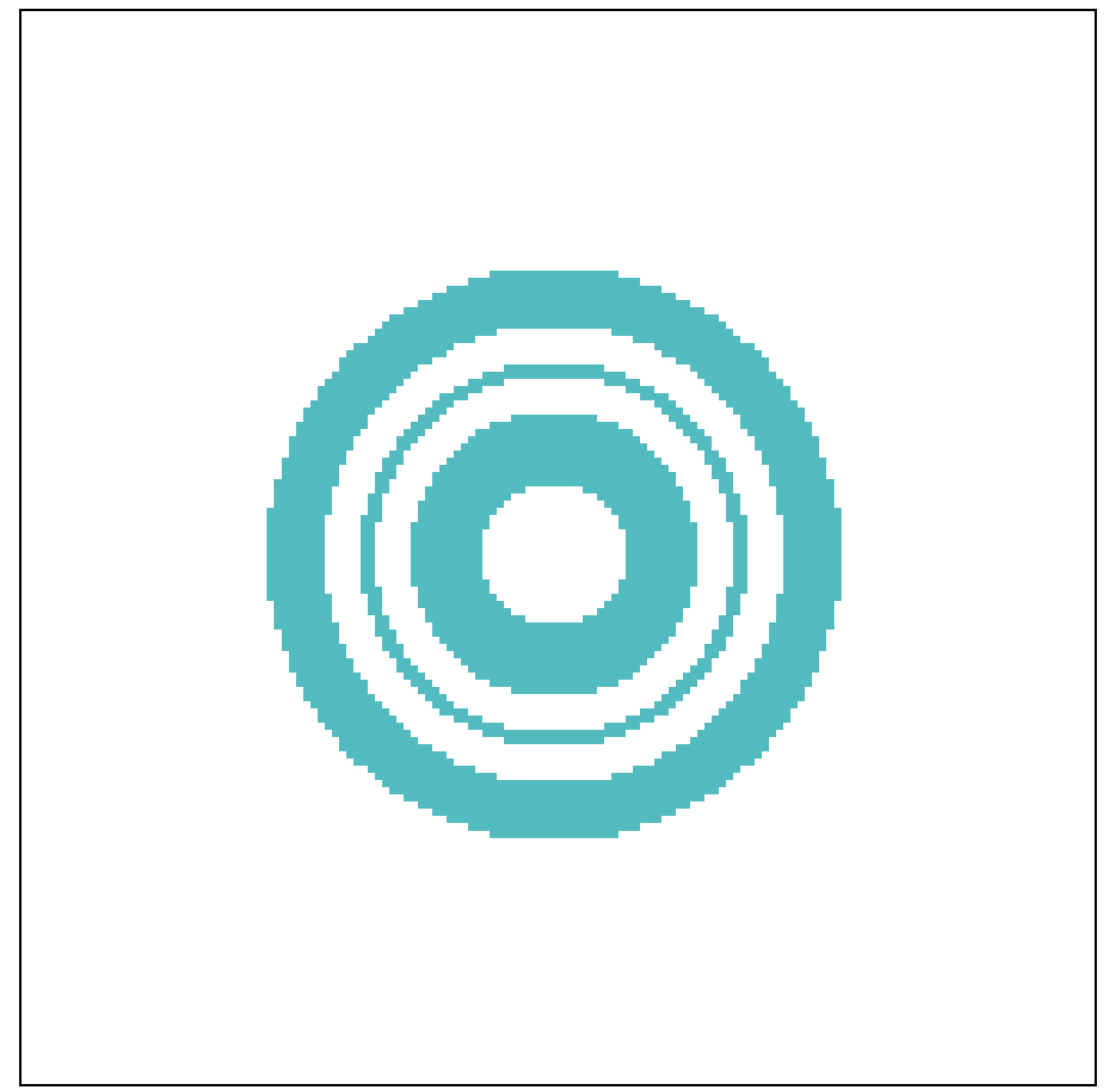} };
                \draw (2.5,15) node{\Large $+$};
                \draw (3.5,15) node{\includegraphics[height=1.4cm]{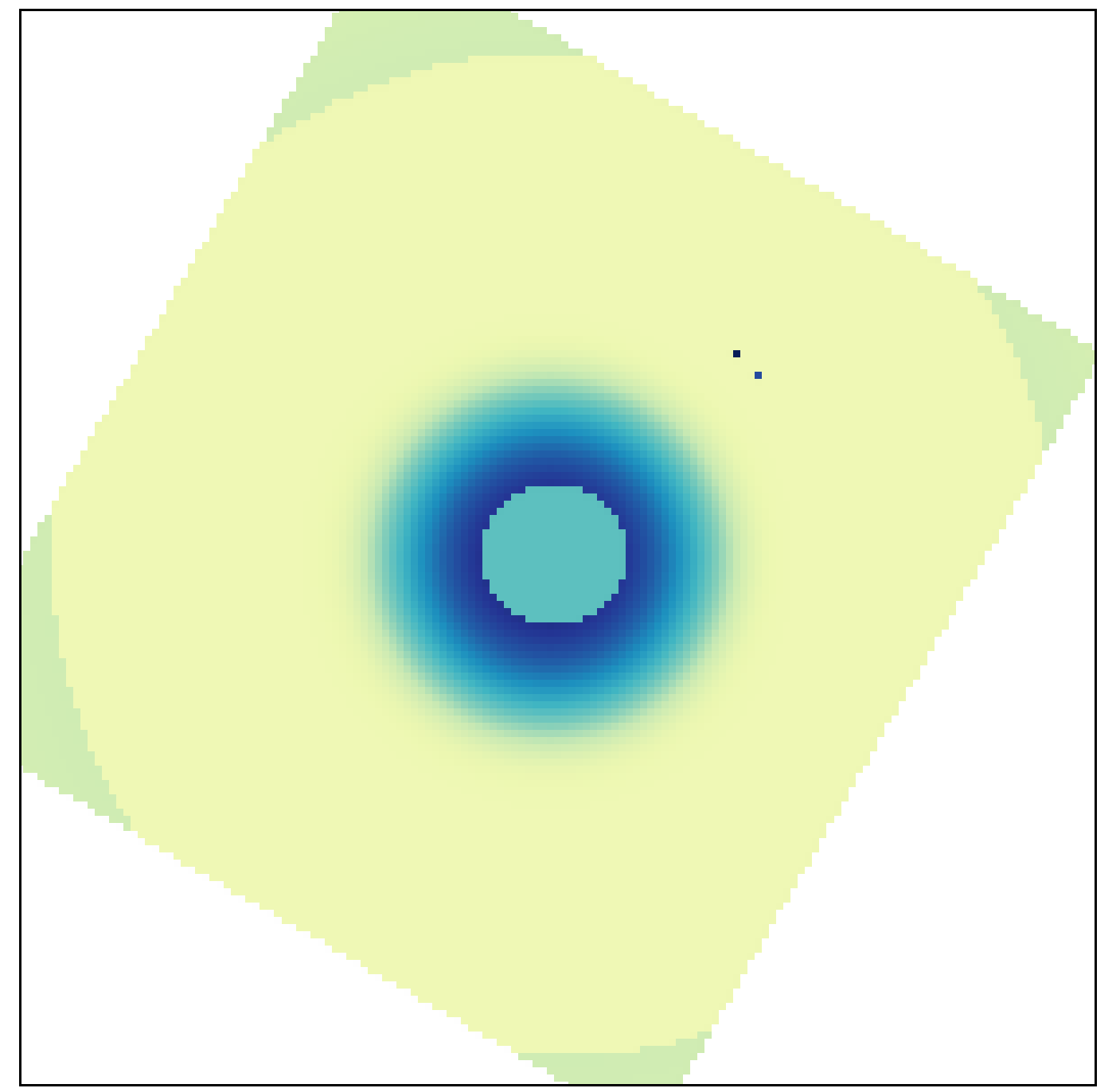} };

                \draw[->, thick] (8.5,16.4) -- (4,13.2) node[midway, above, sloped]{Combination following \eqref{eq:data-approx-model}};
                
                \draw[->, thick] (4,13.2) arc (110:70:14)  node[midway, above]{Related weights \eqref{eq:weights}};
                
                \draw (0,11.5) node[right]{\includegraphics[width=7cm]{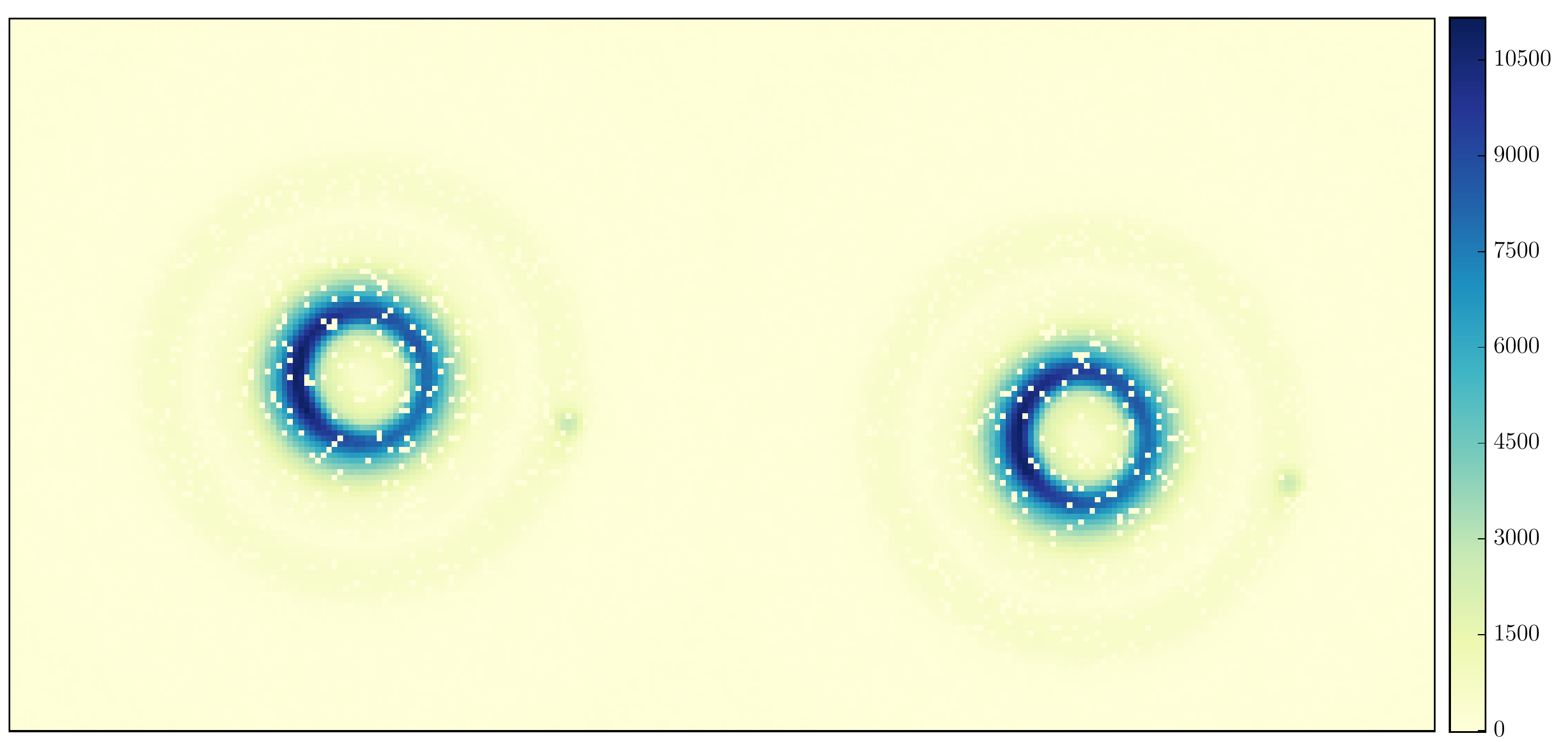} };
                \draw (1,10.5) node[right]{\includegraphics[width=7cm]{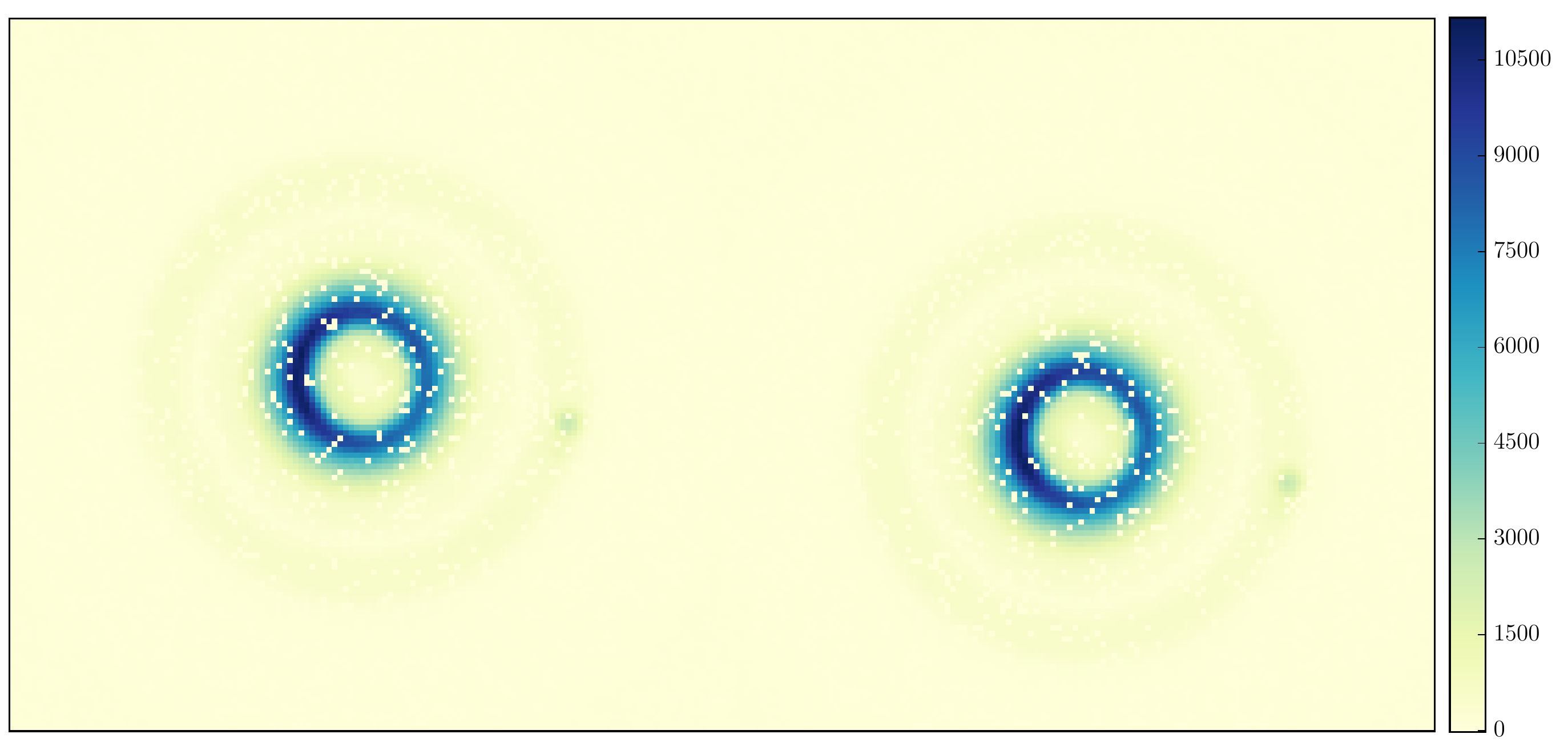} };
                \draw (1.5,10) node[right]{\includegraphics[width=7cm]{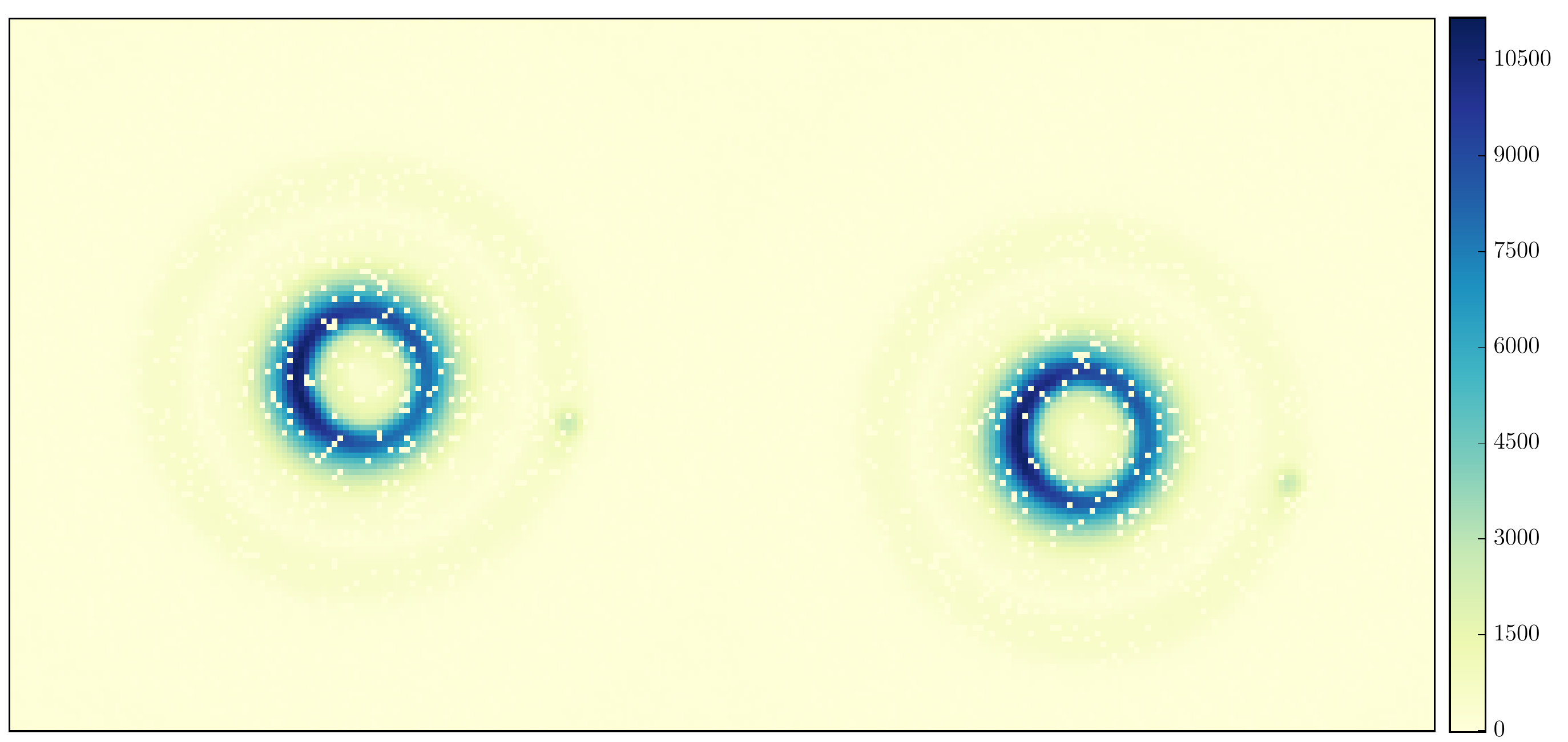} };
                \draw (0.7,9.5) node[rotate=-45]{\Large $\bullet \bullet \bullet$};
                \draw (0.7,12.5) node[rotate=-45]{\Large $\bullet \bullet \bullet$};
                \draw (7.2,12.5) node[rotate=-45]{\Large $\bullet \bullet \bullet$};
                \draw (4, 8) node{$\V{d}$};
                
                \draw (16,11.5) node[left]{\includegraphics[width=7cm]{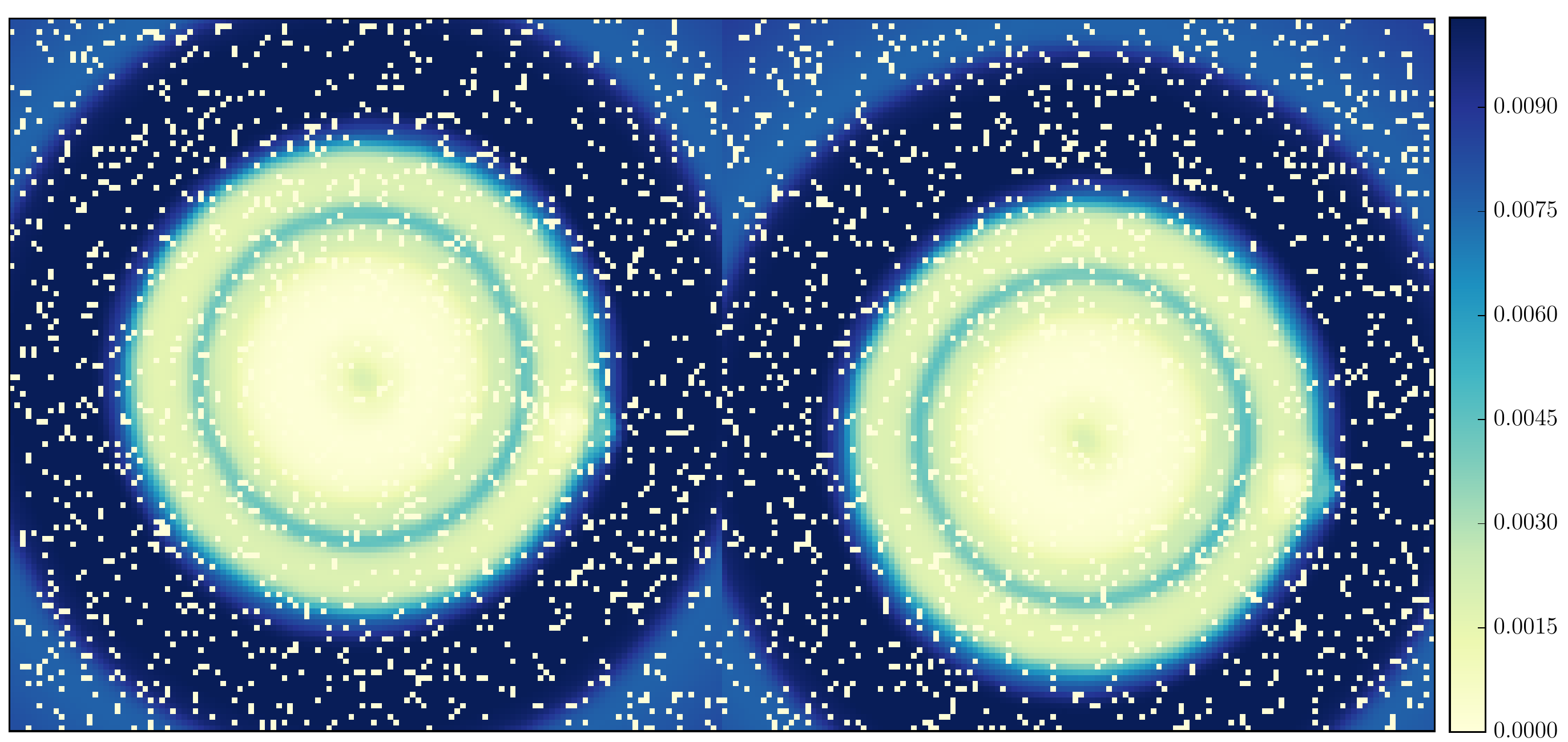} };
                \draw (17,10.5) node[left]{\includegraphics[width=7cm]{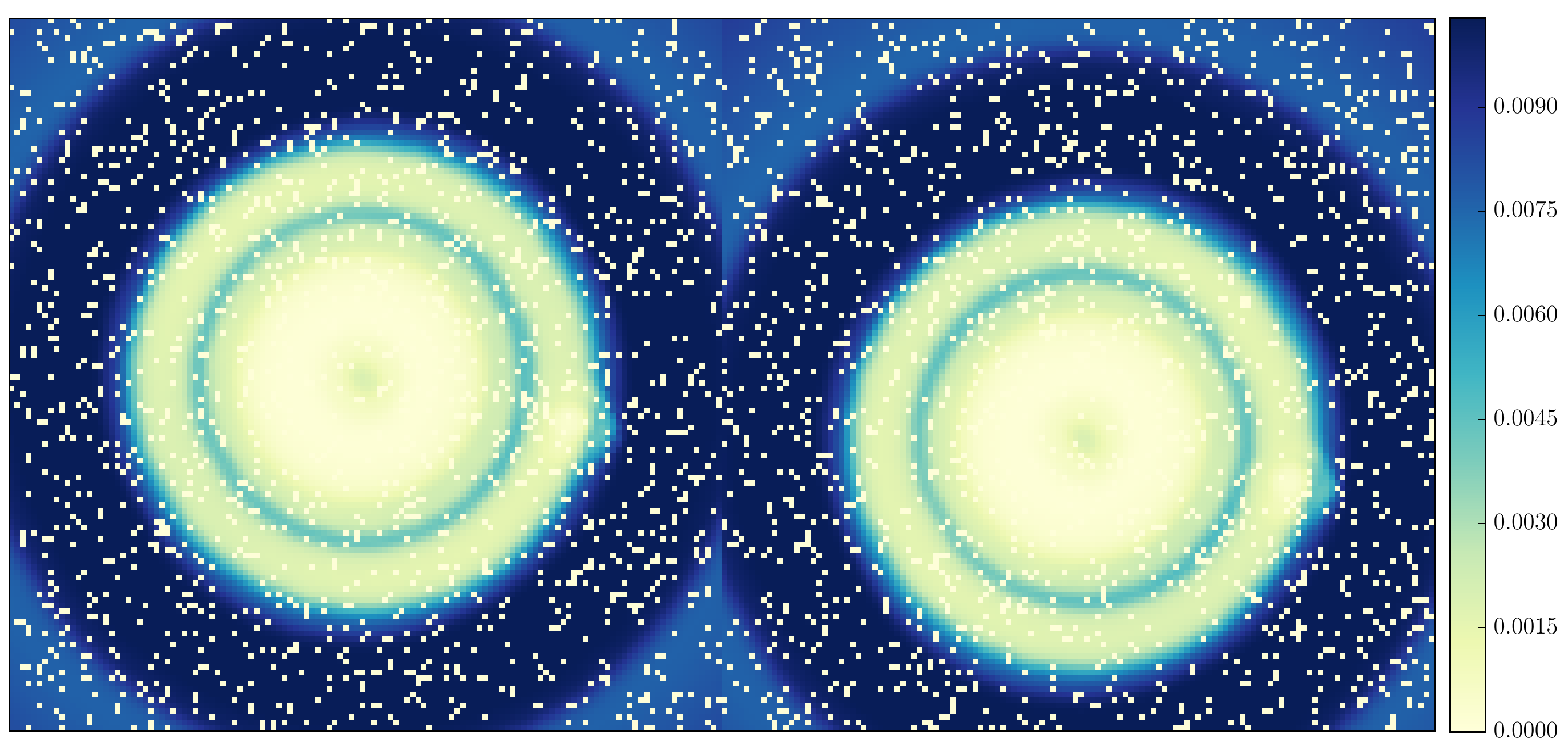} };
                \draw (17.5,10) node[left]{\includegraphics[width=7cm]{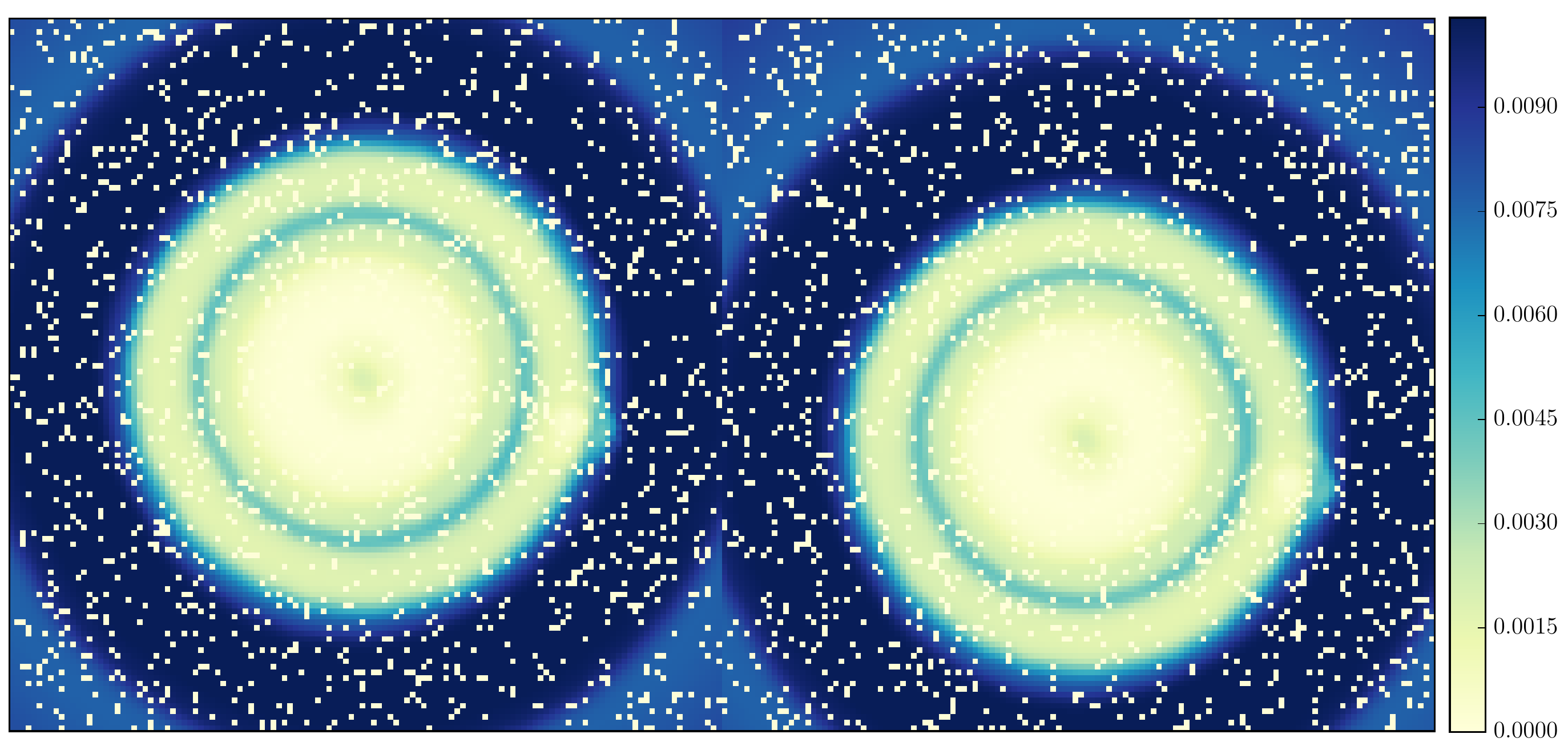} };
                \draw (9.5,9.5) node[rotate=-45]{\Large $\bullet \bullet \bullet$};
                \draw (9.5,12.5) node[rotate=-45, white]{\Large $\bullet \bullet \bullet$};
                \draw (16,12.5) node[rotate=-45]{\Large $\bullet \bullet \bullet$};
                \draw (13.5, 8) node{$\M{W}$};
                
                \draw[->, thick] (4,7.7) -- (4,4.4) node[midway, above, sloped]{Pre-processing};
                
                \draw (4, 4) node[rotate=-90]{\resizebox{0.5cm}{6cm}{$\{$ }};
                \draw (4, -1) node{$\V{d}$};
                
                \draw (0,2.5) node[right]{\includegraphics[width=3cm]{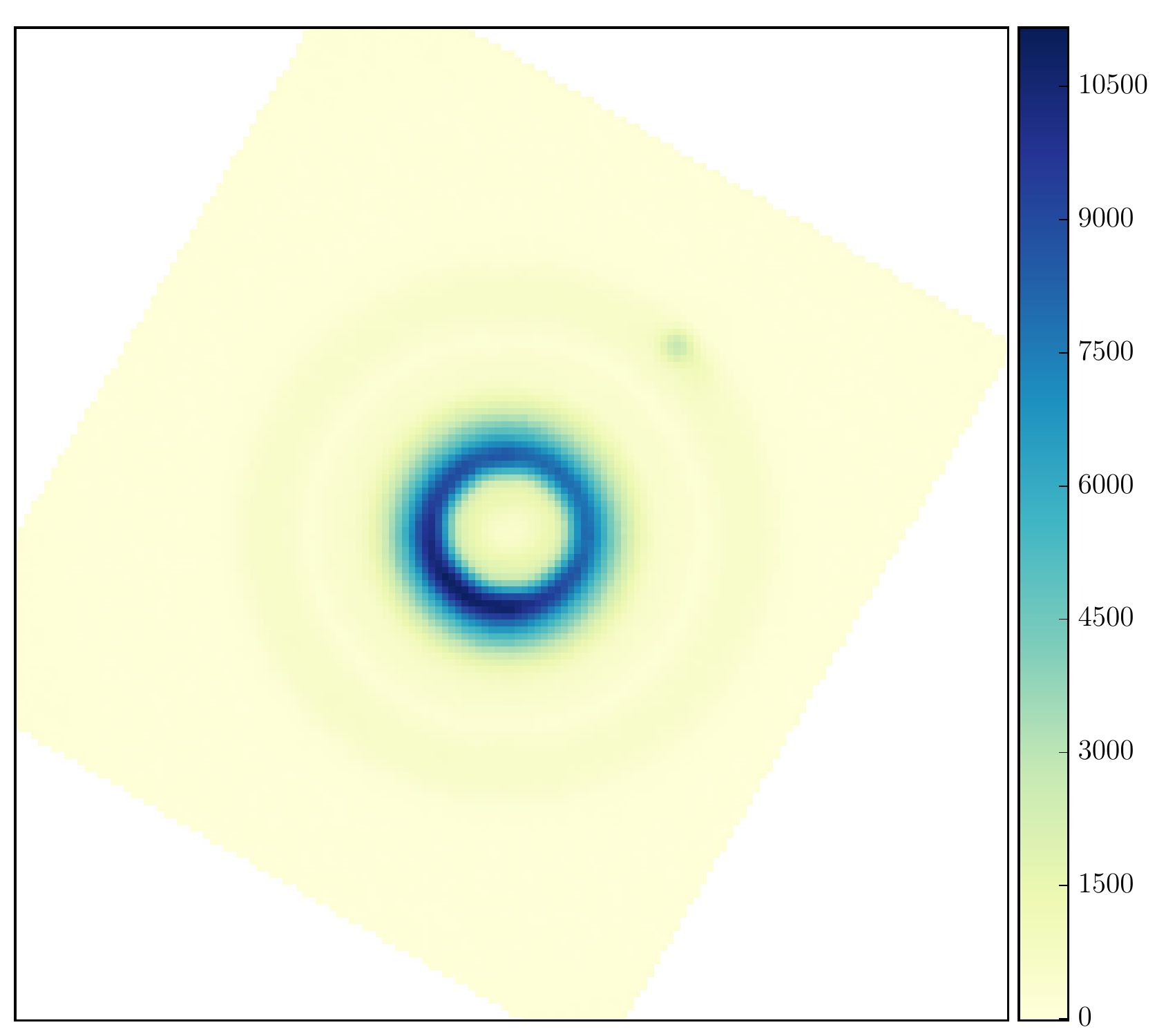} };
                \draw (3.5,2.5) node[right]{\includegraphics[width=3cm]{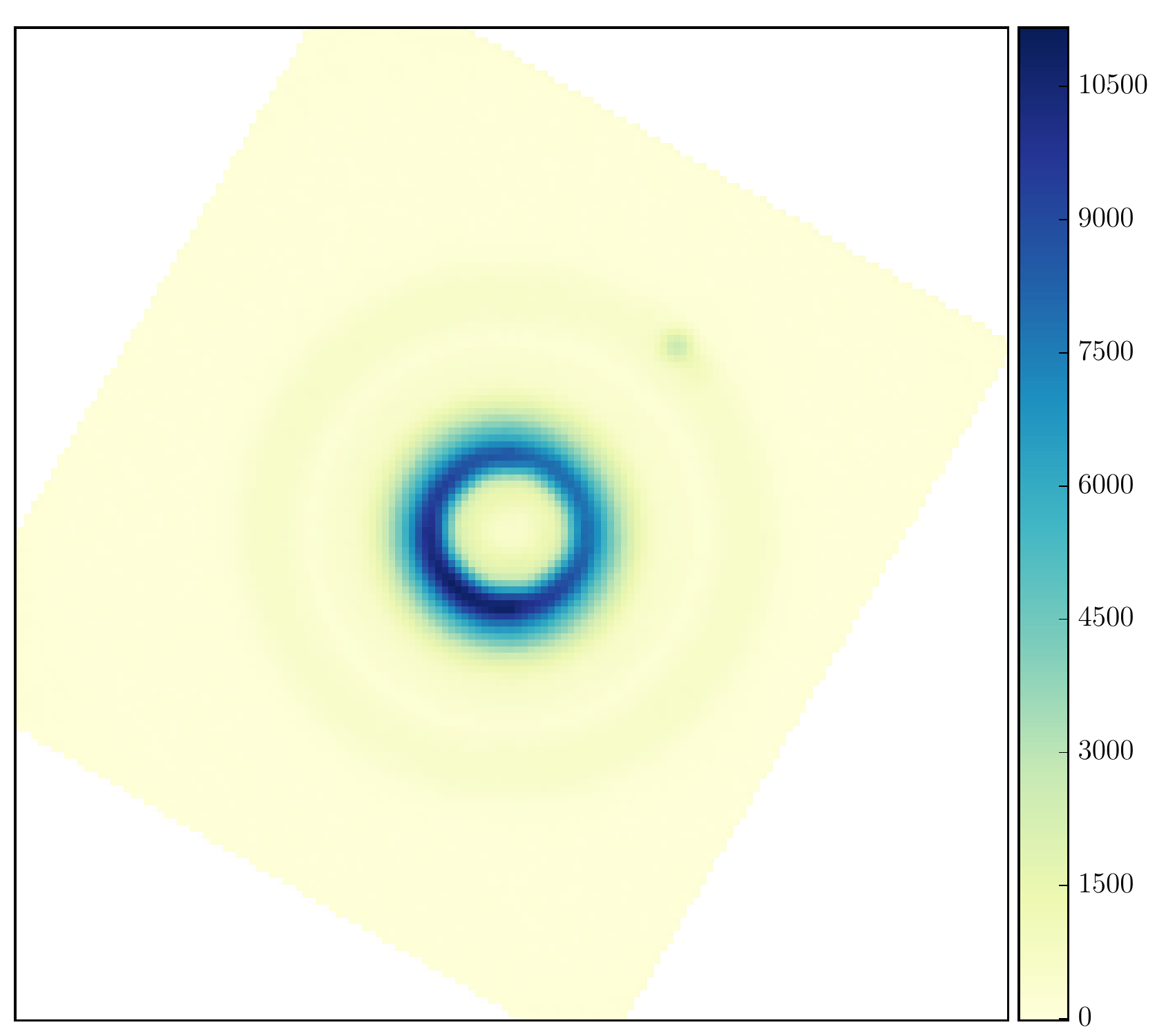} };
                \draw (1,1.5) node[right]{\includegraphics[width=3cm]{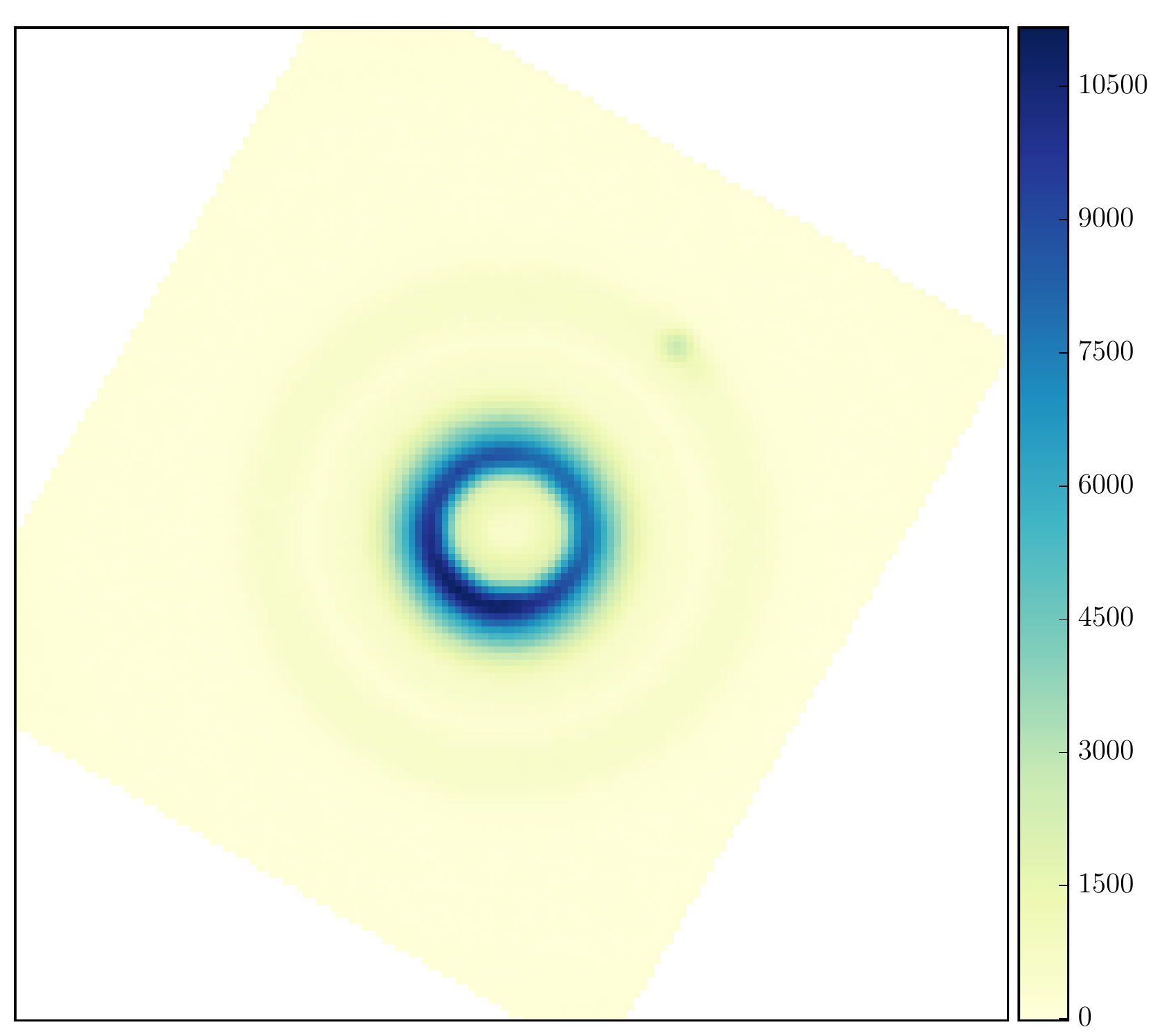} };
                \draw (4.5,1.5) node[right]{\includegraphics[width=3cm]{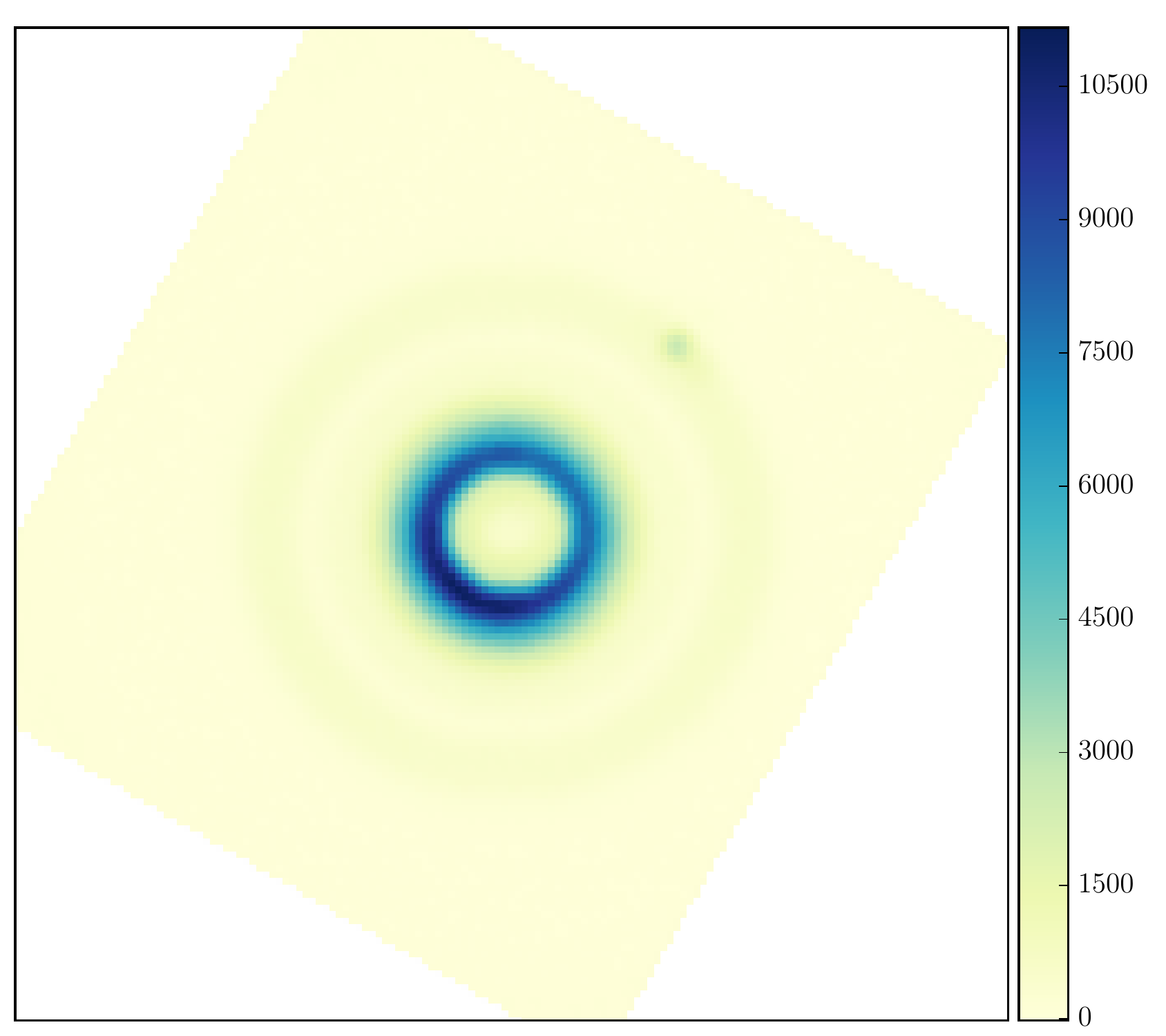} };
                \draw (1.5,1) node[right]{\includegraphics[width=3cm]{CD_1_1_128.pdf} };
                \draw (5,1) node[right]{\includegraphics[width=3cm]{CD_1_2_128.pdf} };
                \draw (0.7,0.7) node[rotate=-45]{\large $\bullet \bullet \bullet$};
                \draw (0.7,3.2) node[rotate=-45]{\large $\bullet \bullet \bullet$};
                \draw (6.7,3.2) node[rotate=-45]{\large $\bullet \bullet \bullet$};
                
                \draw[->, thick] (13.5,7.7) -- (13.5,4.4) node[midway, above, sloped]{Pre-processing};
                
                \draw (13.5, 4) node[rotate=-90]{\resizebox{0.5cm}{6cm}{$\{$ }};
                \draw (13.5, -1) node{$\M{W}$};

                \draw (12.5,2.5) node[left]{\includegraphics[width=3cm]{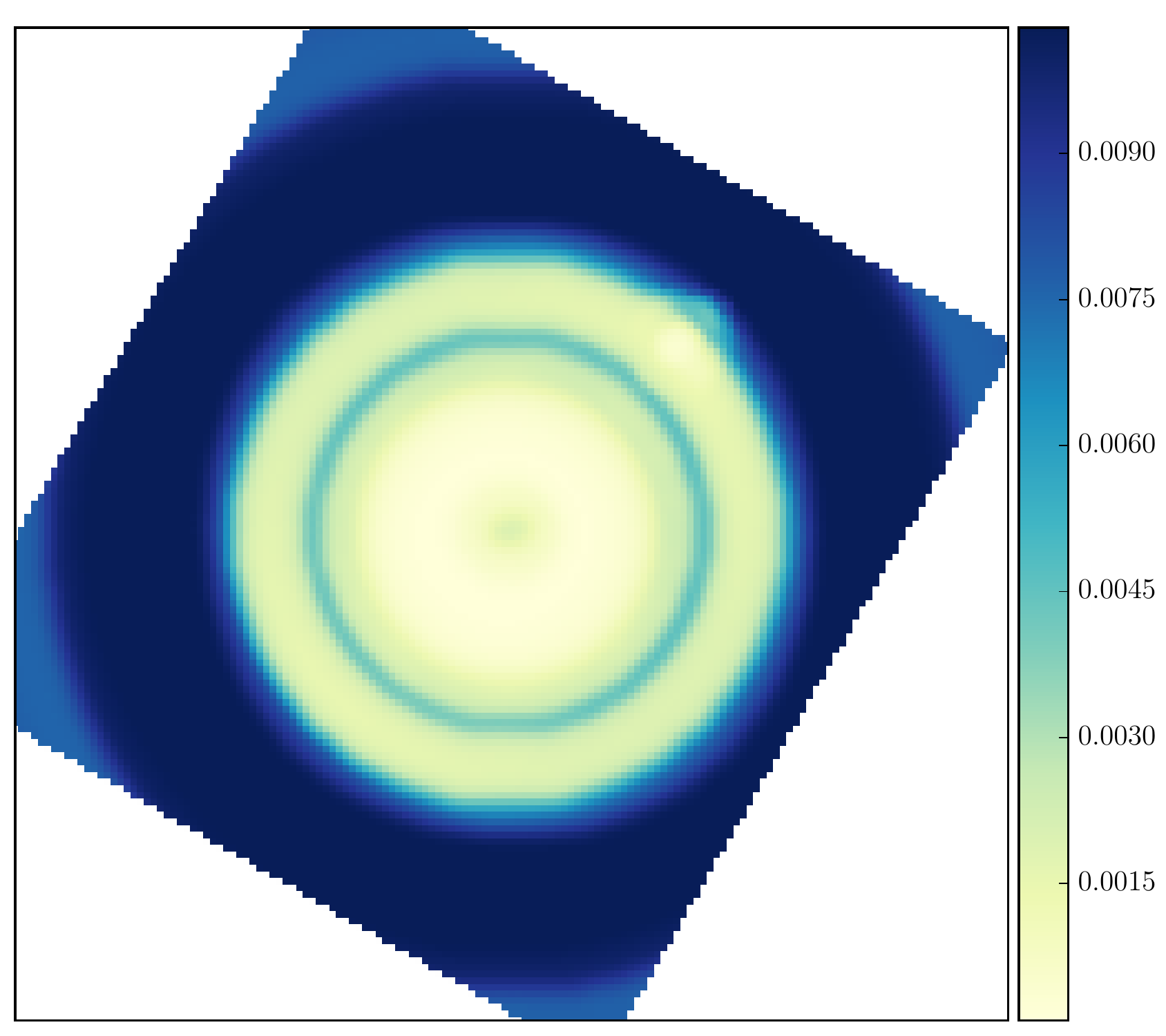} };
                \draw (16,2.5) node[left]{\includegraphics[width=3cm]{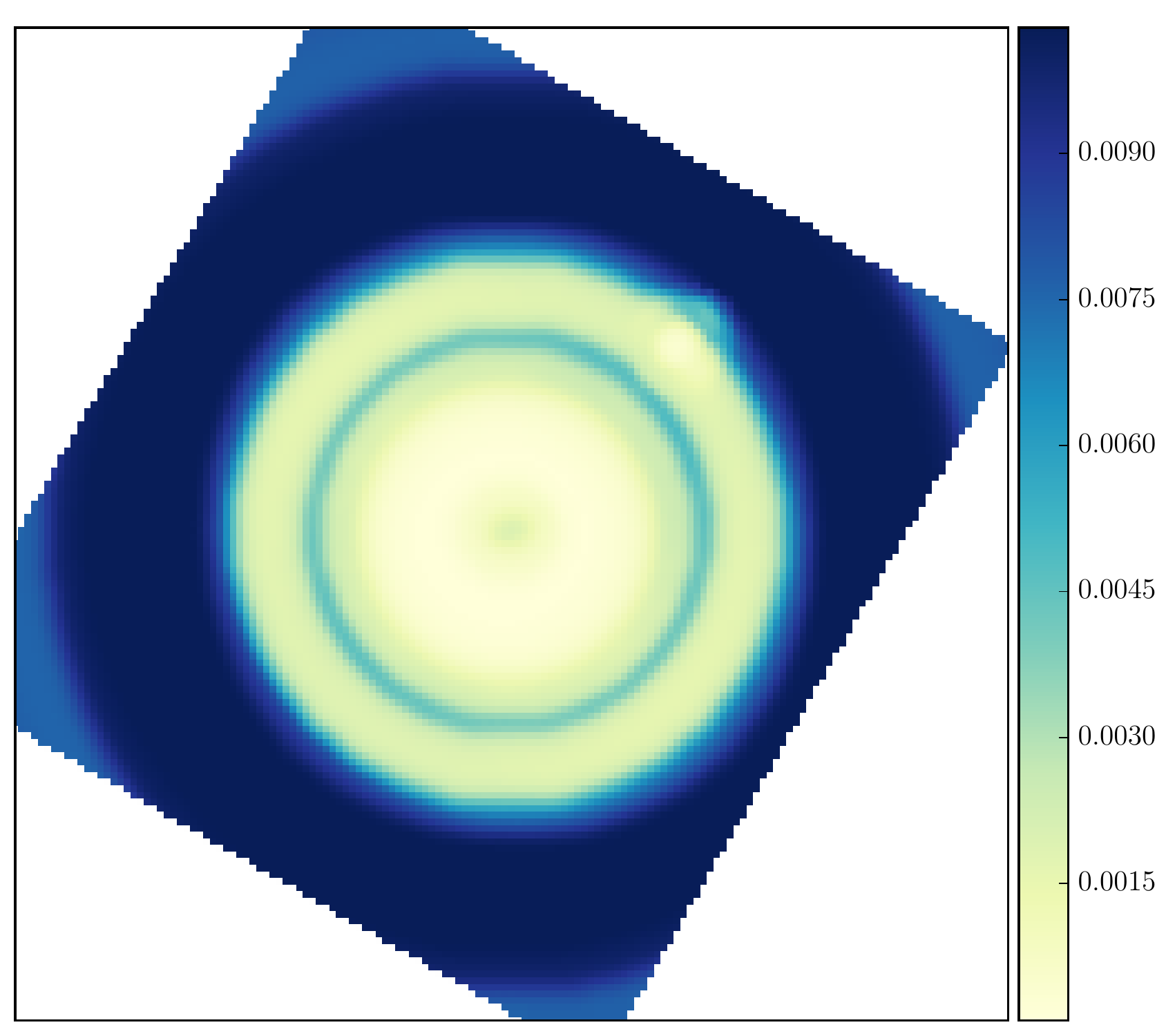} };
                \draw (13.5,1.5) node[left]{\includegraphics[width=3cm]{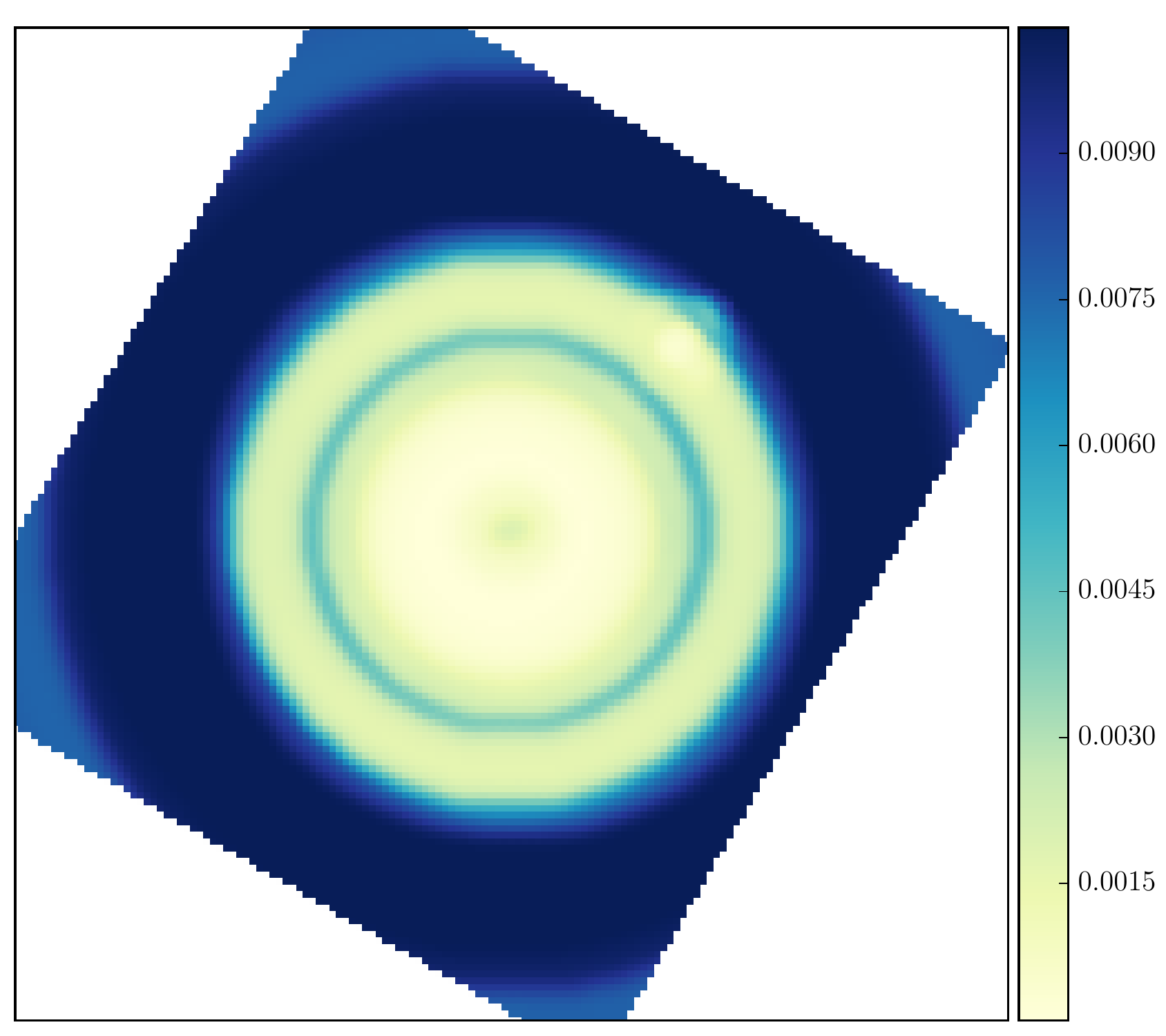} };
                \draw (17,1.5) node[left]{\includegraphics[width=3cm]{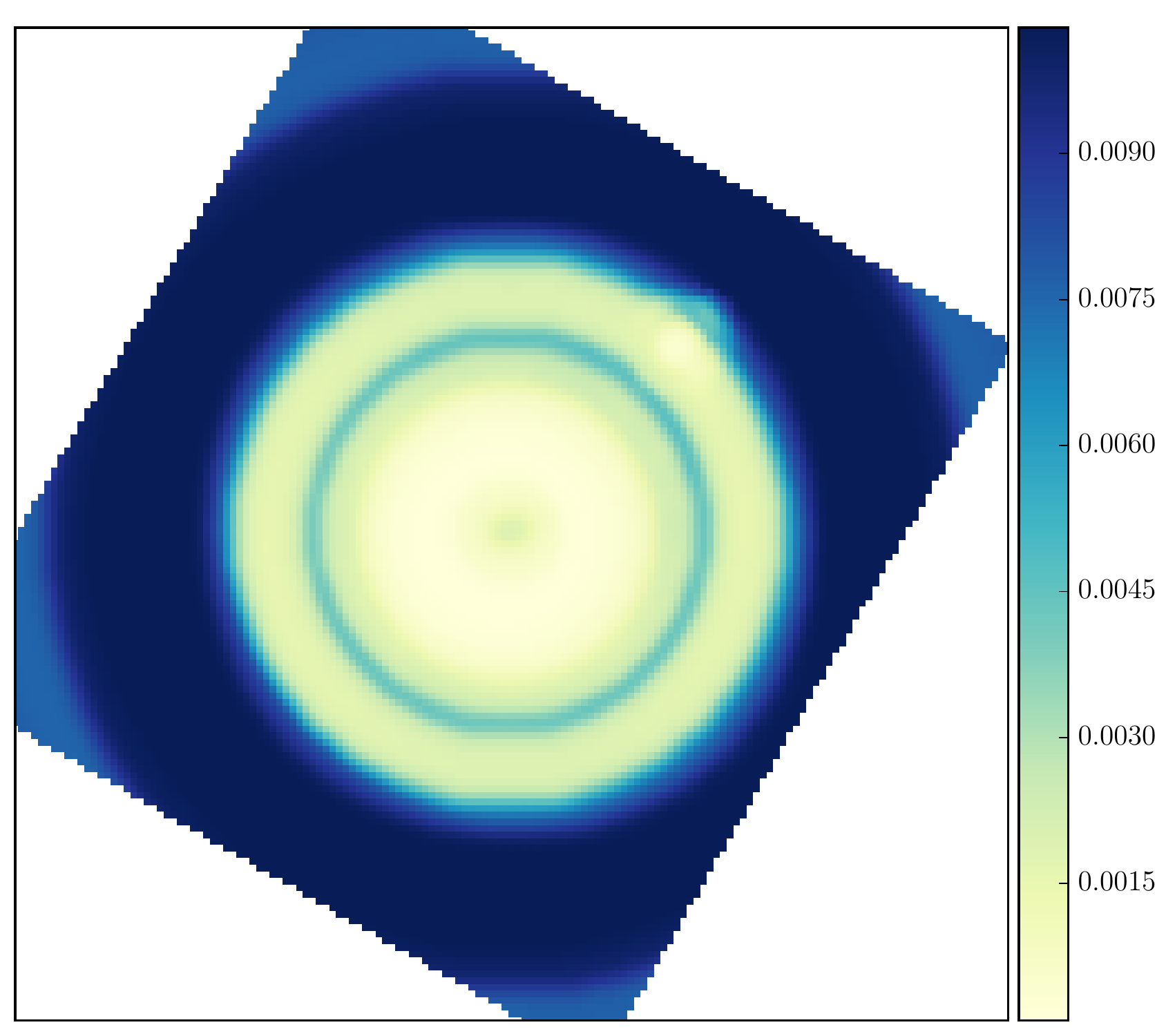} };
                \draw (14,1) node[left]{\includegraphics[width=3cm]{CW_1_1_128.pdf} };
                \draw (17.5,1) node[left]{\includegraphics[width=3cm]{CW_1_2_128.pdf} };
                \draw (10,0.7) node[rotate=-45]{\large $\bullet \bullet \bullet$};
                \draw (10,3.2) node[rotate=-45]{\large $\bullet \bullet \bullet$};
                \draw (16,3.2) node[rotate=-45]{\large $\bullet \bullet \bullet$};
                
                \end{scope}
        \end{tikzpicture}
    }
    \caption{\LD{Schematic describing the process of the data simulation. Starting from synthetic maps $\Tr{\Iu}$, $\Tr{\Ip}$, and $\Tr{\theta}$ based on a disc model, we generate artificial calibrated and pre-processed datasets $\V{d}$ and associated weights $\M{W}$, illustrated here for $\tau^\Tag{disk}=10\%$.}}
    \label{fig:simulateddata}
\end{figure*}

The ${\Iu}^\Tag{disk}$ component is mixed with the unpolarized
${\Iu}^\Tag{star}$ stellar components and a point source companion (star close
to the host star of different brightness). The unpolaized intensity is represented as $\Iu=
{\Iu}^\Tag{star} + {\Iu}^\Tag{disk}$. It is important to keep in mind that
unmixing both disk and stellar unpolarized components is not possible from DPI
data without the diversity introduced by ADI. The $\tau^\Tag{disk}$ value used
to synthetise datasets is thus inaccessible in practice from observational
polarimetric datasets. To assert the case difficulty, one can then use the
total polarization, given for all pixel $n \in \{1, \ldots, N\}$ by:
\begin{equation}
	\tau^\Tag{total}(\V{x}_n)=\frac{\Ip_n}{\Iu_n + \Ip_n },
	\label{eq:tautot}
\end{equation} 
and the Signal-to-Noise Ratio (SNR), given for all pixel $n \in \{1, \ldots,
N\}$  by:
\begin{equation}
	\textrm{SNR}(\V{x}_n)=\frac{\sqrt{K_\alpha}\Ip_n}{\sqrt{(\Iu_n + \Ip_n)/2 + \sro^2}},
	\label{eq:SNR}
\end{equation} 
where $\sro^2$ is the read-out noise variance. The difference between
$\tau^\Tag{total}$ and $\tau^\Tag{disk}$ is that the last one does not take \LDrevised{into}
account the unpolarized star residuals. The figure \ref{fig:SNR_TAU} present
the SNR maps and the maps of total polarization ratio of the synthetic
parameters generated for different $\tau^\Tag{disk}$. At the center, where the
unpolarized star residual are the brightest, the SNR and the total polarization
ratio are the weakest especially in the case of small ${\Iu}^\Tag{disk}$. Yet
the SNR grows with the separation from the star center (like the stellar
contribution or when the polarized contribution of the disk increases.
\LD{Fig.~\ref{fig:simulateddata} present the true simulated maps for $\tau^\Tag{disk}=10\%$.}

Finally, to generate synthetic calibrated data, the Stokes maps are combined
following the expression of the data physical model
\eqref{eq:data-approx-model}, with $K \times M$ noise realizations from a fixed
random seed. \LDrevised{The values of $10\%$ of the pixels, choosen randomly, are replaced 
by zeros to mimic bad pixel behaviours.}
The weights related to each acquisition are simulated at the same time following
\eqref{eq:weights}. The datasets are composed of \LDrevised{$8$} HWP
cycles, with \LDrevised{$2$} acquisitions per positions in each
cycles, \LDrevised{leading to $16$} images per position of Half-Wave Plate (HWP). 
\LDrevised{Thus, the total number of images per datasets is $K=64$.}
 
Several datasets are created with $\tau^\Tag{disk} \in \{3\%, 7\%, 10\%, 15\%,
25\% \}$, corresponding to difficult cases for $\tau^\Tag{disk}< 10\%$ and less
difficult brighter cases above this threshold.

Before producing $K \times 2N$ noise realization on each dataset, the random
seed is reset to the same value. This allow the reproducibility of the
results. In fact this realization is obtained by the multiplication of the
standard deviation of the pixel to a gaussian, centered and reduced gaussian
realization. Since the random seed is the same for each dataset, the
realization is the same for the given pixel, only the standard deviation
changes.

In order to compare the results of the Double Difference to the results of the
RHAPSODIE methods the dataset are pre-processed. The bad pixels are
interpolated; then the left and right part of the images are cut, recentered
and rotated.

\end{appendix}

\end{document}